\documentclass[sigconf]{acmart}
\usepackage{amsmath,amsfonts,amsthm}
\usepackage{graphicx}
\usepackage{textcomp}
\usepackage{xcolor}
\def\BibTeX{{\rm B\kern-.05em{\sc i\kern-.025em b}\kern-.08em
    T\kern-.1667em\lower.7ex\hbox{E}\kern-.125emX}}
    
\usepackage[utf8]{inputenc}
\usepackage{graphicx}
\usepackage{dutchcal}
\usepackage[mathscr]{eucal}
\usepackage{commath}
\usepackage{latexsym}
\usepackage{setspace}
\usepackage[english]{babel}
\usepackage[noend]{algpseudocode}
\usepackage{algorithmicx}
\usepackage{booktabs}
\usepackage{algorithm}
\usepackage[utf8]{inputenc}
\usepackage{nicefrac}
\usepackage{mathtools}
\usepackage{centernot}
\usepackage{makecell}
\usepackage{tabularx}
\usepackage{xcolor}
\usepackage{mathtools}
\usepackage{soul}
\usepackage{svg}
\usepackage{hyperref}
\usepackage{comment}
\usepackage{dsfont}
\usepackage{booktabs}
\usepackage{listings}
\usepackage{enumitem}

\DeclareMathAlphabet{\mathcal}{OMS}{cmsy}{m}{n}
\SetMathAlphabet{\mathcal}{bold}{OMS}{cmsy}{b}{n}
\begin{document}

\newcommand{\bigO}{\mathcal{O}}
\newcommand{\vect}[1]{\boldsymbol{#1}}
\newcommand\normx[1]{\Vert#1\Vert}





\newcommand{\revision}[1]{{#1}}
\newtheorem{comment1}{Reviewer 1's Comment}
\newtheorem{comment2}{Reviewer 2's Comment}
\newtheorem{comment3}{Reviewer 3's Comment}
\newtheorem{mrcomment}{Meta-Reviewer's Comment}

\setlist{nolistsep}

\newtheorem{thm}{Theorem}[section]
\newtheorem{lm}[thm]{Lemma}
\newtheorem{problem}{Problem}

\newcommand{\TriComment}[1]{\hfill$\triangleright$\textit{#1}}

\algrenewcommand\algorithmicrequire{\textbf{Input:}}
\algrenewcommand\algorithmicensure{\textbf{Output:}}

\title{NeuroSketch: Fast and Approximate Evaluation of Range Aggregate Queries with Neural Networks}

\author{Sepanta Zeighami}
\affiliation{\institution{University of Southern California}}
\email{zeighami@usc.edu}

\author{Cyrus Shahabi}
\affiliation{\institution{University of Southern California}}
\email{shahabi@usc.edu}

\author{Vatsal Sharan}
\affiliation{\institution{University of Southern California}}
\email{vsharan@usc.edu}

\setlength{\abovedisplayskip}{0pt}
\setlength{\belowdisplayskip}{0pt}
\setlength{\abovedisplayshortskip}{0pt}
\setlength{\belowdisplayshortskip}{0pt}

\setlength{\textfloatsep}{0pt}
\setlength{\floatsep}{0pt}
\setlength{\intextsep}{0pt}
\setlength{\dbltextfloatsep}{0pt}
\setlength{\dblfloatsep}{0pt}
\setlength{\abovecaptionskip}{0pt}
\setlength{\belowcaptionskip}{0pt}

\begin{abstract}
    Range aggregate queries (RAQs) are an integral part of many real-world applications, where, often, fast and approximate answers for the queries are desired. Recent work has studied answering RAQs using machine learning (ML) models, where a model of the data is learned to answer the queries. However, there is no theoretical understanding of why and when the ML based approaches perform well.  Furthermore, since the ML approaches model the data, they fail to capitalize on any query specific information to improve performance in practice. In this paper, we focus on modeling ``queries'' rather than data and train neural networks to learn the query answers. This change of focus allows us to theoretically study our ML approach to provide a distribution and query dependent error bound for neural networks when answering RAQs. We confirm our theoretical results by developing NeuroSketch, a neural network framework to answer RAQs in practice. Extensive experimental study on real-world, TPC-benchmark and synthetic datasets show that NeuroSketch answers RAQs multiple orders of magnitude faster than state-of-the-art and with better accuracy.
\end{abstract}

\maketitle

\pagestyle{plain}

\section{Introduction}
Range aggregate queries (RAQs) are intrinsic to many real-world applications, e.g., calculating net profit for a period from sales records or average pollution level for different regions for city planing \cite{ma2019dbest}. Due to large volume of data, exact answers can take too long to compute and fast approximate answers may be preferred. In such scenarios, there is a time/space/accuracy trade-off, where algorithms can sacrifice accuracy for time or space. For example, consider a geospatial database containing latitude and longitude of location signals of individuals and, for each location signal, the duration the individual stayed in that location. A potential RAQ on this database, useful for understanding the popularity of different Points of Interests, is to calculate the average time spent by users in an area. Approximate answers within a few minutes of the exact answer can be acceptable in such applications. We use this scenario as our running example.

Research on RAQs has focused on improving the time/space/accuracy trade-offs. Various methods such as histograms, wavelets and data sketches (see \cite{graham2012synopses} for a survey) have been proposed to model the data for this purpose. Recent efforts use machine learning (ML) \cite{ma2019dbest, thirumuruganathan2019approximate, hilprecht2019deepdb} to improve the performance. Such approaches learn models of the data to answer RAQs. Experimental results show ML-based methods outperform non-learning methods in practice.

Nonetheless, there is no theoretical understanding of when and why an ML based approach performs well. This is because modeling data makes it difficult to reason about the performance of specific queries. That is, some queries may be easier to answer than others, e.g., average value of one attribute may be constant for different query ranges, while that of another attribute might change drastically. 
Furthermore, modelling the data misses the opportunity to utilize information about queries in practice. For instance, patterns in query answers can be used to learn a compact representation of the data \textit{with respect to the queries}, improving the performance, while there may be no such patterns within the entire dataset.

In this paper, instead of learning \textit{data models}, we propose to learn \textit{query models}. 
In our example of calculating the average visit duration for a POI, the input to a query model is the POI location and the model is trained to output the average visit duration for the POI. Query modeling skips learning explicitly the data distribution and instead learns query answers, so that we can explicitly relate errors in modeling to errors in query answering. Nevertheless, this is non-trivial and requires a detailed study of modelling errors. To the best of our knowledge, no existing attempt in the literature theoretically relates data and query properties to the error of a learned model when answering RAQs.

We utilize neural networks as our query model. Specifically, we consider training a neural network that takes as input an RAQ and outputs the answer to the query. We theoretically study this approach, and provide, for the first time, a \textit{Data distribution and Query Dependent error bound} (hereafter referred to as DQD bound) for neural networks when answering RAQs. DQD bound theoretically relates properties of the data distribution and the RAQ to the accuracy a neural network can achieve when answering the query. 

In our theoretical analysis, we consider \texttt{AVG}, \texttt{COUNT} and \texttt{SUM} queries, assume the database is a collection of i.i.d samples from a data distribution and make a suitable Lipschitz assumption on the query and data distribution. We then use VC-sampling theory and our novel result on neural network approximation power to show the existence of a neural network that can answer the queries on the database with bounded error. The bound gets tighter (i.e., more accurate neural networks can be learned) as the data size, or query range, increases. Alternatively, a smaller neural network can be used to answer queries with a fixed desired error when the data size, or query range, increases. Intuitively, this is a result of the reduction in variance (due to sampling) of query answers when the database is larger, because more data points are sampled from the data distribution. Furthermore, our results utilize the Lipschitz property to provide a complexity measure that quantifies the difficulty of answering a query from a data distribution. Using the complexity measure, our results show settings where existence of a small neural network with low query answering error is guaranteed.

To confirm our theoretical results, we design \textit{NeuroSketch}, a neural network framework that answers RAQs orders of magnitude faster than state-of-the-art and with better accuracy. NeuroSketch uses DQD results to allocate more model capacity to queries that are difficult to answer, thereby reducing error without increasing query time. While DQD provides a theoretical grounding for NeuroSketch, in practice NeuroSketch is not limited to some of the assumptions we made to prove DQD bounds, for example, it can answer more general RAQs, such as \texttt{STD} and \texttt{MEDIAN}.


To summarize, our major contributions are:
\begin{itemize}
    \item We present the first theoretical analysis for using ML to answer RAQs. This includes a novel analysis framework, a novel use of VC-sampling theory and a novel result on neural network approximation power.
    \item We show theoretically how data distribution, data size, query range and aggregation function are related to the neural network error when answering RAQs. This opens the possibility for a query optimizer that, for a data distribution, decides when to build and use a neural network for query processing.
    \item To confirm our theoretical results, we design {\em NeuroSketch}, the first neural network framework to answer generic RAQs.
    \item Extensive experiments show that NeuroSketch provides orders of magnitude gain in query time and better accuracy over state-of-the-art, (DBEst \cite{ma2019dbest} and DeepDB \cite{hilprecht2019deepdb}) on real-world, TPC-benchmark and synthetic datasets.
\end{itemize}

We present our problem definition in Sec.~\ref{sec:def}, DQD bound in Sec.~\ref{sec:theory}, NeuroSketch in Sec.~\ref{sec:practice}, our empirical study in Sec.~\ref{sec:exp}, related work in Sec.~\ref{sec:rel_work} and conclude in Sec.~\ref{sec:conc}.
\vspace{-0.3cm}
\section{Problem Definition}\label{sec:def}
\textbf{Problem Setting}. Consider a dataset $D$ with $n$ records and $\bar{d}$ attributes, $A_1$, ..., $A_{\bar{d}}$. Assume records of $D$ are random i.i.d samples from a data distribution $\chi$ and $A_i\in [0, 1]$ with probability 1 for all $1\leq i\leq \bar{d}$ (otherwise the attributes can be normalized). We first consider the following SQL query and discuss extensions to general RAQs in Sec.\ref{sec:general}. 

\begin{tabular}{c}
\centering
\begin{lstlisting}
SELECT $\texttt{AGG}$($A_m$) FROM $D$ WHERE
$c_1\leq A_1< c_1+r_1$ AND ... AND $c_{\bar{d}} \leq A_{\bar{d}}  < c_{\bar{d}}+r_{\bar{d}} $
\end{lstlisting}
\end{tabular}

For any $i$, $c_i$ and $c_i+r_i$ are the lower and upper bounds on the attribute $A_i$. $c_i$ and $r_i$ can be $0$ and $1$, respectively, in which case there are no restrictions on the values of $A_i$ in the query. We say that an attribute is \textit{not active} in the query in that case, and is \textit{active} otherwise. \texttt{AGG} is a user defined \textit{aggregation function}, with examples including \texttt{SUM}, \texttt{AVG} and \texttt{COUNT} aggregation functions. $A_m$ is called the \textit{measure attribute}, where $m$ is an integer between 1 and $\bar{d}$. Let $\mathbf{c}=(c_1, ..., c_{\bar{d}})$ and $\mathbf{r}=(r_1, ..., r_{\bar{d}})$ be $\bar{d}$-dimensional vectors. We call the pair $\mathbf{q}=(\mathbf{c}, \mathbf{r})$ a \textit{query instance}. Different query instances correspond to different range predicates for the measure attribute $A_m$ and aggregation function \texttt{AGG}. We define the function $f_D(.)$ so that for a query $\mathbf{q}$, $f_D(\mathbf{q})$ is the answer to the above SQL statement. We call $f_D:[0, 1]^{d}\rightarrow\mathbb{R}$ a \textit{query function}, where $d=2\bar{d}$ is the dimensionality of the query function. Furthermore, we define $\mathcal{Q}=\{(\mathbf{c}, \mathbf{r})\in [0, 1]^{d}, c_i+r_i\leq 1\forall i\}$ as the set of all possible queries. 

\begin{figure}
    \centering
    \begin{minipage}{0.49\columnwidth}
    \centering
    \includegraphics[width=0.9\textwidth]{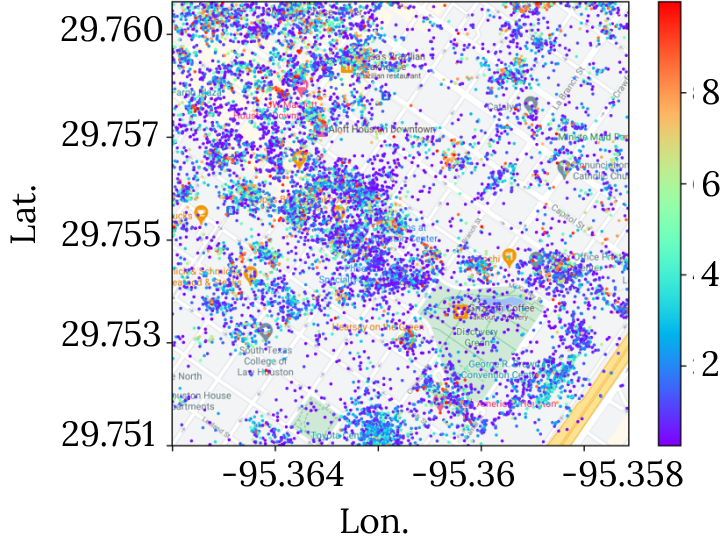}
    \end{minipage}
    \begin{minipage}{0.49\columnwidth}
    \centering
    \includegraphics[width=0.9\textwidth]{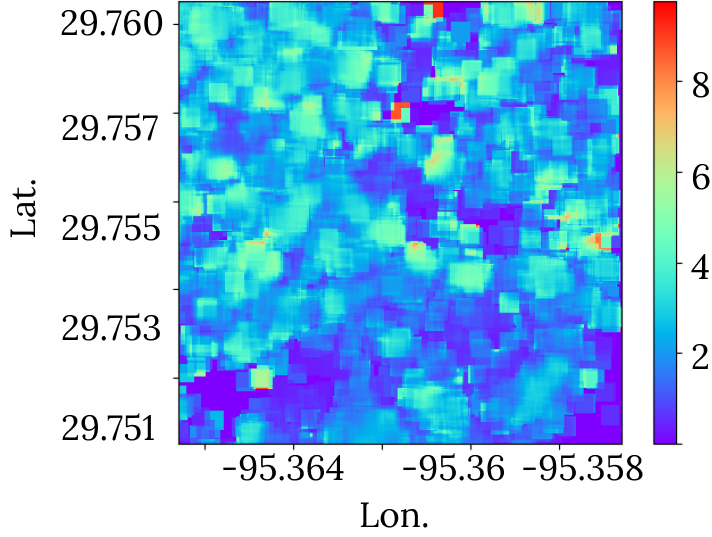}
    \end{minipage}
    \caption{(left) Database of location signals. (right) Avg. visit duration query function. Color shows visit duration in hours}
    \label{fig:visit_query_function}
\end{figure}

\begin{example}\label{ex:visit}
Consider a database of user location reports and the duration a user stayed in the reported location, shown in  Fig.~\ref{fig:visit_query_function} (left). On this database, consider the RAQ of returning average visit duration of users in a 50m$\times$50m rectangle with bottom left corner at the geo-coordinate $(c_1, c_2)$. The query function, $f_D(c_1, c_2):=f_D(c_1, c_2, 50m, 50m)$, takes as input the geo-coordinate of the rectangle and outputs the average visit duration of data points in the rectangle (we have fixed $r_1$ and $r_2$ to 50m in this example). Fig.~\ref{fig:visit_query_function} (right) plots $f_D(c_1, c_2)$, which shows, $f_D(-95.3615, 29.758, 50m, 50m)=9$, i.e., for query instance (-95.3615, 29.758, 50m, 50m) the answer is 9. 
\end{example}

\vspace{-0.2cm}

\textbf{Neural Networks to Answer RAQs} We learn a neural network, $\hat{f}(.; \theta)$, to approximate the query function, $f_D(.)$. The neural network takes as input an RAQ, $\mathbf{q}$. The model forward pass outputs an answer, $\hat{f}(\mathbf{q}; \theta)$. The goal is to train a neural network so that its answer to the query, $\hat{f}(\mathbf{q}; \theta)$, is similar to the ground-truth, $f_D(\mathbf{q})$. If such a neural network is small and can be evaluated fast, we can use the neural network to directly answer the RAQ efficiently and accurately, by performing a forward pass of the model. 

\textbf{Problem Statement}. Let $\Sigma(\hat{f})$ be the space complexity of the neural network, which is the amount of space required to store all its parameters. Let $\tau(\hat{f})$ be its query time complexity, which is the time it takes to perform a forward pass of the neural network. We study the error, $\lVert f_D-\hat{f}\rVert$, in answering queries, where we mostly consider the 1-norm, defined as $\lVert f_D-\hat{f}\rVert_1=\int_{\mathbf{q}\in \mathcal{Q}}|f_D(\mathbf{q})-\hat{f}(\mathbf{q})|$ or $\infty$-norm, defined as $\lVert f_D-\hat{f}\rVert_{\infty}=\sup_{\mathbf{q}\in \mathcal{Q}}|f_D(\mathbf{q})-\hat{f}(\mathbf{q})|$. The problem studied in this paper is learning to answer range aggregate queries with time and space constraints, formulated as follows.
\begin{problem}\label{prob:RAQ}
Given a query function $f_D$, class of possible neural networks, $\mathcal{F}$, and time and space requirements $t$ and $s$, find 
$$\arg\min_{\hat{f}\in \mathcal{F}} \lVert f_D-\hat{f}\rVert \text{ s.t. }\Sigma(\hat{f})\leq s \text{, } \tau(\hat{f})\leq t.$$
\end{problem}

\textbf{Notation}. Bold face letters, e.g., $\mathbf{c}$, denote vectors, and subscripts denote the elements of a vector, e.g., $c_i$ is the $i$-th element of $\mathbf{c}$.

\vspace{-0.2cm}
\section{DQD Bound for Neural Networks Answering RAQs}\label{sec:theory}
We theoretically study the relationship between the accuracy a neural network can achieve when answering RAQs and data and query properties. Sec.~\ref{sec:dqd}, states our Data distribution and Query Dependent error bound (DQD bound) when considering \texttt{SUM} and \texttt{COUNT} aggregation functions, and discusses its implications. We prove the bound in Sec.~\ref{sec:theory:proof}. We present results for \texttt{AVG} query function in Sec.~\ref{sec:theory:other} and discuss how our techniques can be generalized to other query functions and modelling choices.  

 
\subsection{DQD Bound Statement}\label{sec:dqd}
\subsubsection{Incorporating Data Distribution}\label{sec:dqd:notation} The data distribution, $\chi$, underlying a database, $D$, impacts the difficulty of answering queries on the database with a neural network. For instance, in Example~\ref{ex:visit}, if all users have the same visit duration for all their visits, the query function $f_D(c_1, c_2)$ will be constant, and thus can be easily modeled. On the other hand, the skewness in the data distribution, as depicted in Fig.~\ref{fig:visit_query_function}, can make answering queries more difficult as the query function $f_D(c_1, c_2)$ changes drastically from one location to another. Importantly, this is a property of the data distribution, $\chi$, and not only of the observed database $D$. For instance, we expect similar skewness in observations if we collect more user data (i.e., as $D$ grows), or if location data are collected from a different period of time not covered in $D$ (i.e., a different sample of $\chi$). Thus, by incorporating data distribution in our analysis, we are able to study the impact of data size as well as properties intrinsic to the distribution (that will be unaffected by the randomness in observations) on answering RAQs. To do so, (1) we need to capture the dependence of query answers on data distribution and (2) find a means of measuring the complexity of modeling query answers when data follows a certain distribution.

To capture the dependence on data distribution, we define \textit{distribution query function}, $f_\chi(\mathbf{q})$, as the expected value of the query function, i.e., $f_\chi(\mathbf{q})=\mathbb{E}_{D\sim \chi}[f_D(\mathbf{q})]$, where $D$ is sampled from data distribution, $\chi$. We refer to the query function, $f_D(\mathbf{q})$, as \textit{observed query function} to distinguish it from distribution query function. 

To capture the difficulty of modeling a function, we use the $\rho$-Lipschitz property. A function, $f$, is $\rho$-Lipschitz if $|f(\mathbf{x})-f(\mathbf{x}')|\leq \rho\lVert \mathbf{x}-\mathbf{x}'\rVert_1$, for all $\mathbf{x}$ and $\mathbf{x}'$ in the domain of the function, where we consider $\rho$-Lipschitz property in 1-norm. \revision{Intuitively, $\rho$ captures the magnitude of correlation between $\mathbf{x}$ and $f(\mathbf{x})$. It bounds how much $f(\mathbf{x})$ can change with a change in $\mathbf{x}$. If $\rho$ is large, $f$ can change abruptly even with a small change in $\mathbf{x}$. This makes the function more difficult to approximate, as more model parameters will be needed to account for all such possible abrupt changes.}

Combining the above, we propose to use the Lipschitz constant of the normalized Distribution Query function, referred to as \textit{LDQ}, as a measure of difficulty of answering RAQs. LDQ is the Lipschitz constant of the function $f(\mathbf{q})=\frac{f_\chi(\mathbf{q})}{n}=\frac{1}{n}\mathbb{E}_{D\sim \chi}[f_D(\mathbf{q})]$. We normalize the distribution query function by data size to account for its change in magnitude when data size increases (for sum and count queries, magnitude of $f_D(\mathbf{q})$ increases as data size increase). LDQ is a property of $\chi$ and $f_D$. For ease of reference, we often implicitly assume a given data distribution $\chi$ and refer to LDQ as a property of a query function.

\subsubsection{Theorem Statement}\label{sec:dqd:statement}
Let $f^{S}_D$ and $f^{C}_D$ be query functions with aggregation functions \texttt{SUM} and \texttt{COUNT}, respectively, and let $\rho_S$ and $\rho_C$ be their respective LDQs. For $i\in \{S, C\}$, we study the time, space and accuracy of a neural network, $\hat{f}$, when approximating $f_D^i$, as formalized below.

\begin{theorem}[DQD Bound]\label{thm:sum_count_all}
For $i\in \{S, C\}$, there exists a neural network $\hat{f}$ with space and query time complexity $\Tilde{O}(d(\varkappa\rho d \varepsilon_1^{-1}+1)^d)$, where $\Tilde{O}$ hides logarithmic factors, s.t.
\begin{align*}
\mathop{\mathds{P}}_{D\sim\chi}\left[\frac{1}{n}\normx{\hat{f}-f_D^i}_1\geq \varepsilon_1+\varepsilon_2\right]\leq \varkappa_2^{d+1}d\varepsilon_2^{-d}\exp{(-\varkappa_2^{-1}\varepsilon_2^2n)},    
\end{align*}
Where $\varkappa_1$ and $\varkappa_2$ are universal constant.
\end{theorem}

Proof of Theorem~\ref{thm:sum_count_all} is presented in Sec.~\ref{sec:theory:proof}. Here, we discuss the theorem statement and its implications.

\textbf{A Confidence/Error Analysis}. DQD bound relates, with a desired probability (i.e., confidence level), error a neural network can achieve when answering RAQs to its query time and space complexity through data dependent properties. The error is scaled by data size, $n$, to account for the change in the magnitude of query answers when data size changes. Parameter $\varepsilon_1$ allows trading-off accuracy for space or time complexity and $\varepsilon_2$ allows trading-off accuracy for confidence in the bound. The probability is over sampling a database from the data distribution. 
That is, DQD states that, when observing a database $D$ that follows a distribution $\chi$, with high probability, there exists a neural network that can answer RAQs on $D$ and achieve the specified time/space/accuracy trade-off. 

\textbf{Distribution Dependent Complexity Measure}. DQD bound establishes LDQ of the query function as a measure of complexity when answering RAQs with neural networks. It implies that query time will be faster when LDQ is small. LDQ is a property of the data distribution and the query in question. Thus, Theorem~\ref{thm:sum_count_all} allows us to quantify how easy or difficult it is to approximate query answers for a data distribution using a neural network. 
We provide specific examples of LDQ for different data distributions in Sec.~\ref{sec:dqd:dist_dep} and empirically confirm impact of LDQ on query answering in Sec.~\ref{sec:theory_example}. 

\textbf{Faster on Larger Databases}. Let the confidence in the DQD bound be $\delta=\varkappa_2^{d+1}d\varepsilon_2^{-d}\exp{(-\varkappa_2^{-1}\varepsilon_2^2n)}$. Fixing the value of $\delta$, we observe that $n$ and $\varepsilon_2$ are negatively correlated, where increasing data size $n$ leads to reduction in $\varepsilon_2$. That is, for a fixed confidence parameter, the error of a neural network decreases as data size increases. Now let $\varepsilon=\varepsilon_1+\varepsilon_2$ be the total neural network error. Also fixing $\varepsilon$ in addition to $\delta$ but allowing $\varepsilon_1$ to vary, we observe that increase in data size results in smaller query time and space complexity, for a fixed neural network error and confidence level. Thus, DQD bound shows the counter-intuitive result that when answering queries with a neural network query time can be lowered by increasing the database size. We empirically confirm this phenomenon in Sec.~\ref{sec:theory_example}. Intuitively, this happens because when data size is larger the model only needs to learn the patterns in the data distribution, while for smaller databases, the observed database can be different from the data distribution and the model has to memorize all the points, making it more challenging.  

\textbf{Low-Error Cases}. DQD bound shows that a neural network can answer queries fast and accurately if the data size is large and LDQ of a query function is small. Thus, DQD bound shows scenarios when using a neural network can provide good performance and presents a property of data distribution that can guarantee low error for a neural network framework when answering RAQs. Nonetheless, it does not preclude neural networks from performing well in other scenarios, which requires further theoretical investigation. 

\textbf{Achieving Zero Error}. For a fixed and small data size, even if neural network size is allowed to approach infinity, the DQD bound provides a non-zero error bound. That is, letting neural network size go to infinity by reducing $\varepsilon_1$ to zero does not achieve total zero error (we empirically verify this in Sec.~\ref{sec:theory_example}), as the total error in that case will be equal to $\varepsilon_2$ (which can be large depending on $n$). This is because $f_D$ can be discontinuous even though $f_\chi$ is assumed to be Lipschitz continuous, so that no neural network can approximate it exactly. Points of discontinuity can be seen in Fig.~\ref{fig:visit_query_function} (right), where the query answer can suddenly change. Such points of discontinuity happen when the query boundary matches a data point, because in such cases, arbitrarily small changes to the query boundary can change the query answer. As data size increases, $f_D$ behaves more like a continuous function (because $f_\chi$ is Lipschitz continuous), so the achievable error by a neural network goes down. \revision{Note that techniques that create a discontiuous function approximator, e.g., quantizating the query space, can potentially help a neural network achieve zero error, as a large enough neural network can memorize a fininte set of points exactly \cite{yun2019small}. However, our DQD bound is for queries over space of reals (i.e., approximation of infinite set of points), and without input preprocessing or quantization.}



\subsubsection{Impact of Distribution and LDQ}\label{sec:dqd:dist_dep} The model complexity needed to answer RAQs depends on data distribution through LDQ of $f_{D}^S$ and $f_{D}^C$. We provide examples of LDQ for different distributions.
 
\begin{example}\label{example:unif1d}
Let $\chi$ be a 1-dimensional uniform distribution. By definition, we have $f_{\chi}^C(c_1, r_1)=n\mathds{P}_{p\sim \chi}[p\in(c_1, r_1)]$ , where $(c_1, r_1)$ defines a query range (see Sec.~\ref{sec:def}) and $p$ is a data point sampled from $\chi$. $\chi$ is uniform so $\mathds{P}_{p\sim \chi}[p\in(c_1, r_1)]=r_1$. Differentiating and using the definition, $\frac{\partial f_{\chi}^C(c_1, r_1)}{\partial c_1} = 0$ and $\frac{\partial f_{\chi}^C(c_1, r_1)}{\partial r_1} = n$, so that $\frac{1}{n}f_{\chi}^C(c_1, r_1)$ is $\rho$-Lipschitz with $\rho=1$. A similar result also holds for $\frac{1}{n}f_{\chi}^S(c_1, r_1)$. The small Lipschitz constant matches the intuition that uniform distribution is easy to approximate. 
\end{example} 

\begin{example}
Let $\chi$ be a 1-dimensional Gaussian distribution with standard deviation $\sigma$ and $\mu=0$, we have that 
\begin{align*}
    |\frac{\partial \mathds{P}_{p\sim \chi}[p\in(c_1, r_1)]}{\partial c_1}|&=|\frac{1}{\sigma \sqrt{2\pi}}(e^{-\frac{1}{2}(\frac{c_1+r_1}{\sigma})^2}-e^{-\frac{1}{2}(\frac{c_1}{\sigma})^2})|\\
    &\leq \frac{2}{\sigma \sqrt{2\pi}}
\end{align*} 
and that 
$$|\frac{\partial \mathds{P}_{p\sim \chi}[p\in(c_1, r_1)]}{\partial r_1}|=|\frac{1}{\sigma \sqrt{2\pi}}e^{-\frac{1}{2}(\frac{c_1+r_1}{\sigma})^2}|\leq \frac{1}{\sigma \sqrt{2\pi}}$$
so that $\frac{1}{n}f_{\chi}^C(c_1, r_1)$ is $\rho$-Lipschitz with $\rho=\frac{3}{\sigma\sqrt{2\pi}}$. Thus, for smaller $\sigma$ the function becomes more difficult to model, as the neural network has to model a sharp change in the function. 
\end{example}


\subsubsection{Measuring Complexity in Practice}\label{sec:dqd:practice} DQD bound can help decide whether to use neural networks to answer RAQs, or to design complexity aware algorithms for practical use-cases (as we do in Sec.~\ref{sec:practice}). Such use-cases require measuring LDQ, which can be difficult in practice. For two queries $\mathbf{q}$ and $\mathbf{q}'$, the Lipschitz constant bounds the \textit{maximum} change in the function, $f$, normalized by distance, $\frac{|f(\mathbf{q})-f(\mathbf{q}')|}{\normx{\mathbf{q}-\mathbf{q}'}}$. Since this maximum is calculated over all query pairs, it is difficult to estimate. Furthermore, it depends on the data distribution itself, while we only have access to samples from the distribution. In practice, we observed that the \textit{Average} Query function Change, AQC, can be used as a proxy for LDQ. Specifically, we define $AQC=\frac{1}{{|Q| \choose 2}}\sum_{\mathbf{q}, \mathbf{q}'\in Q}\frac{|f(\mathbf{q})-f(\mathbf{q}')|}{\normx{\mathbf{q}-\mathbf{q}'}}$, where $Q\subseteq \mathcal{Q}$ is a set of queries sampled from all possible queries. We experimentally verify the usefulness of this complexity measure in Sec.~\ref{exp:ablation}. 

\subsection{DQD Bound Proof}\label{sec:theory:proof}
\subsubsection{Analysis Framework}\label{sec:theory:framwork}
For a neural network $\hat{f}$ when modelling a query function, $f_D$, we decompose its error, $\Delta = \frac{1}{n}\normx{f_D-\hat{f}}_1$, into two terms, \textit{approximation error} and \textit{sampling error}:
\begin{align}\label{eq:framework}
    \Delta\leq
    \underbrace{\frac{1}{n}\normx{f_{\chi}-\hat{f}}_1}_{\text{approximation error, }\Delta_{a}}+\underbrace{\frac{1}{n}\normx{f_{\chi}-f_D}_1}_{\text{sampling error, }\Delta_{s}}
\end{align}
Approximation error, $\Delta_a$, quantifies how accurately the neural network can approximate the distribution query function. $\Delta_a$ depends on the space/time complexity of the neural network. For instance, larger neural networks have more representation power and can approximate a distribution query function more accurately. Sampling error, $\Delta_s$, quantifies the difference, due to sampling, between the distribution and observed query functions. $\Delta_s$ depends on data size: the more data sampled, the more similar observed and distribution query functions will be (latter is the expected value of the former). We bound each term separately in Secs.~\ref{sec:theory:appx} and \ref{sec:theory:sampling}. Sec.~\ref{sec:theory:combine} combines the results which yields Theorem~\ref{thm:sum_count_all}.


\vspace{-0.1cm}
\subsubsection{Bounding Approximation Error}\label{sec:theory:appx}
For a desired bound on approximation error, $\Delta_a$, we characterize the time/space complexity required for a neural network to achieve the error bound. Universal function approximation theorem \cite{pinkus1999approximation, changcun2020relu} guarantees existence of a neural network of \textit{arbitrary} time/space complexity that can achieve any desired error value, but does not show its time/space complexity. 
Recent work (e.g., \cite{lu2021deep,yarotsky2020phase,petersen2018optimal}) study \textit{number} of neural network parameters needed to achieve a desired error. However, number of neural network parameters cannot be related to its space complexity, because magnitude of the parameters can be unbounded, thus leading to unbounded storage cost even for a fixed number of parameters. We present the following theorem, showing the required time/space complexity to achieve a desired error bound, $\varepsilon_1$ (see Sec.~\ref{sec:rel_work} for a comprehensive discussion of related work).

\vspace{-0.1cm}
\begin{theorem}\label{thm:nn_appx_error}
Given a $\rho$-Lipschitz function $f$, there exists a neural network, $\hat{f}$, with space and time complexity
$\Tilde{O}(d(\varkappa\rho d \varepsilon_1^{-1}+1)^d)$
, where $\Tilde{O}$ hides logarithmic factors in $\rho$, $d$ and $k$, such that 
\begin{enumerate}[label=(\alph*)]
    \item $\lVert f-\hat{f}\rVert_1\leq \varepsilon_1$.
    \item Furthermore, if $d\leq 3$, $\lVert f-\hat{f}\rVert_{\infty}\leq \varepsilon_1$,
\end{enumerate}
Where $\varkappa$ is a universal constant.
\end{theorem}
\vspace{-0.2cm}Theorem~\ref{thm:nn_appx_error} (a) bounds $\Delta_a$ by considering $f_\chi$ as the function, $f$, in the theorem statement. Theorem~\ref{thm:nn_appx_error} (b) provides a stronger guarantee that can provide an $\infty$-norm DQD bound in low dimensions. For conciseness, we have not stated that version of DQD bound since the ideas are similar.
Theorem~\ref{thm:nn_appx_error} is a step towards characterizing neural network approximation power in a data management context. We expect tighter characterizations to be possible, especially for high dimensions. Our theoretical framework for DQD bound can readily benefit from such tighter characterizations. Nonetheless, $d$ is small for many practical applications when answering RAQs. For instance, in Example~\ref{ex:visit} that mimics a real-world use-case, the query function is 4-dimensional.

\begin{figure}
    \includegraphics[width=0.5\columnwidth]{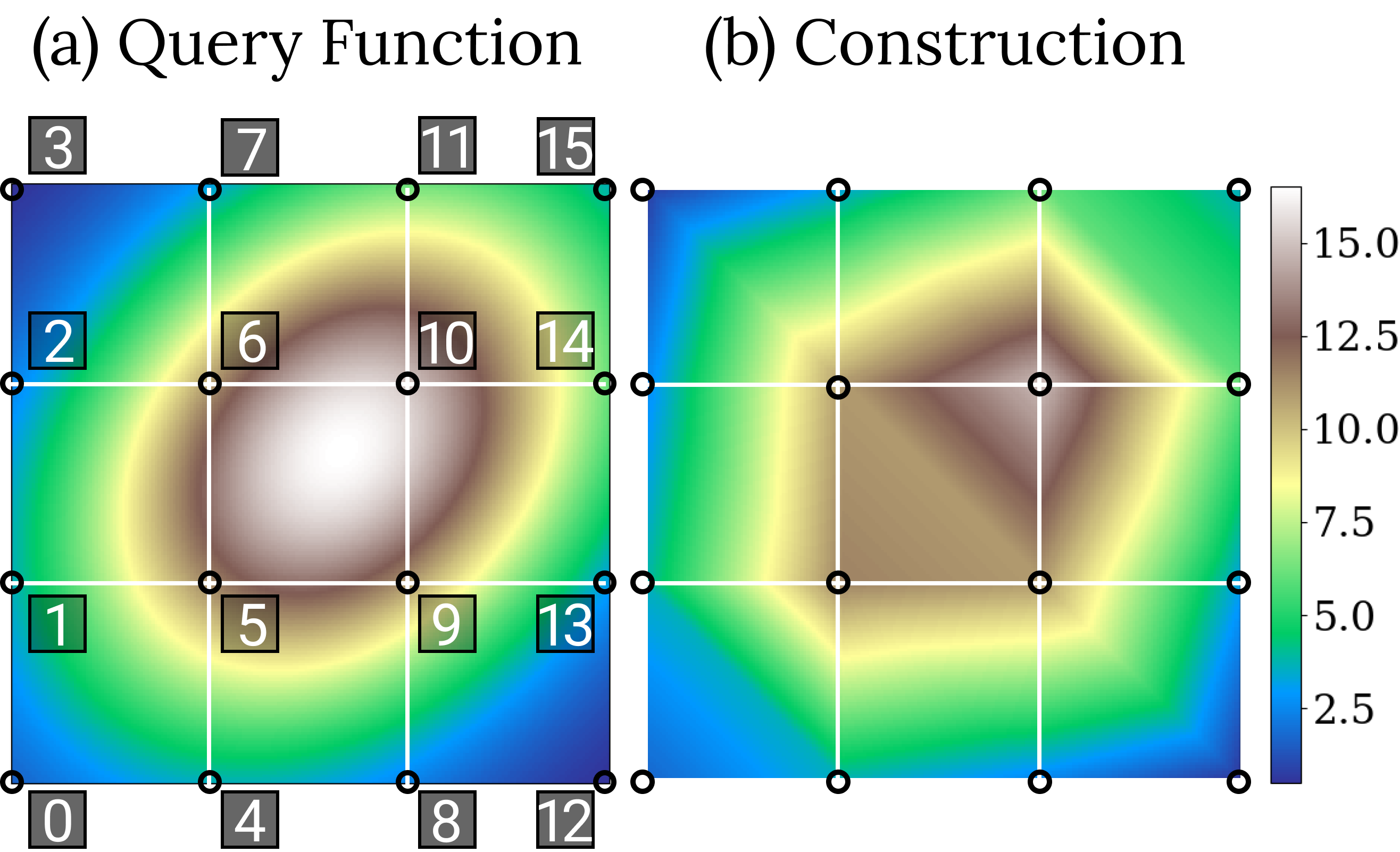}
    \includegraphics[width=0.46\columnwidth]{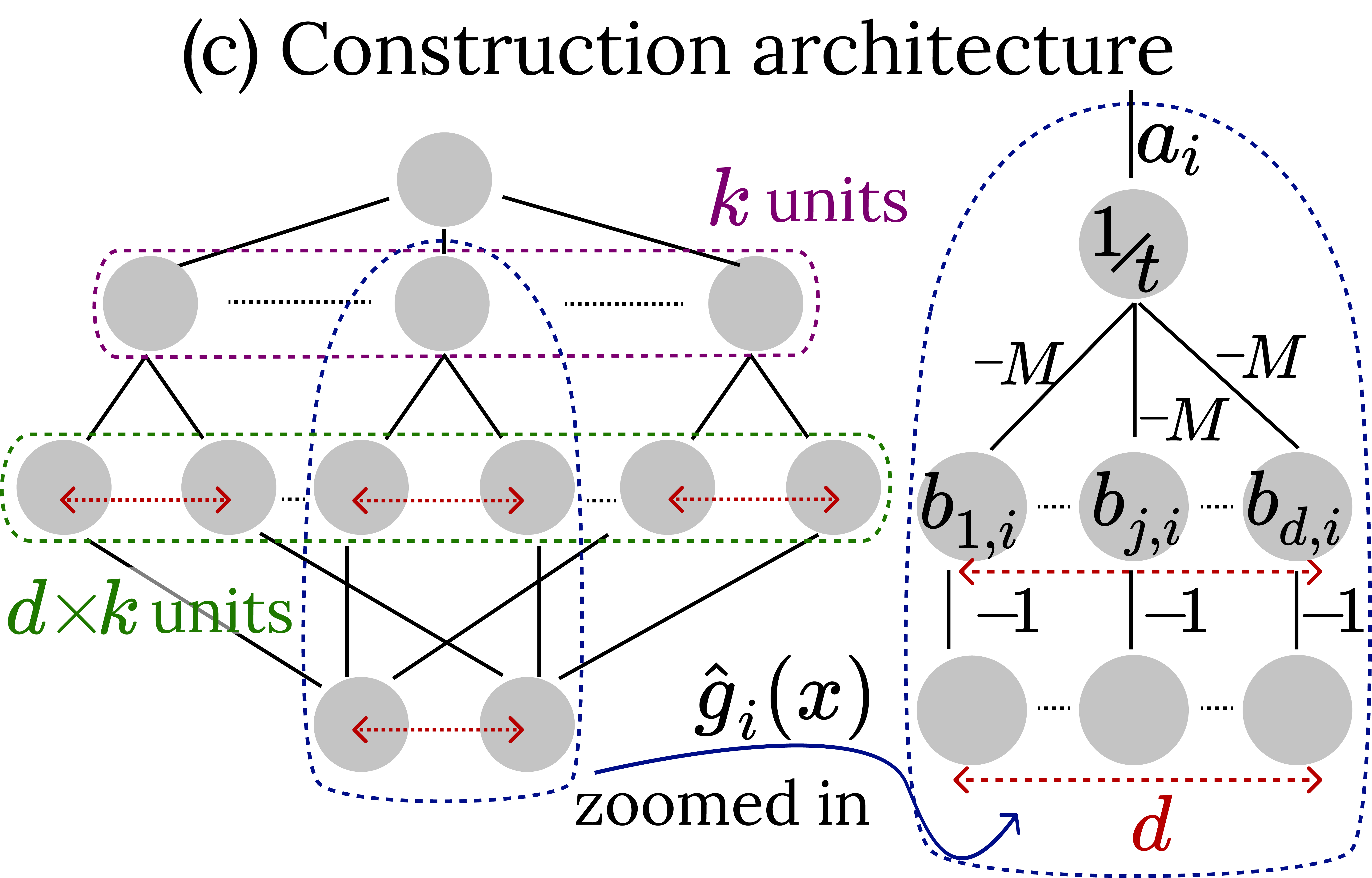}
    \caption{Constructed neural network and its architecture. Values on edges and nodes show edge weight and unit bias.}
    \label{fig:construction}
\end{figure}

\textit{Proof Sketch of Theorem \ref{thm:nn_appx_error}}. 
\revision{We uniformly partition the space into cells and construct a neural network that estimates cell vertices exactly. This \textit{memorization} property is used to bound error within each cell. For instance, Fig.~\ref{fig:construction} (a) shows the distribution query function for a \texttt{COUNT} query with fixed range $r=0.1$ on a two-dimensional Gaussian data distribution. A 3x3 grid on input space creates 16 vertices, shown in Fig.~\ref{fig:construction} (a). Our construction ensures that the error for these 16 vertices is zero, as shown in Fig.~\ref{fig:construction}~(b).}

\revision{\textbf{Network Architecture}. We construct a ReLU neural network, $\hat{f}$, with two hidden layers, shown in Fig.~\ref{fig:construction}~(c). $\hat{f}$ can be written as a summation of $k$ smaller units, called g-units. Each g-unit ensures that a cell vertex is memorized correctly and $k$ controls neural network size. The $i$-th g-unit, $\hat{g}_i$ for $1\leq i\leq k$, is constructed as shown in Fig.~\ref{fig:construction} (c). It has $d$ inputs, $d$ units in its first layer and 1 unit in its second layer. Each input is only connected to one of the units in the first layer with weight -1. All units in the first layer are connected to the unit in the second layer, and their weight is $-M$, where $M$ is a constant at least equal to 1. The $j$-th unit, $1\leq j\leq d$ in first layer has bias $b_{j, i}$ and the unit in second layer has bias $\frac{1}{t}$ for an integer $t$. The output of the second unit is multiplied by a parameter $a_i$. 
Then, the neural network is $\hat{f}(\boldsymbol{x}) = \sum_{i=1}^k\hat{g}_i(\boldsymbol{x})+b$, where $b$ is the bias of the third layer. The tunable parameters of the neural network are $a_i$, $b_{j, i}$, and $b$ for $1\leq i\leq k$ and $1\leq j\leq d$.}

\begin{figure}
    \includegraphics[width=1\columnwidth]{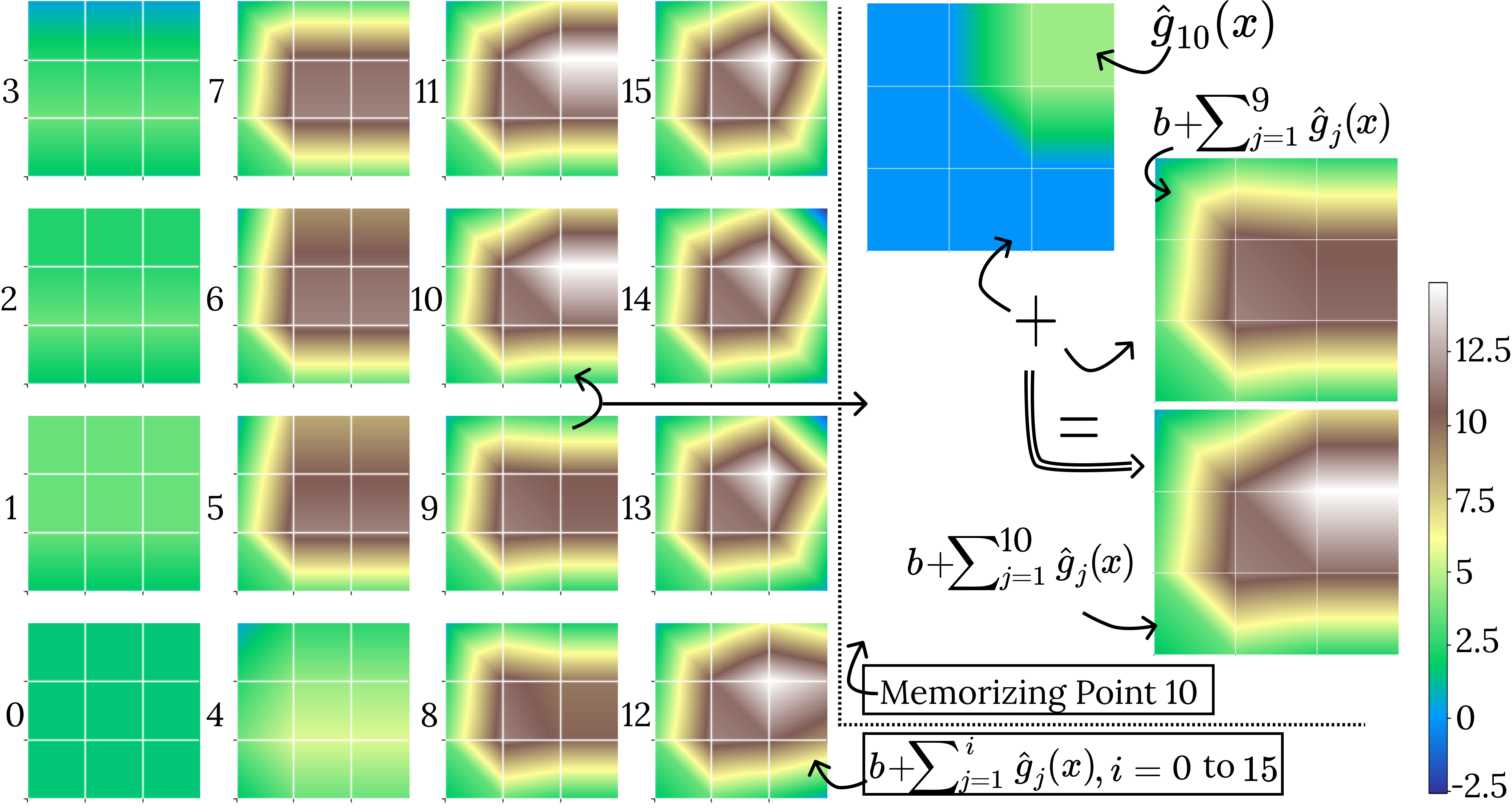}
    \caption{Neural Network Construction Steps}
    \label{fig:const_full_body}
\end{figure}

\revision{\textbf{Network Parameters}. 
Let the set of cell vertices in the uniform grid be $P=\{(i_1, ..., i_d)/t, i_r\in\mathbb{Z}, 0\leq i_r\leq t\}$, for an integer $t$ so that $k=|P|=(t+1)^d$ (recall that input space is $[0, 1]^d$). Also let $\boldsymbol{\pi}^i$ be the base $t+1$ representation of an integer $i$ written as a vector, i.e., $\boldsymbol{\pi}^i=(\pi^i_1, ..., \pi^i_d)$ so that $i=\sum_{r=1}^d\pi^i_r(t+1)^{d-r}$. For example, when $t=3$, $\pi^{6}=(1, 2)$, since $6=1(t+1)+2$. 
Note that $\frac{\boldsymbol{\pi}^i}{t}\in P$ and $\langle\frac{\boldsymbol{\pi}^0}{t}, ..., \frac{\boldsymbol{\pi}^{k-1}}{t}\rangle$  is an ordering of cell vertices.  
Alg.~\ref{alg:construction} enumerates using this ordering over the cell vertices and sets, at the $i$-th iteration, the parameters of the $i$-th g-unit so that $\frac{\boldsymbol{\pi}^i}{t}$ is correctly memorized. It calculates, $\hat{y}$, the estimate of the neural network for point $\frac{\boldsymbol{\pi}^i}{t}$ based on the g-units set before the $i$-th iteration (line \ref{alg:construction:current_sum}). Then it sets the parameter of the $i$-th g-unit to account for the difference between $\hat{y}$ and the true value, $f(\frac{\boldsymbol{\pi}^i}{t})$. Fig.~\ref{fig:const_full_body} shows this process in our example. On the left, Fig.~\ref{fig:const_full_body} shows, at the end of each iteration $i$, the function $b+\sum_{j=1}^i\hat{g}_j(\boldsymbol{x})$ (define $\sum_{j=1}^0\hat{g}_j(\boldsymbol{x})=0$). On the right it shows that at the 10-th iteration, the model sets $\hat{g}_{10}$ to memorize the 10-th point correctly. Alg.~\ref{alg:construction}  and g-unit architecture are designed so that when the 10-th point is memorized, the neural network value for the previously memorized points does not change.}


\begin{algorithm}[t]
\begin{algorithmic}[1]
\Require A function $f$, a parameter $t$
\Ensure Neural network $\hat{f}$
\State $b\leftarrow f(0)$
\For{$i\leftarrow 1$ \textbf{to} $(t+1)^d-1$}  
    \State $\hat{y} \leftarrow b+\sum_{j=1}^{i-1}\hat{g}_{j}(\frac{\boldsymbol{\pi}^i}{t})$\label{alg:construction:current_sum}
    \For{$r\leftarrow 1$ \textbf{to} $d$}  
        \State $b_{r, i}\leftarrow \frac{\pi^i_r}{t}$
    \EndFor
    \State $a_i \leftarrow t(f(\frac{\boldsymbol{\pi}^i}{t})-\hat{y})$\label{alg:construction:set_a}
\EndFor
\State \Return $\hat{f}$
\end{algorithmic}
\caption{Neural Network Construction}\label{alg:construction}
\end{algorithm}

\revision{\textbf{Proving the Bound.} 
We provide proof sketch for Theorem~\ref{thm:nn_appx_error} (a), using lemmas formally stated and proven in Sec.~\ref{appx:proof}. 
Proof for Theorem~\ref{thm:nn_appx_error} (b) is similar. Lemma~\ref{lemma:correct_mem} states that $\hat{f}(\mathbf{x})$ achieves zero error at cell vertices, i.e.,
\begin{align}\label{eq:mem}
    f(\mathbf{x})=\hat{f}(\mathbf{x}),\forall\mathbf{x}\in P.
\end{align}
Furthermore, $f$ is $\rho$-Lipschitz so its change is bounded within each cell. That is, for $\mathbf{x}, \mathbf{x}'\in C^i$, where $C^i=\{\frac{\boldsymbol{\pi^i}}{t}+\mathbf{z}, \mathbf{z}\in[0, \frac{1}{t}]^d\}$ is the subset of input space in the $i$-th cell, the Lipschitz property implies
\begin{align}\label{eq:rho_lipschitz}
    |f(\mathbf{x})-f(\mathbf{x}')|\leq \frac{\rho d}{t}.
\end{align}
Lemma~\ref{prop:change} proves that $\hat{f}$ change is bounded within each cell, i.e.
\begin{align}\label{eq:nn_change}
    |\hat{f}(\mathbf{x})-\hat{f}(\mathbf{x'})|\leq \phi(d,\rho,t, \mathbf{x}, \mathbf{x'})
\end{align}
for some function $\phi$ specified in Lemma~\ref{prop:change}. $\phi$ depends on $\mathbf{x}$ and $\mathbf{x'}$ since the bound is different depending on where in space $\mathbf{x}$ and $\mathbf{x'}$ are. Using triangle inequality with Eq.~\ref{eq:rho_lipschitz} and \ref{eq:nn_change}, we have
\begin{align}\label{eq:error_change}
|\hat{f}(\mathbf{x})-f(\mathbf{x})-(\hat{f}(\mathbf{x'})-f(\mathbf{x'}))|\leq \frac{d\rho}{t}+\phi(d,\rho,t, \mathbf{x}, \mathbf{x'}).
\end{align}
Letting $\mathbf{x}'=\frac{\boldsymbol{\pi}^i}{t}$ in Eq.~\ref{eq:error_change} and using Eq.~\ref{eq:mem}, we obtain
\begin{align}\label{eq:bound_error_phi}
|\hat{f}(\mathbf{x})-f(\mathbf{x})|\leq \frac{d\rho}{t}+\phi(d,\rho,t, \mathbf{x}, \frac{\boldsymbol{\pi}^i}{t}).
\end{align}
Lemma~\ref{prop:error} shows that integrating right hand side of Eq.~\ref{eq:bound_error_phi} over $\mathbf{x}$ and across cells yields $\frac{3\rho d}{t}$ so we bound the 1-norm error as
\begin{align}
\normx{\hat{f}-f}_1\leq \frac{3\rho d}{t}.
\end{align}
Lemma~\ref{lemma:space} shows that space and time complexity of $\hat{f}$ is $\Tilde{O}(kd)$. 
Setting $\varepsilon_1=\frac{3\rho d}{t}$ and $\varkappa=3$, recalling that $k=(t+1)^d$, and substituting $k=(\varkappa\rho d \varepsilon_1^{-1}+1)^d$ in the space/time complextiy experssion proves Theorem~\ref{thm:nn_appx_error} (a). Lemma proofs require a detailed study of neural network behaviour, see Sec.~\ref{appx:proof}
.}

\vspace{-0.15cm}
\subsubsection{Bounding Sampling Error}\label{sec:theory:sampling}
We present the following theorem that bounds the sampling error with high probability. 

\begin{theorem}\label{cor:sum_count_err}
Let $f^{C}_\chi$ and $f^{S}_\chi$ be distribution query functions for \texttt{COUNT} and \texttt{SUM} aggregation functions and $f^{C}_D$ and $f^{S}_D$ the corresponding observed query functions for a database, $D$, of $n$ points in $d$ dimensions sampled from $\chi$. For $i\in\{S, C\}$, 
\begin{align*}
\mathop{\mathds{P}}_{D\sim\chi}\left[\frac{1}{n}\normx{f^{i}_\chi-f^{i}_D}_\infty>\varepsilon_2\right]\leq
\varkappa^{d+1}d\varepsilon_2^{-d}\exp{(-\varkappa^{-1}\varepsilon_2^2n)},\end{align*}
Where $\varkappa$ is a universal constant.
\end{theorem}

Theorem~\ref{cor:sum_count_err} 
provides a high probability bound on $\Delta_s$ in Eq.~\ref{eq:framework}. 
The proof of Theorem~\ref{cor:sum_count_err} uses VC sampling theory, which presents a novel use of VC theory for the database literature. VC theory helps us understand the impact of the distribution a database follows on operations performed (e.g., answering RAQs) on the database. In fact, Theorem~\ref{cor:sum_count_err} is independent of our use of learned models, and simply characterizes impact of sampling when answering RAQs on a database that follows a certain data distribution. This is different from the typical use of VC theory in machine learning, where the goal is to study generalization of a trained model to unseen testing data. We present a proof sketch for the case of  \texttt{COUNT}. Proof for \texttt{SUM} is similar, but uses a generalization of VC-dimension. 

\textit{Proof Sketch of Theorem~\ref{cor:sum_count_err} for \texttt{COUNT}}. We start by rewriting the query function. Define the indicator function $h$ as 
\begin{align*}
    h_{\mathbf{q}}^C(\mathbf{p})=
    \begin{cases}
    1 & \text{if}\;\;\forall i,  c_i\leq p_i < c_i+r_i\\
    0              & \text{otherwise}.
\end{cases}
\end{align*}
So $f_D(\mathbf{q})=\sum_{\mathbf{p}\in D}h_{\mathbf{q}}^C(\mathbf{p})$ and $f_\chi(\mathbf{q})=nE_{\mathbf{p}\sim \chi}[h_{\mathbf{q}}(\mathbf{p})]$. Let $\mathcal{H}^C=\{h_{\mathbf{q}}^C, \forall \mathbf{q}\}$, so to bound error $\sup_{\mathbf{q}}\frac{1}{n}|f_D(\mathbf{q})-f_\chi(\mathbf{q})|$, we bound
\begin{align}\label{eq:bound}
\sup_{h\in\mathcal{H^C}}|\frac{1}{n}\sum_{\mathbf{p}\in D}h(\mathbf{p})-\mathbb{E}_{\mathbf{p}\sim \chi}[h(\mathbf{p})]|.
\end{align}
VC-dimension of $\mathcal{H}^C$ is known to be $2d$~\cite{shalev2014understanding} \revision{(see Lemma \ref{prop:q_func_vc}
)}, so applying VC theory bounds~\cite{anthony1999neural} \revision{(stated in Theorem~\ref{thm:vc}
)} to Eq.~\ref{eq:bound} proves the theorem. \qed

\vspace{-0.2cm}
\subsubsection{Completing the Proof}\label{sec:theory:combine}
Let $\varepsilon_1$ and $\varepsilon_2$ be the two error parameter, and let $\hat{f}$ be the neural network in Theorem~\ref{thm:nn_appx_error} that achieves error $\varepsilon_1$. Furthermore, let $E_1$ be the event $\frac{1}{n}\normx{f^{i}_\chi-f^{i}_D}_\infty
\leq\varepsilon_2$ holds for a random $D$ sampled from $\chi$. Observe that if $E_1$ holds, by triangle inequality, the event $E_2$ defined as $\frac{1}{n}\normx{\hat{f}-f^{i}_D}_1\leq\varepsilon_2+\varepsilon_1$ also holds. Thus, $\mathds{P}[E_1]\leq \mathds{P}[E_2]$. Taking the complement of both event, and observing that probability of complement of $E_1$ is bounded by Theorem~\ref{cor:sum_count_err} yields Theorem~\ref{thm:sum_count_all}. 

\vspace{-0.2cm}
\subsection{Other Query Functions and Model Choices}\label{sec:theory:other}
\vspace{-0.1cm}
Proof of DQD bound for \texttt{SUM} and \texttt{COUNT} aggregation functions decomposes the error into approximation error and sampling 
error. Theorem~\ref{thm:nn_appx_error}, which bounds the approximation error, is independent of the aggregation function used and applies to any function. To utilize the theoretical framework for other query functions, we need to bound the corresponding sampling error (Theorem~\ref{cor:sum_count_err} is specific to \texttt{SUM} and \texttt{COUNT}). In Sec.~\ref{sec:theory:avg}, we discuss this for \texttt{AVG} aggregation function and provide a general discussion for other query functions in Sec.~\ref{sec:theory:other_q}. In Sec.~\ref{sec:theory:other_model} we discuss the applicability of our analysis framework to other modeling choices.

\subsubsection{\texttt{AVG} Aggregation Function}\label{sec:theory:avg}
Our study of \texttt{AVG} aggregation function is a variation of that of \texttt{SUM} and \texttt{COUNT}. We discuss the differences, then present our sampling error bound. 

First, we consider a variation of distribution query function, defined as $\bar{f}^{A}_\chi(\mathbf{q})=\frac{f^{S}_\chi(\mathbf{q})}{f^{C}_\chi(\mathbf{q})}$. 
which we found to be easier to theoretically study ($\bar{f}^{A}_\chi$ is not the expected answer to \texttt{AVG} query, but expected answer to \texttt{SUM} query divided by expected answer to \texttt{COUNT} query). Since it depends on data distribution, it still allows us to study impact of data distribution on query answering. Second, we define LDQ as the Lipschitz constant of $\bar{f}^{A}_\chi$. LDQ in this case is not normalized by data size (as it was for \texttt{SUM} and \texttt{COUNT} in Sec.~\ref{sec:dqd:notation}), since magnitude of query answers for \texttt{AVG} do no change as data size changes. Third, for small query ranges few points in the database may match the query, even if data size is large. In such cases, for \texttt{AVG} aggregation function, the observed query function will be a poor estimate of the distribution query function. For \texttt{COUNT} or \texttt{SUM} query functions, few data points in a range means that both \texttt{SUM} and \texttt{COUNT} values are small, but this is not the case for the \texttt{AVG} function whose distribution query answer is independent of the number of points sampled in the range. To capture this dependence on query range, we define $\mathcal{Q}_\xi=\{\mathbf{q}, s.t., f^{C}_\chi(\mathbf{q})\geq \xi\}$. Our bound depends on $\xi$, which captures the probability of observing a point in a range. 

\vspace{-0.15cm}
\begin{lemma}\label{lemma:avg_bound_error}
Recall that $f^{A}_D(\mathbf{q})=\frac{f^{S}_D(\mathbf{q})}{f^{C}_D(\mathbf{q})}$ is the \texttt{AVG} query function. Let $\operatorname{err}(\mathbf{q})=\frac{|\bar{f}^{A}_\chi(\mathbf{q})-f^{A}_D(\mathbf{q})|}{|\bar{f}^{A}_\chi(\mathbf{q})|+1}$. We have
\begin{align*}
    \mathop{\mathds{P}}_{D\sim\chi}\left[{\sup}_{\mathbf{q}\in \mathcal{Q}_\xi}\operatorname{err(\mathbf{q})}\geq \varepsilon\right]&\leq \\\varkappa^{d+1} d&
    \left(\frac{1+\varepsilon}{\xi\varepsilon}\right)^d\exp{\left(-\varkappa^{-1}(\frac{\xi\varepsilon}{1+\varepsilon})^2n\right)},
\end{align*}
Where $\varkappa$ is a universal constant.
\end{lemma}

\vspace{-0.15cm}
\textit{Proof Sketch}. Proof applies Theorem~\ref{cor:sum_count_err} to numerator and denominator of \texttt{AVG} query function (Sec.~\ref{sec:proof:avg_bound_error}
). \qed

Combining Lemma~\ref{lemma:avg_bound_error} and Theorem~\ref{thm:nn_appx_error} show similar discussions to Sec.~\ref{sec:dqd:statement} on dependence on data distribution and size also apply to \texttt{AVG} queries. Lemma~\ref{lemma:avg_bound_error} also shows impact of query range. 

\textbf{More Accurate on Larger Ranges}. Impact of query range is modeled through the parameter $\xi$. Larger $\xi$ means the bound applies to larger ranges, where the confidence in the bound increases with $\xi$. Fixing the confidence level, observe that $\xi$ and $\varepsilon$ are negatively correlated. Increasing the query ranges considered reduces the sampling error. Thus, if LDQ of the query function is small (approximation error is low) and query range is large (sampling error is low), a neural network can answer \texttt{AVG} RAQs accurately and efficiently. LDQ can be calculated similar to examples in Sec.~\ref{sec:dqd:statement}. 


\vspace{-0.05cm}
\subsubsection{Other Query Functions}\label{sec:theory:other_q} Bounding sampling error for queries with \texttt{COUNT}, \texttt{SUM} or \texttt{AVG} aggregation functions but different range predicates (e.g., circular predicate $(\mathbf{c}, \mathbf{r})$ matching points $\mathbf{p}$, $\normx{\mathbf{p}-\mathbf{c}}_{2}\leq \mathbf{r}$) can be done similar to proof of Theorem~\ref{cor:sum_count_err} (only finding range predicate's VC-dimension needs further study). However, applicability of VC theory depends on the aggregation function. 

\vspace{-0.05cm}
\subsubsection{DQD for Query Modelling Approaches}\label{sec:theory:other_model} 
Our analysis framework allows for providing DQD bounds for other \textit{query modeling} approaches, where we define query modelling as an approach that directly models the query answers. 
%
%
Furthermore, our analysis of sampling error (Theorem~\ref{cor:sum_count_err}, Lemma~\ref{lemma:avg_bound_error}) does not depend on modeling choices and is generic to query modeling approaches. Thus, insights about the role of data size can be applicable to other query modeling approaches. For instance, consider answering count queries on uniformly distributed data in range [0, 1], as in Example~\ref{example:unif1d}. For data size $n$, as data size increases, the number of data points in a query $(c_1, r_1)$ becomes more similar to $r_1\times n$, which is the expected number of points that fall in any range of length $r_1$. Thus, one can estimate the answer to count query with a model $\hat{g}$ defined as $\hat{g}(c_1, r_1)=n\times r_1$. Answering queries with $\hat{g}$ takes constant time (it's a single operation), and its accuracy improves as data size increases, as supported by Theorem~\ref{cor:sum_count_err}. 
\section{NeuroSketch}\label{sec:practice}

\begin{figure}
    \centering
    \includegraphics[width=0.93\columnwidth]{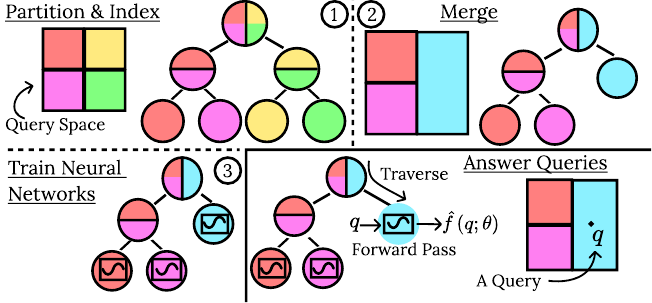}
    \caption{NeuroSketch Framework}
    \label{fig:neurosketch_overview}
\end{figure}


DQD bound formalizes how complexity of answering RAQs relates to data and query properties. In this section, we present a novel \textit{complexity-aware}  neural network framework, NeuroSketch, that utilizes results from DQD bound to allocate model capacity. 
We first present an overview of NeuroSketch, then discuss its details and finally discuss how it can be used in real-world database systems together with our DQD bound.



\vspace{-0.15cm}
\subsection{NeuroSketch Overview}
The key idea behind NeuroSketch design is that, even on the same database, some queries can be more difficult to answer than others (e.g., larger ranges vs. smaller ranges, see Sec.~\ref{sec:theory:avg}). By allocating more model capacity to queries that are more difficult, we can improve the performance. We do so by partitioning the query space and training independent neural networks for each partition. The partitioning allows diverting model capacity to harder queries, which our DQD bound allows us to quantify. By creating models specialized for a specific part of the query space, \textit{query specialization} allows us to control how model capacity is used across query space.


Fig.~\ref{fig:neurosketch_overview} shows an overview of NeuroSketch. During a pre-precessing step, (1) we partition and index the query space using a kd-tree. The partitioning is done based on our query specialization principle, with the goal of training a specialized neural network for different parts of the query space. (2) To account for the complexity of the underlying function in our partitioning, we merge the nodes of the kd-tree that are \textit{easier} to answer based on our DQD bound, so that our model only has to specialize for the certain parts of the space that are estimated to be more difficult. (3) After some nodes of the kd-tree have been merged, we train a neural network for all the remaining leaves of the kd-tree. Finally, to answer queries at query time, we traverse the kd-tree to find the leaf node a query falls inside, and perform a forward pass of the neural network.


\vspace{-0.15cm}
\subsection{NeuroSketch Details}
Training  NeuroSketch uses a training query set $Q\subseteq\mathcal{Q}$. $Q$ can be sampled from $\mathcal{Q}$ according to a workload distribution, or can be a uniform sample in the absence of any workload information. We do not assume access to workload information, but our framework can take advantage of the query workload if available. 

\begin{algorithm}[t]
\begin{algorithmic}[1]
\Require A kd-tree node $N$, tree height $h$ and dimension, $i$ to split the node, $N$ on 
\Ensure A kd-tree with height $h$ rooted at $N$
\If{$h=0$} 
    \State \Return 
\EndIf
\State $N.val\leftarrow$ median of $N.Q$ along $i$-th dimension
\State $N.dim\leftarrow i$
\State $Q_{left}=\{\mathbf{q}|\mathbf{q}\in N.Q, q[N.dim]\leq N.val\}$
\State $Q_{right}=\{\mathbf{q}|\mathbf{q}\in N.Q, q[N.dim]>N.val\}$
\For{$x\in \{left, right\}$}
    \State $N_x$ $\leftarrow$ new node
    \State $N_x.Q\leftarrow Q_x$
    \State $N.x\leftarrow N_x$\TriComment{Adding $N_x$ as left or right child of $N$}
    \State $get\_index(N_x, h-1, (N.dim+1)\mod d)$
\EndFor
\end{algorithmic}
\caption{$partition\_\&\_index(N, h, i)$}\label{alg:get_index}
\end{algorithm}

\begin{algorithm}[t]
\begin{algorithmic}[1]
\Require kd-tree root node $N$ and desired number of partitions $s$
\Ensure kd-tree with $s$ leaf nodes
\Repeat
    \ForAll{Leaf nodes $N$}
        \State $AQC_N\leftarrow \frac{1}{{|N.Q| \choose 2}}\sum_{\mathbf{q}, \mathbf{q}'\in N.Q, \mathbf{q}\neq \mathbf{q}'}\frac{|f_D(\mathbf{q})-f_D(\mathbf{q}')|}{\normx{\mathbf{q}-\mathbf{q}'}}$\label{alg:merge:measure_complexity}
    \EndFor
    \State $N\leftarrow$ the leaf node with smallest $ACQ_N$
    \State $N.marked\leftarrow true$
    \ForAll{Sibling leaf nodes $N_1, N_2$}
        \If{$N_1.marked=N_2.marked=true$}
        \State Merge $N_1$ and $N_2$\label{alg:merge:merge}
        \EndIf
    \EndFor
\Until{There are $s$ leaf nodes}
\end{algorithmic}
\caption{$merge(N, s)$}\label{alg:maerge}
\end{algorithm}

\textbf{Partitioning \& Indexing}. 
To partition the space, we choose partitions that are smaller where the queries are more frequent and larger where they are less frequent. This allows us to divert more model capacity to more frequent queries, thereby boosting their accuracy if workload information is available. We achieve this by partitioning the space such that all partitions are equally probable. To do so, we build a kd-tree on our query set, $Q$, where the split points in the kd-tree can be considered as estimates of the median of the workload distribution (conditioned on the current path from the root) along one of its dimensions. We build the kd-tree by specifying a maximum height, $h$, and splitting every node until all leaf nodes have height $h$, which creates $2^h$ partitions. Splitting of a node $N$ is done based on median of one of the dimensions of the subset, $N.Q$, of the queries, $Q$, that fall in $N$. 
Alg.~\ref{alg:get_index} shows this procedure. To build an index with height $h$ rooted at a node, $N_{root}$ (note that $N_{root}.Q=Q$), we call $partition\_\&\_index(N_{root}, h, 0)$. We note that other partitioning methods (e.g., clustering the queries to perform partitioning) are also possible, but we observed kd-tree to be a simple practical solution with little overhead that performed well.

\textbf{Merging}. We merge some of kd-tree leaves using DQD bound. As discussed in Sec.~\ref{sec:dqd:practice}, LDQ can be difficult to measure in practice, so we use AQC as a proxy, as shown in Alg.~\ref{alg:maerge}. At each iteration, we first measure the approximation complexity for the leaf nodes, in line~\ref{alg:merge:measure_complexity}, where the approximation complexity, $AQC_N$ for a leaf node $N$ is calculated based on queries that fall in the node $N$. Then, we mark the node with the smallest $AQC_N$ for merging. When two sibling leaf nodes are marked, they are merged together, as shown in line~\ref{alg:merge:merge}. The process continues until the number of remaining leaf nodes reaches the desired threshold. In practice, we observed that the quantity $AQC_N$ is correlated with the error of the neural networks, which empirically justifies this design choice (see Sec.~\ref{exp:ablation}).

\begin{algorithm}[t]
\begin{algorithmic}[1]
\Require A dataset $D$, a kd-tree node $N$
\Ensure Neural network $\hat{f}$ for node $N$
\State Initialize the parameters, $\theta$, of a neural network $\hat{f}(.; \theta)$\label{alg:sgd:init}
\Repeat
    \State Sample, $Q_{batch}$, a subset of $N.Q$
    \State Update $\theta$ in direction $-\nabla_\theta\sum_{\mathbf{q}\in Q_{batch}}\frac{(\hat{f}(\mathbf{q};\theta)-f_D(\mathbf{q}))^2}{|Q_{batch}|}$\label{alg:sgd:update}
\Until{convergence}
\State \Return $\hat{f}$
\end{algorithmic}
\caption{Model Training}\label{alg:sgd}
\end{algorithm}
\begin{algorithm}[t]
\begin{algorithmic}[1]
\Require kd-tree root node $N$ and query $\mathbf{q}$
\Ensure Answer to $\mathbf{q}$
\While{$N$ is not leaf}
    \If{$q[N.dim] \leq N.val$}
        \State $N \leftarrow N.left$
    \Else
        \State $N \leftarrow N.right$
    \EndIf
\EndWhile
\Return $N.model.forward\_pass(\mathbf{q})$
\end{algorithmic}
\caption{$answer\_query(N, \mathbf{q})$}\label{alg:get_answer}
\end{algorithm}

\textbf{Training Neural Networks}. We train an independent model for each of the remaining leaf nodes after merging. For a leaf node, $N$, the training process is a typcial supervised learning procedure and shown in Alg.~\ref{alg:sgd} for completeness. The answer to queries for training, used in line~\ref{alg:sgd:update} of Alg.~\ref{alg:sgd}, can be collected through any known algorithm, where a typical algorithm iterates over the points in the database, pruned by an index, and for a candidate data point checks whether it matches the RAQ predicate or not. This is a pre-processing step and is only performed once to train our model. The process is embarrassingly parallelizable across training queries, if preprocessing time is a concern. \revision{Furthremore, if the data is disk resident, we keep partial \texttt{SUM}/\texttt{COUNT} answers for each training query while scanning data from disk, so a single scan of data is sufficient  (similar to building disk-based indexes) to collect training query answers. Once trained, NeuroSketch is much smaller than data and expected to fit in memory, so it will be much faster than disk-based solutions}. We use Adam optimizer \cite{kingma2014adam} for training and train a fully connected neural network for each of the partitions. The architecture is the same for all the partitions and consists of $n_l$ layers, where the input layer has dimensionality $d$, the first layer consists of $l_{first}$ units, the next layers have $l_{rest}$ units and the last layer has 1 unit. We use relu activation for all layers (except the output layer). $n_l$, $l_{first}$ and $l_{rest}$ are hyper-parameters of our model. Although approaches in neural architecture search \cite{zoph2016neural} can be applied to find them, they are computationally expensive. Instead, we do a grid search to find the hyper-parameters so that NeuroSketch satisfies the space and time constraints in Problem~\ref{prob:RAQ} while maximizing its accuracy.

\textbf{Answering Queries}.
As shown in Alg.~\ref{alg:get_answer}, to answer a query, $\mathbf{q}$, first, the kd-tree is traversed to find the leaf node that the query $\mathbf{q}$ falls into. The answer to the query is a forward pass of the neural network corresponding to the leaf node.

\vspace{-0.4cm}
\subsection{General RAQs and Real-World Application}\label{sec:general}
\textbf{General RAQs}. NeuroSketch can be used for more general RAQs than defined in Sec.~\ref{sec:def}. An RAQ consists of a range predicate, and an aggregation function $\texttt{AGG}$. In NeuroSketch, we make no assumption on the aggregation function \texttt{AGG} and our empirical results evaluated NeuroSketch on \texttt{SUM}, \texttt{AVG}, \texttt{COUNT}, \texttt{MEDIAN} and \texttt{STD}. We consider range predicates that can be represented by a \textit{query instance} $\mathbf{q}$, and a binary \textit{predicate function}, $P_f(\mathbf{q}, \mathbf{x})$, that takes as inputs a point in the database, $\mathbf{x}$, $\mathbf{x}\in D$, and the query instance $\mathbf{q}$, and outputs whether $\mathbf{x}$ matches the predicate or not. Then, given a predicate function and an aggregation function, range aggregate queries can be represented by the query function $f_D(\mathbf{q})=\texttt{AGG}(\{\mathbf{x}:\mathbf{x} \in D, P_f(\mathbf{x}, \mathbf{q})=1\})$. We avoid specifying how the predicate function should be defined to keep our discussion generic to arbitrary predicate functions, but some examples follow. To represent the RAQs of the form discussed in Sec.~\ref{sec:def}, $\mathbf{q}$ can be defined as lower and upper bounds on the attributes and $P_f(\mathbf{q}, \mathbf{x})$ defined as the WHERE clause in Sec.~\ref{sec:def}. We can also have $P_f(\mathbf{q}, \mathbf{x}) = x[1]>x[0]\times q[0]+q[1]$, so that $P_f(\mathbf{q}, \mathbf{x})$ and $\mathbf{q}$ define a half-space above a line specified by $\mathbf{q}$. For many applications, WHERE clauses in SQL queries are written in a parametric form \cite{sql_microsoft, sql_nodepostgres, sql_dataworld}  (e.g., WHERE $X_1> ?param1$ OR $X_2> ?param2$, where $?param$ is the common SQL syntax for parameters in a query). Such queries can be represented as query functions by setting $\mathbf{q}$ to be the parameters of the WHERE clause. 

\textbf{NeuroSketch and DQD in Practice}. Possible RAQs correspond to various query function and NeuroSketch learns different models for different query functions. This follows the \textit{query specialization} design principle, where a specialized model is learned to answer a query function well. A query processing engine can be used to decide which query functions to use NeuroSketch for. 
\revision{This can happen both on the fly, when answering queries, and during database maintenance. During maintenance, DQD bound can be used to decide which queries to build NeuroSketch for (e.g., for queries with small LDQs). Moreover, after NeuroSketch is built for a query function, DQD can be used to decide whether to use NeuroSketch for a specific query instance or not on the fly. For instance, queries with large ranges (that NeuroSketch answers accurately according to DQD) can be answered by NeuroSketch, while queries with smaller ranges can be asked directly from the database.} 

\begin{figure*}[t]
\begin{minipage}{.21\textwidth}
   \centering
    \includegraphics[width=\textwidth]{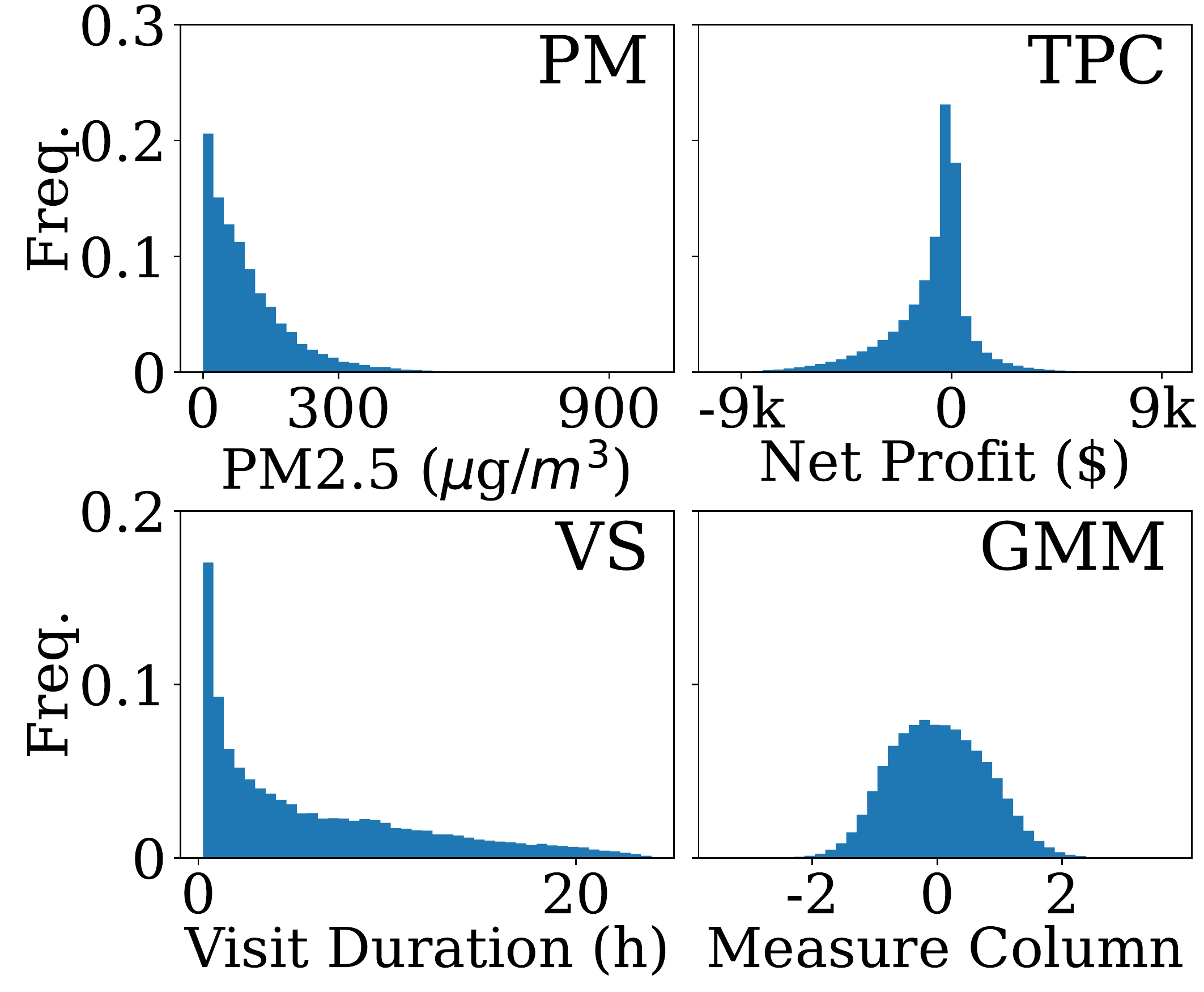}
    \caption{Measure col. distribution (shared y-axis)}
    \label{fig:measure_dist}
\end{minipage}
\begin{minipage}{.78\textwidth}
        \centering
    	\includegraphics[width=\textwidth]{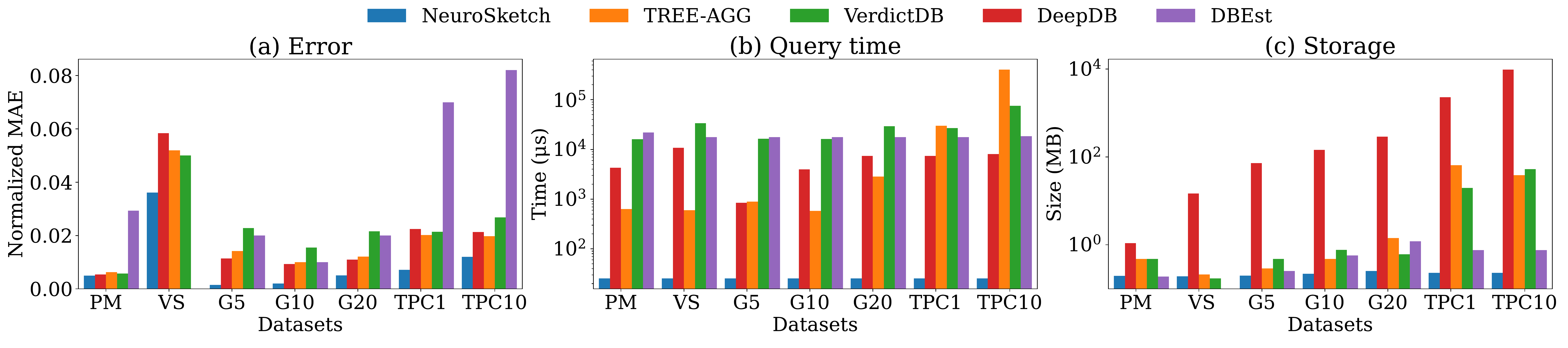}
        \caption{RAQs on different datasets}
        \label{fig:exp:range_agg_data}
\end{minipage}
\end{figure*}

\begin{table}[t]
    \centering
  \begin{tabular}{c|c|c|c|c}
        \textbf{Dataset} &  G5, G10, G20 &PM \cite{liang2015assessing}& TPC1, TPC10 \cite{nambiar2016the} &VS \\\hline
        \textbf{\# Points} & $10^5$ &4.17$\times10^4$ & 2.65$\times10^6$, 2.65$\times10^7$ & $10^5$ \\\hline
        \textbf{Dim} & 5, 10, 20 &  4 & 13&3 
    \end{tabular}    \caption{Dataset information}
    \label{tab:dataset_size}
\end{table}

\vspace{-0.2cm}
\section{Empirical Study}\label{sec:exp}
\vspace{-0.05cm}
\subsection{Experimental Setup}\label{exp:setup}
\vspace{-0.05cm}
\textbf{System Setup.} Experiments are performed on a machine with Ubuntu 18.04 LTS, an Intel i9-9980XE CPU (3GHz), 128GB RAM and a GeForce RTX 2080 Ti NVIDIA GPU.

\textbf{Datasets}. Table~\ref{tab:dataset_size} shows the datasets used in our experiments, with details discussed below. \revision{Fig.~\ref{fig:measure_dist} shows the histogram of measure column values used in the experiments.}

\textit{PM}. PM \cite{liang2015assessing} contains Fine Particulate Matter (PM2.5) measuring air pollution and other statistics (e.g., temperature) for locations in Beijing. Similar to \cite{ma2019dbest}, PM2.5 is the measure attribute. 

\textit{TPC-DS}. We used TPC-DS \cite{nambiar2016the}, \revision{a synthetic benchmark dataset}, with scale factors 1 and 10, respectively referred to as TPC1 and TPC10. Since we study RAQs, we use the numerical attributes in store\_sales table as our dataset, and net\_profit as measure attribute. 

\textit{Veraset}. As was used in our running example, we use Veraset dataset, which contains anonymized location signals of cell-phones across the US collected by Veraset \cite{veraset}, a data-as-a-service company. Each location signal contains an anonymized id, timestamp and the latitude and longitude of the location. We performed stay point detection \cite{ye2009mining} on this dataset (to, e.g., remove location signals when a person is driving), and extracted location visits where a user spent at least 15 minutes and for each visit, also recorded its duration. 100,000 of the extracted location visits in downtown Houston were sampled to form the dataset used in our experiments, which contains three columns: latitude, longitude and visit duration. We let visit duration to be the measure attribute.

\begin{figure*}[t]
\begin{minipage}[t]{.33\textwidth}
         \centering
        \includegraphics[width=\textwidth]{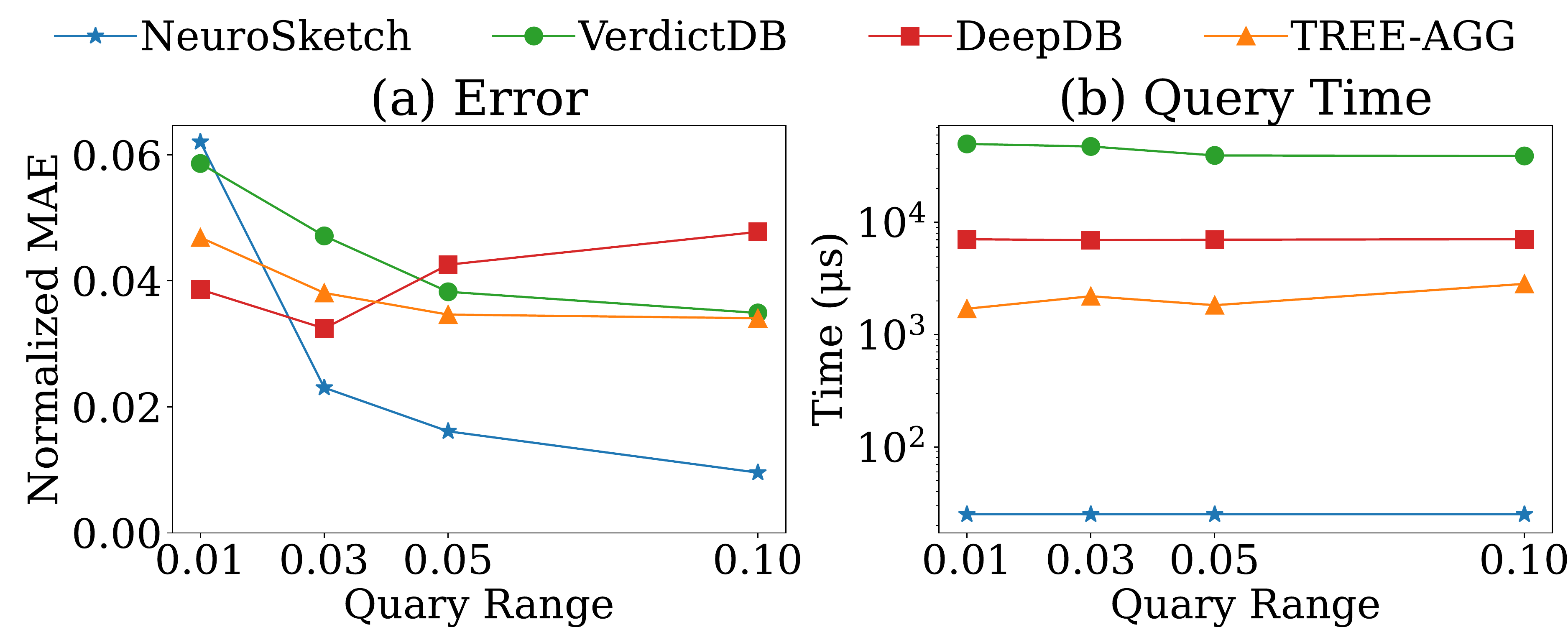}
        \caption{Varying query range}
        \label{fig:exp:range_agg_qrange}
    \end{minipage}
    \begin{minipage}[t]{.33\textwidth}
        \centering
    	\includegraphics[width=1\textwidth]{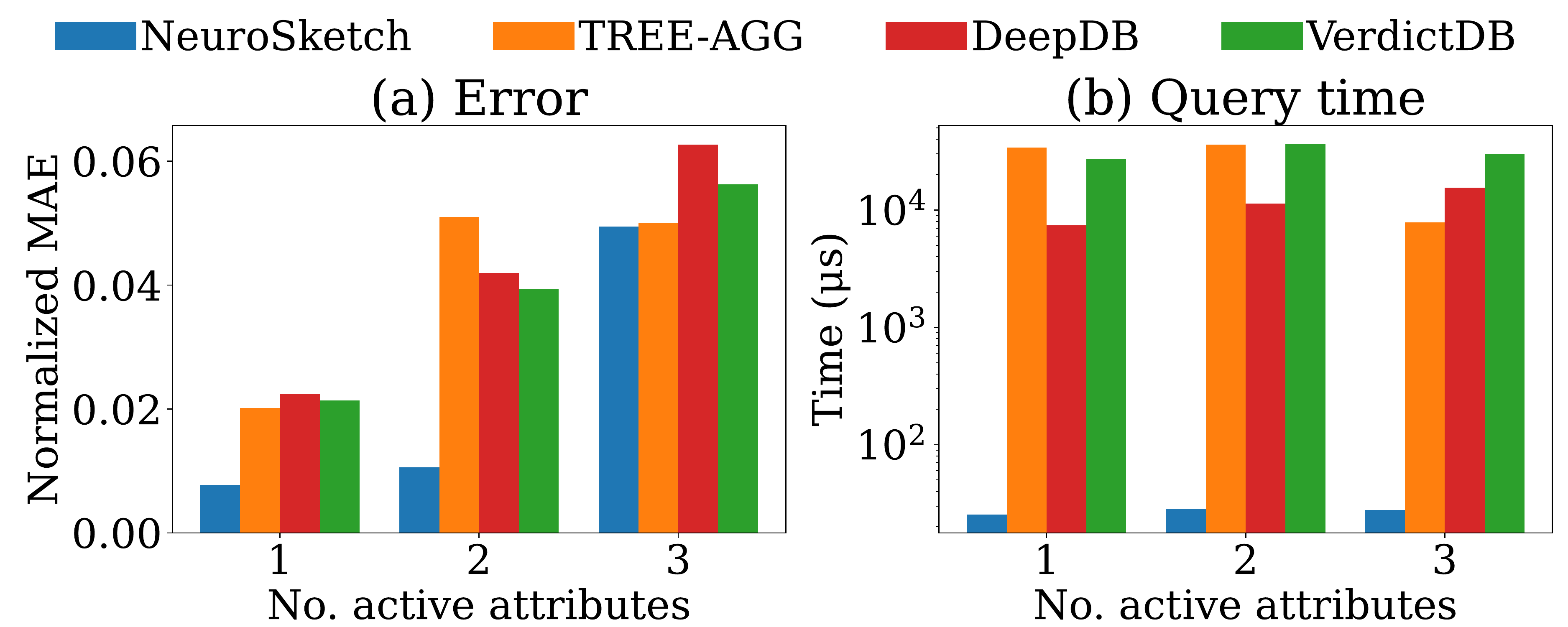}
        \caption{Varying no. of active attributes}
        \label{fig:exp:range_agg_dim}
    \end{minipage}  
    \begin{minipage}[t]{.33\textwidth}
        \centering
        \includegraphics[width=\textwidth]{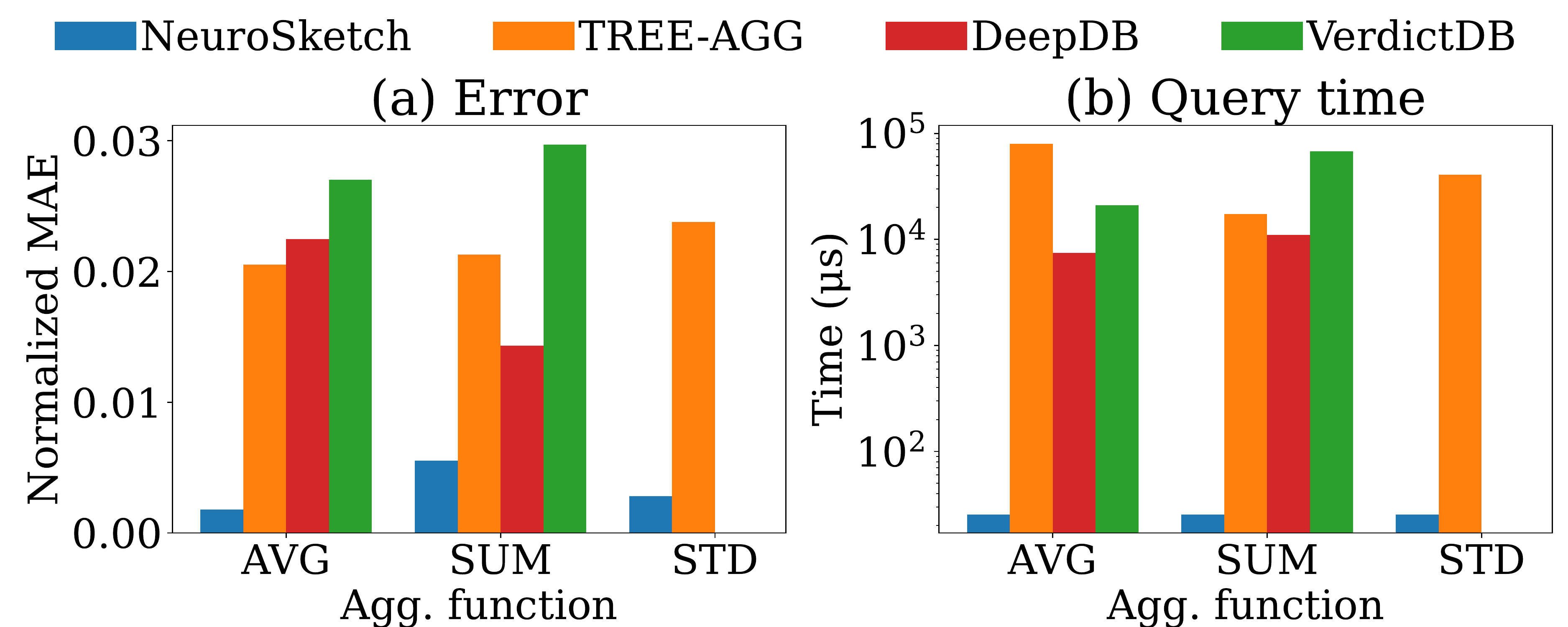}    	
        \caption{Varying agg. function}
        \label{fig:exp:range_agg_aggf}
    \end{minipage}
    
\end{figure*}

\textit{GMMs}. We study data dimensionality with synthetic 5, 10 and 20 dimensional data from Gaussian mixture models (GMM) (100 components, random mean and co-variance), referred to as G5, G10 and G20. GMMs are often used to model real data distribution~\cite{reynolds2009gaussian}. 

\textbf{Query Distribution}. Our experiments consider query functions consisting of \texttt{AVG}, \texttt{SUM}, \texttt{STDEV} (standard deviation) and \texttt{MEDIAN} aggregation functions together with two different predicate functions. First, similar to \cite{ma2019dbest}, our experiments show the performance on the predicate function defined by the WHERE clause in Sec.~\ref{sec:def}. We consider up to 3 active attributes in the predicate function. To generate a query instance with $r$ active attributes, we first select, uniformly at random, $r$ activate attributes (from a total of $d$ possible attributes). Then, for the selected active attributes, we randomly generate a range. Unless otherwise stated, the range for each active attribute is uniformly distributed. This can be thought of as a more difficult scenario for {NeuroSketch} as it requires approximating the query function equally well over all its domain, while also giving a relative advantage to other baselines, since they are unable to utilize the query distribution. Unless otherwise stated, for all datasets except Veraset, we report the results for one active attributes and use \texttt{AVG} aggregation function. For Veraset, we report the results setting latitude and longitude as active attributes. Second, to show how NeuroSketch can be applied to application specific RAQs, in Sec.~\ref{sec:exp:workload}, we discuss answering the query of median visit duration given a \textit{general} rectangle on Veraset dataset.

\textbf{Measurements}. In addition to query time and space used, we report the normalized absolute error for a query in the set of test queries, $T$, defined as $\frac{|f_D(\mathbf{q})-\hat{f}_D(\mathbf{q}, \theta)|}{\frac{1}{|T|}\sum_{\mathbf{q}\in T}|f_D(\mathbf{q})|}$. \revision{We ensure that none of the test queries are in the training set}. The error is normalized by average query result magnitude to allow for comparison over different data sizes and datasets when the results follow different scales.

\textbf{Learned Baselines}. \revision{We use DBEst \cite{ma2019dbest} and DeepDB \cite{hilprecht2019deepdb} as the state-of-the-art model-based AQP engines. Both algorithms learn data models to answer RAQs. We use the open-source implementation of DBEst available at \cite{dbest_imp} and DeepDB at \cite{deepdb_imp}. For DBEst, we perform a gird search on its MDN architecture (number of layers, layer width, number of Gaussian componenets) and optimize it per dataset. For DeepDB we optimize its RDC threshold for each dataset.}
We do not use \cite{thirumuruganathan2019approximate} as a baseline, which samples new data points at query time from a learned model to answer queries because the results in \cite{thirumuruganathan2019approximate} show worse accuracy and same query time (\cite{thirumuruganathan2019approximate} improves storage) compared with sampling directly from the data (which we have included as baseline). We also modified \revision{NeuroCard} \cite{yang2020neurocard}, a learned cardinality estimation method to answer RAQs, but we observed the modified approach to perform worse than DeepDB on RAQs. We do not present the results for \cite{yang2020neurocard}, since it is not designed for RAQs and performed worse than DeepDB.

\textbf{Sampling-based Baselines}. We use VerdictDB \cite{park2018verdictdb} as our sampling-based baseline, using its publicly available implementation \cite{verdict_imp}. We also implemented a sampling-based baseline designed specifically for range aggregate queries, referred to as TREE-AGG. In a pre-processing step and for a parameter $k$, TREE-AGG samples $k$ data points from the database uniformly. Then, for performance enhancement and easy pruning, it builds an R-tree index on the samples, which is well-suited for range predicates. At query time, by using the R-tree, finding data points matching the query is done efficiently, and most of the query time is spent on iterating over the points matching the predicate to compute the aggregate attribute required. For both TREE-AGG and VerdictDB, we set the number of samples so that the error is similar to that of DeepDB.

\textbf{NeuroSketch Training and Evaluation}. NeuroSketch training is performed in Python 3.7 and Tensorflow 2.1, with implementation publicly available at \cite{neurodb_imp}. Model training is done on GPU. Models are saved after training. For evaluation, a separate program written in C++ and running on CPU loads the saved model, and for each query performs a forward pass on the model. Model evaluation is done with C++ and on CPU, without any parallelism for any of the algorithms. Unless otherwise stated, model depth is set to 5 layers, with the first layer consisting of 60 units and the rest of 30 units. The height of the kd-tree is set to 4, and parameter $s=8$ so that the kd-tree has 8 leaf nodes after merging. 

\vspace{-0.4cm}
\subsection{Baseline Comparisons}
\subsubsection{Results Across Datasets} Fig.~\ref{fig:exp:range_agg_data}  (a) shows the error on different datasets, where NeuroSketch provides a lower error rate than the baselines. Fig.~\ref{fig:exp:range_agg_data} (b) shows that NeuroSketch achieves this while providing multiple orders of magnitude improvement in query time. NeuroSketch has a relatively constant query time because, across all datasets, NeuroSketch's architecture only differs in its input dimensionality, which only impacts number of parameters in the first layer of the model and thus changes model size by very little. Due to our use of small neural networks, we observe that model inference time for NeuroSketch is very small and in the order of few microseconds, while DeepDB \revision{and DBEst} answers queries multiple orders of magnitude slower. \revision{DBEst does not support multiple active attributes and thus its performance is not reported for VS}. The results on G5 to G20 show the impact of data dimensionality on the performance of the algorithms. As was suggested by our theoretical results, for NeuroSketch, the error increases as dimensionality increases. A similar impact can be seen for DeepDB, manifesting itself in increased query time. Furthermore, the R-tree index of TREE-AGG often allows it to perform better than the other baselines, especially for low dimensional data.  \revision{Finally, Fig.~\ref{fig:exp:range_agg_data} (c) shows the storage overhead of each methods.  NeuroSketch answers queries accurately by taking less than one MB space, while DeepDB's storage overhead increases with data size, to more than one GB.}

\vspace{-0.3cm}
\subsubsection{Results Across Different Workloads}\label{sec:exp:workload}
We use TPC1 and VS to study impact of query workload on performance of the algorithms. Unless otherwise stated results are on TPC1. \revision{Due to its poor performance on TPC1 and not supporting multiple active attributes (for VS queries), we exclude DBEst from the experiments here.}

\textbf{Impact of Query Range}. We set the query range to $x$ percent of the domain range, for $x\in\{1, 3, 5, 10\}$ and present the results in Fig.~\ref{fig:exp:range_agg_qrange}. The error of NeuroSketch increases for smaller query ranges, as our theoretical results suggest. As mentioned before, this is because for smaller ranges NeuroSketch needs to memorize where exactly each data point \revision{is}, rather than learning the overall distribution of data points. Nevertheless, NeuroSketch provides better accuracy than the baselines for query ranges at least 3 percent, and performs queries orders of magnitude faster for all ranges. If more accurate answers are needed for smaller ranges, increasing the model size of NeuroSketch can improve its accuracy at the expense of query time (see Sec.~\ref{exp:veraset}).

\textbf{Impact of No. of Active Attributes}. In Fig.~\ref{fig:exp:range_agg_dim}, we vary the number of active attributes in the range predicate from one to three. Accuracy of all the algorithms drops when there are more active attributes, with NeuroSketch outperforming the algorithms both in accuracy and query time. Having more active attributes is similar to having smaller ranges, since fewer points will match the query predicate. Thus, our theoretical results explain the drop in accuracy. 

\textbf{Impact of Aggregation Function}. Fig.~\ref{fig:exp:range_agg_aggf} shows how different aggregation functions impact performance of the algorithms. NeuroSketch is able to outperform the algorithms for all aggregation functions. VerdictDB and DeepDB implementation did not support \texttt{STDEV} and no result is reported for \texttt{STDEV} for these methods.

\textbf{Median Visit Duration Query Function}. We consider the query of median visit duration given a \textit{general} rectangular range. The predicate function takes as input coordinates of two points $\mathbf{p}$ and $\mathbf{p}'$, representing the location of two non-adjacent vertices of the rectangle, and an angle, $\phi$, that defines the angle the rectangle makes with the x-axis. Given $\mathbf{q}=(\mathbf{p}, \mathbf{p}', \phi)$, the query function returns median of visit duration of records falling in the rectangle defined by $\mathbf{q}$. This is a common query for real-world location data, and data aggregators such as SafeGraph \cite{safegraph} publish such information.

Table \ref{tab:median_visit_res} shows the results for this query function. Neither DeepDB nor DBEst can answer this query. The predicate function is not supported by those methods, and extending those methods to support them is not trivial. On the other hand, NeuroSketch can answer this query function, with similar performance to other queries on VS dataset. Although VerdictDB can be extended to support this query function, the current implementation does not support the aggregation function, so we do not report the results on VerdictDB.

\begin{figure*}
\begin{minipage}{0.4\textwidth}
  \centering
        \includegraphics[width=\textwidth]{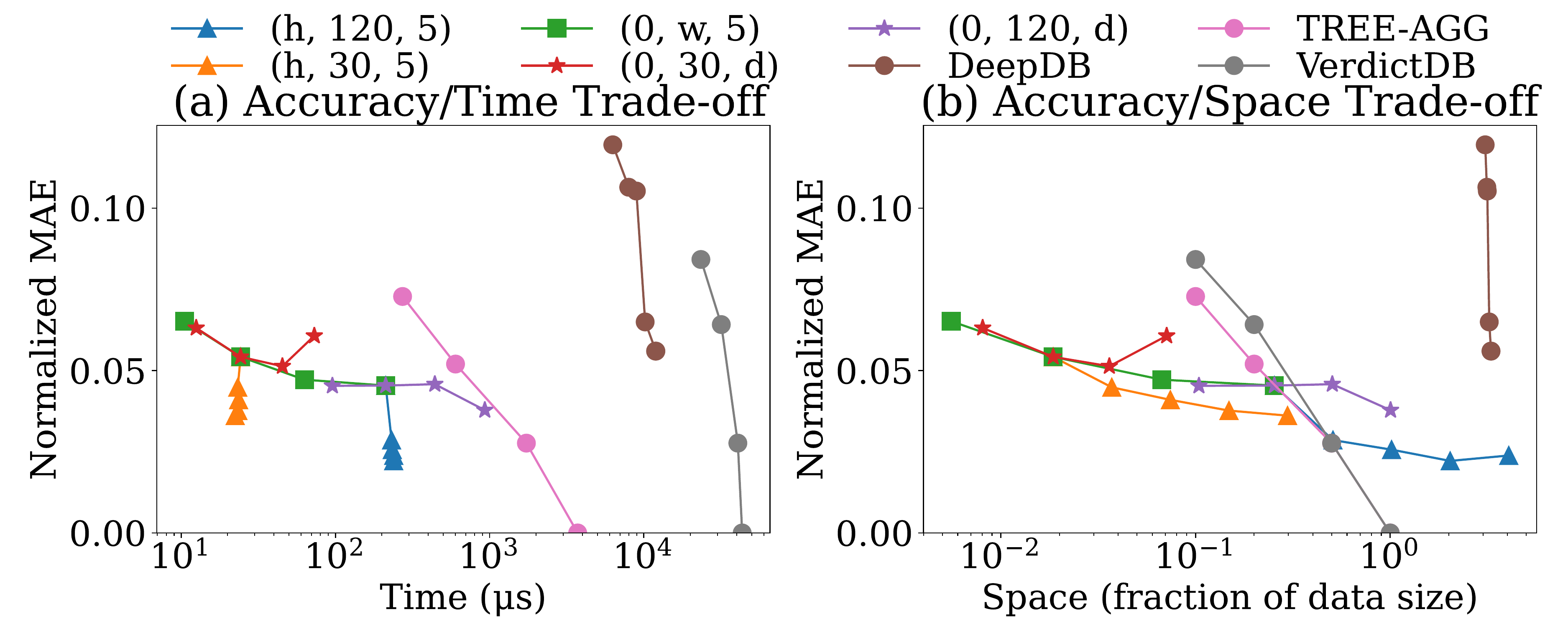}    	   
    	\caption{Time/Space/Accuracy Trade-Off with \\Different Model Architectures}
        \label{fig:exp:RAQ_tradeoff}    
\end{minipage}
\hfill
\begin{minipage}{0.2\textwidth}
\centering
\includegraphics[width=\textwidth]{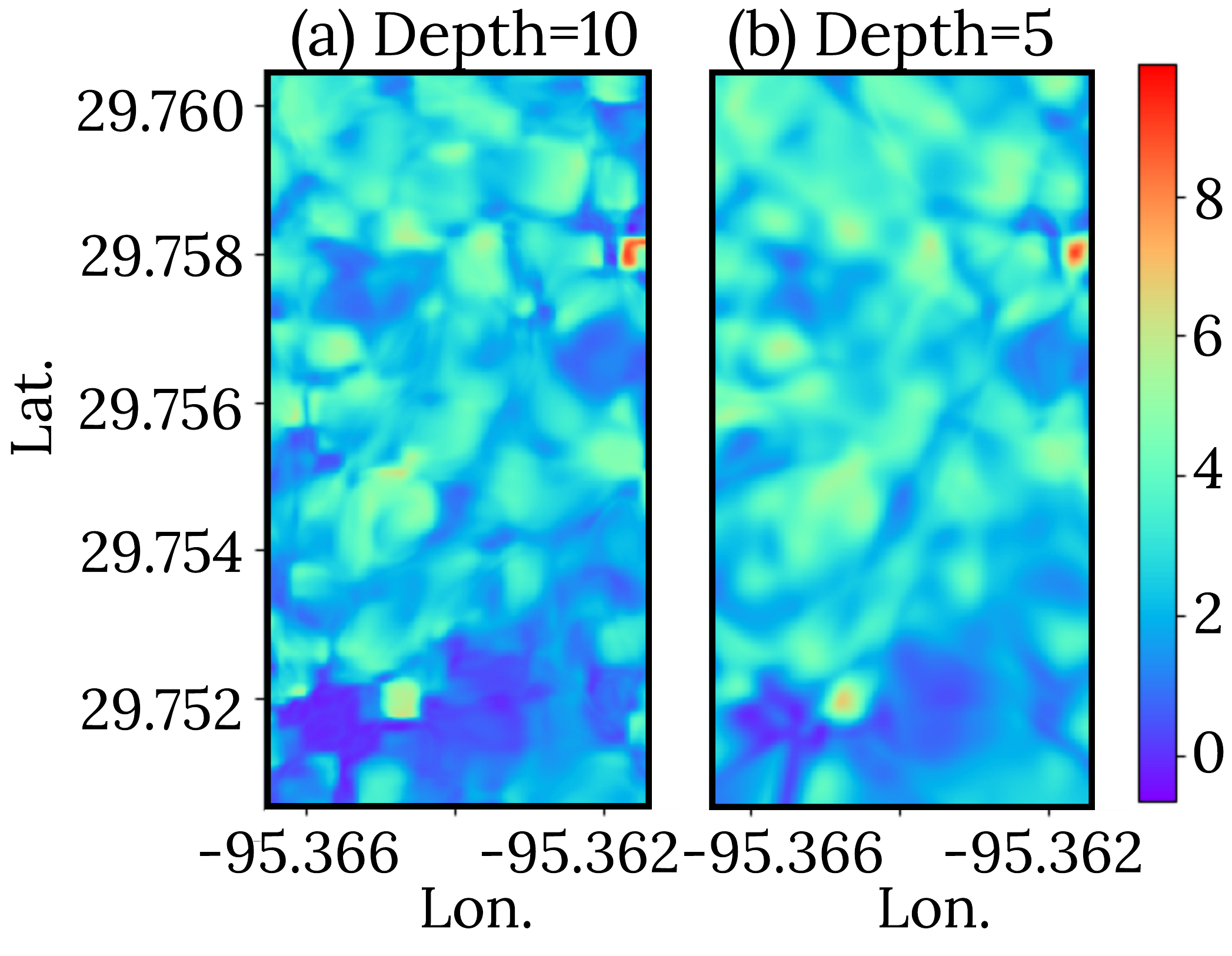}
    \caption{Learned NeuroSketch Visualization}
    \label{fig:learned_query_functions}
\end{minipage}
\hfill
\begin{minipage}{0.35\textwidth}
        \centering
    	\includegraphics[width=\textwidth]{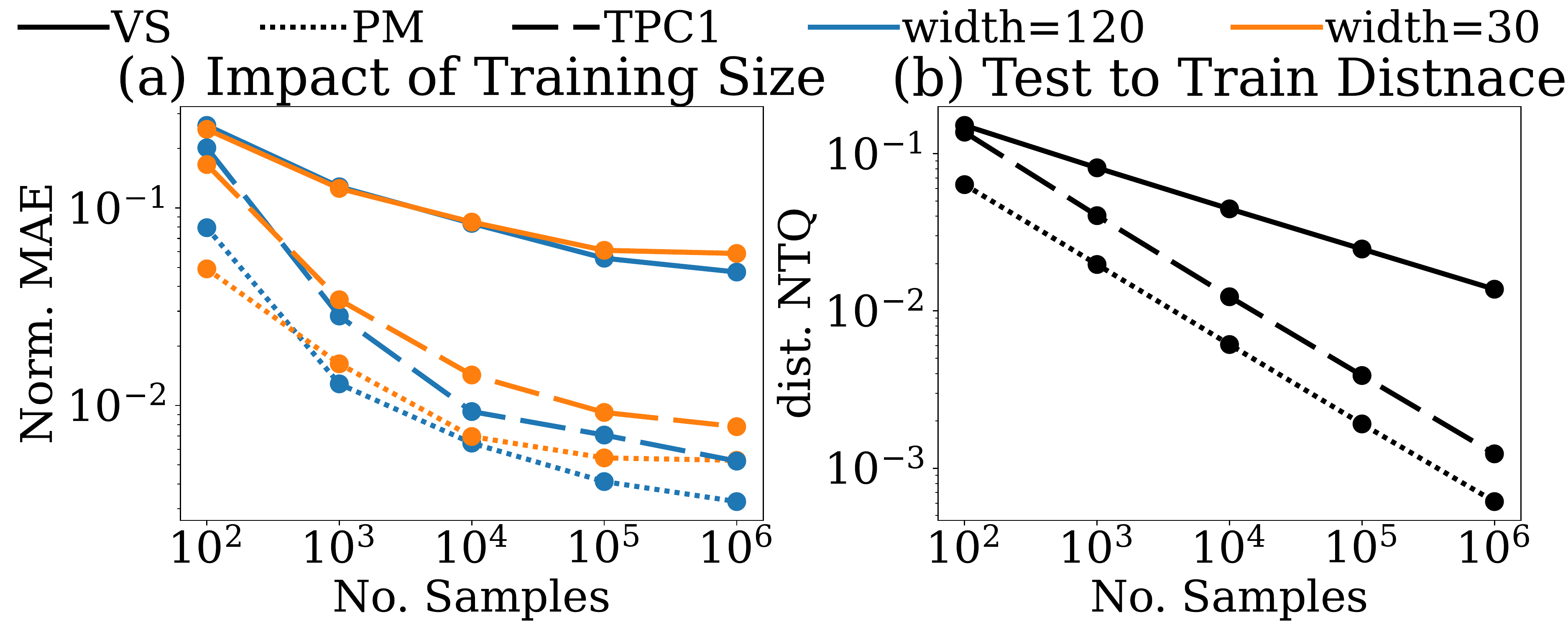}
        \caption{Generalization Study}
        \label{fig:exp:generalization}    
\end{minipage}
\end{figure*}

\begin{table}[t]
    \centering
    \begin{tabular}{c|c|c|c}
        Metric &  NeuroSketch  & \begin{tabular}{@{}c@{}}{TREE-} \\ {AGG}\end{tabular} & \begin{tabular}{@{}c@{}}{DeepDB \&} \\ {VerdictDB}\end{tabular}  \\\hline
        Norm. MAE & 0.045 & 0.052 & N/A  \\\hline
        Query time ($\mu s$) & 25 & 601  & N/A
    \end{tabular}
    \caption{Median visit duration for general rectangles}
    \label{tab:median_visit_res}
\end{table}

\vspace{-0.4cm}
\subsection{Model Architecture Analysis}\label{exp:veraset}
\vspace{-0.1cm}
\subsubsection{Time/Space/Accuracy Trade-Offs of Model Architectures\nopunct}\label{sec:exp:system:tradeoffs}\hfill\\ \textbf{Setup}. We study different time /space/accuracy trade-offs achievable by NeuroSketch and other methods in Fig.~\ref{fig:exp:RAQ_tradeoff} based on different system parameters. For NeuroSketch, we vary number of layers (referred to as depth of the neural network), $d$, number of units per layer (referred to as width of the neural network), $w$, and height of the kd-tree, $h$, to see their impact on its time/space/accuracy (we avoid merging kd-tree nodes here, and study the impact of merging separately in Sec.~\ref{exp:ablation}). Fig.~\ref{fig:exp:RAQ_tradeoff} shows several possible combinations of the hyperparameters. For each line in Fig.~\ref{fig:exp:RAQ_tradeoff}, NeuroSketch is run with two of the hyperparameters kept constant and one changing. The line labels are of the form $(height, width, depth)$, where two of \textit{height, width} or \textit{depth} have numerical values and are the constant hyperparameters for that particular line. Furthermore, the value of one of \textit{height}, \textit{width} or \textit{depth} is $\{d, w, h\}$ and is the variable hyperparameter for the plotted line. For example, line labelled (h, 120, 5) means the experiments for the corresponding line are with a NeuroSketch architecture with 120 number of units per layer, 5 layers and each point plotted corresponds to a different value for the kd-tree height, and label (0, 30, d) means the experiments are run with varying depth of the neural network, with kd-tree height 0 (i.e. only one partition) and the neural width network is 30. The hyperparameter values are as follows. For lines (h, 120, 5) and (h, 30, 50), kd-tree height is varied from 0 to 4, for the line labelled (0, w, 5) neural network width is $\{15, 30, 60, 120\}$ and for lines (0, 120, d) and (0, 30, d) neural network depth is $\{2, 5, 10, 20\}$.

TREE-AGG and VerdictDB are plotted for sampling sizes of 100\%, 50\%, 20\% and 10\% of data size. \revision{For DeepDB, we report results for  RDC thresholds in [0.1, 1] (minimum error is at RDC threshold=0.3. Error increases for values less than 0.1 or more than 1)}.

\textbf{Results}. Fig.~\ref{fig:exp:RAQ_tradeoff} (a) shows the trade-off between query time and accuracy. NeuroSketch performs well when fast answers are required but some accuracy can be sacrificed, while if accuracy close to an exact answer is required, TREE-AGG can perform better. Furthermore, Fig.~\ref{fig:exp:RAQ_tradeoff} (b) shows the trade-off between space consumption and accuracy. Similar to time/accuracy trade-offs, we observe that when the error requirement is not too stringent, NeuroSketch can answer queries by taking a very small fraction of data size. Finally, NeuroSketch outperforms DeepDB in all the metrics. Furthermore, comparing TREE-AGG with VerdictDB shows that, on this particular dataset, the sampling strategy of VerdictDB does not improve upon uniform sampling of TREE-AGG while the R-tree index of TREE-AGG improves the query time over VerdictDB.

Moreover, Fig~\ref{fig:exp:RAQ_tradeoff} shows the interplay between different hyperparameters of NeuroSketch. We see that increasing depth and width of the neural networks improves the accuracy, but after a certain accuracy level the improvement plateaus and accuracy even worsens if depth of the neural network is increased but the width is too small (i.e., the red line). Nevertheless, using partitioning method allows for further improving the time/accuracy trade-off as it improves the accuracy at almost no cost to query time. We also observe that kd-tree improves the space/accuracy trade-off, compared with increasing the width or depth of neural networks. This shows that our paradigm of query specialization is beneficial, as learning multiple specialized models each for a different part of the query space performs better than learning a single model for the entire space. We discuss these results in the context of our DQD bound in Sec.~\ref{sec:theory_example}.

\subsubsection{Visualizing NeuroSketch for Different Model Depth} Fig.~\ref{fig:learned_query_functions} shows the function NeuroSketch has learned for our running example, for two neural networks with the same architecture, but with depths 5 and 10. Comparing Fig.~\ref{fig:learned_query_functions} with Fig.~\ref{fig:visit_query_function}, we observe that NeuroSketch learns a function with similar patterns as the ground truth but the sharp drops in the output are smoothened out. We also observe that the learned function becomes more similar to the ground truth as we increase the number of parameters. Note that the neural networks are of size about 9\% and 3.8\% of the data size.

\vspace{-0.3cm}
\subsection{NeuroSketch Generalization Analysis}\label{sec:exp:generalization}
\begin{figure*}
\centering
\begin{minipage}{0.64\textwidth}
        \centering
    	\includegraphics[width=0.33\textwidth]{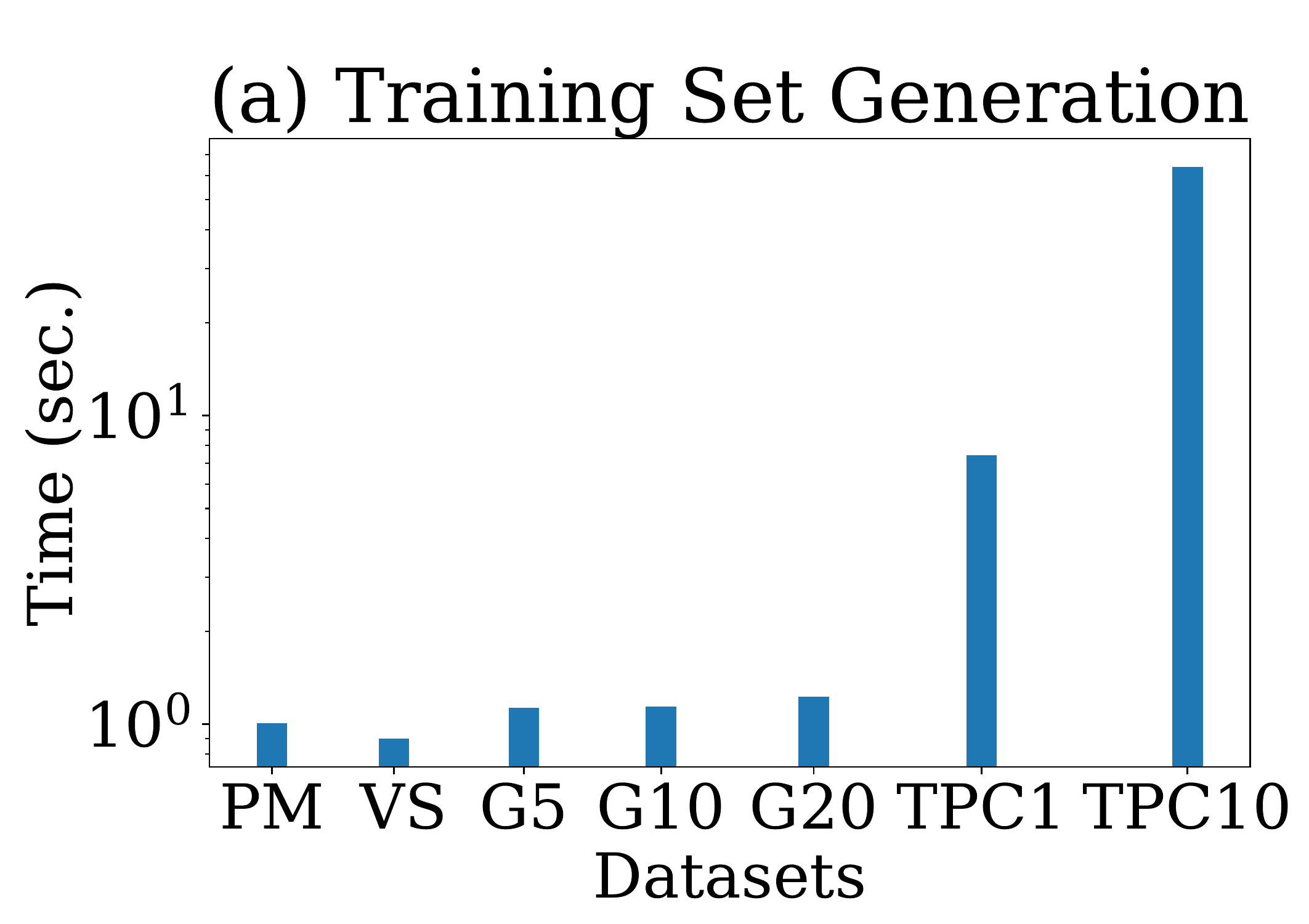}
    	\includegraphics[width=0.582352941\textwidth]{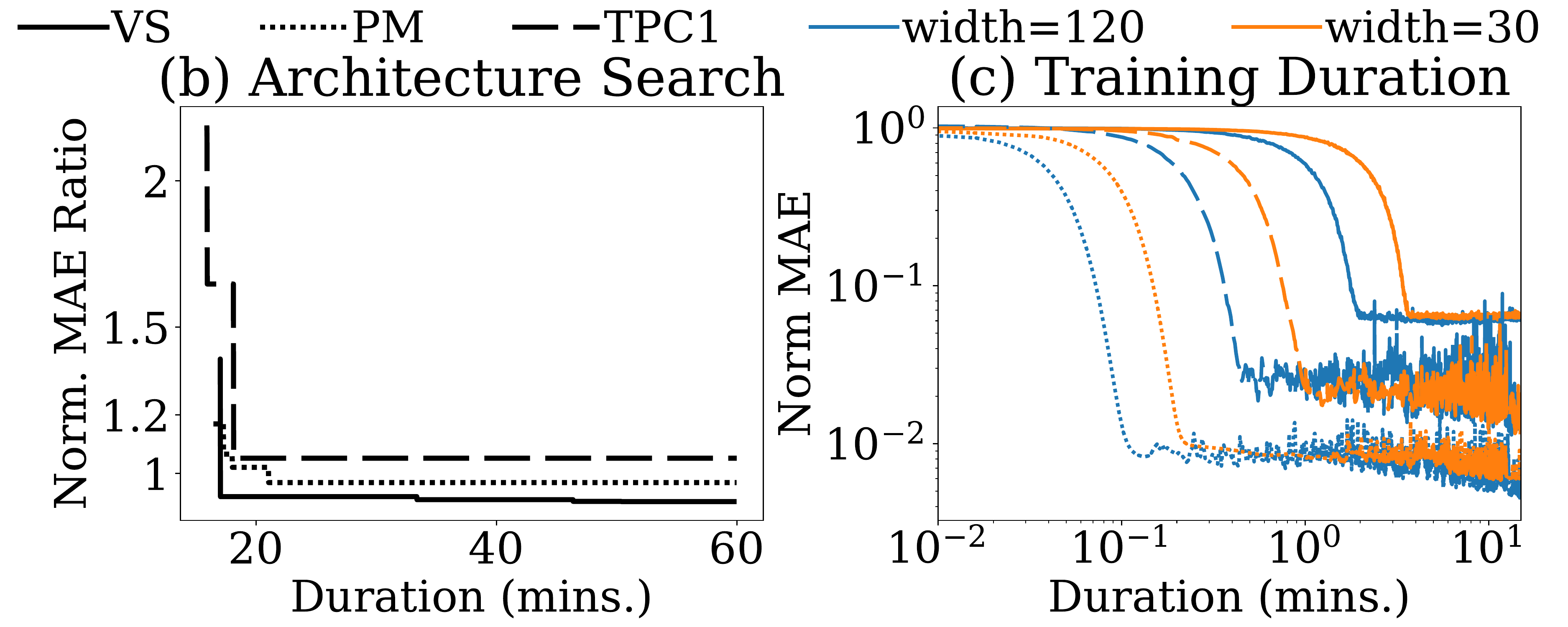}
        \caption{Preprocessing Time Study}
        \label{fig:exp:RAQ_training}    
\end{minipage}
\hfill
\begin{minipage}{0.35\textwidth}
    \centering
    \includegraphics[width=0.9\textwidth]{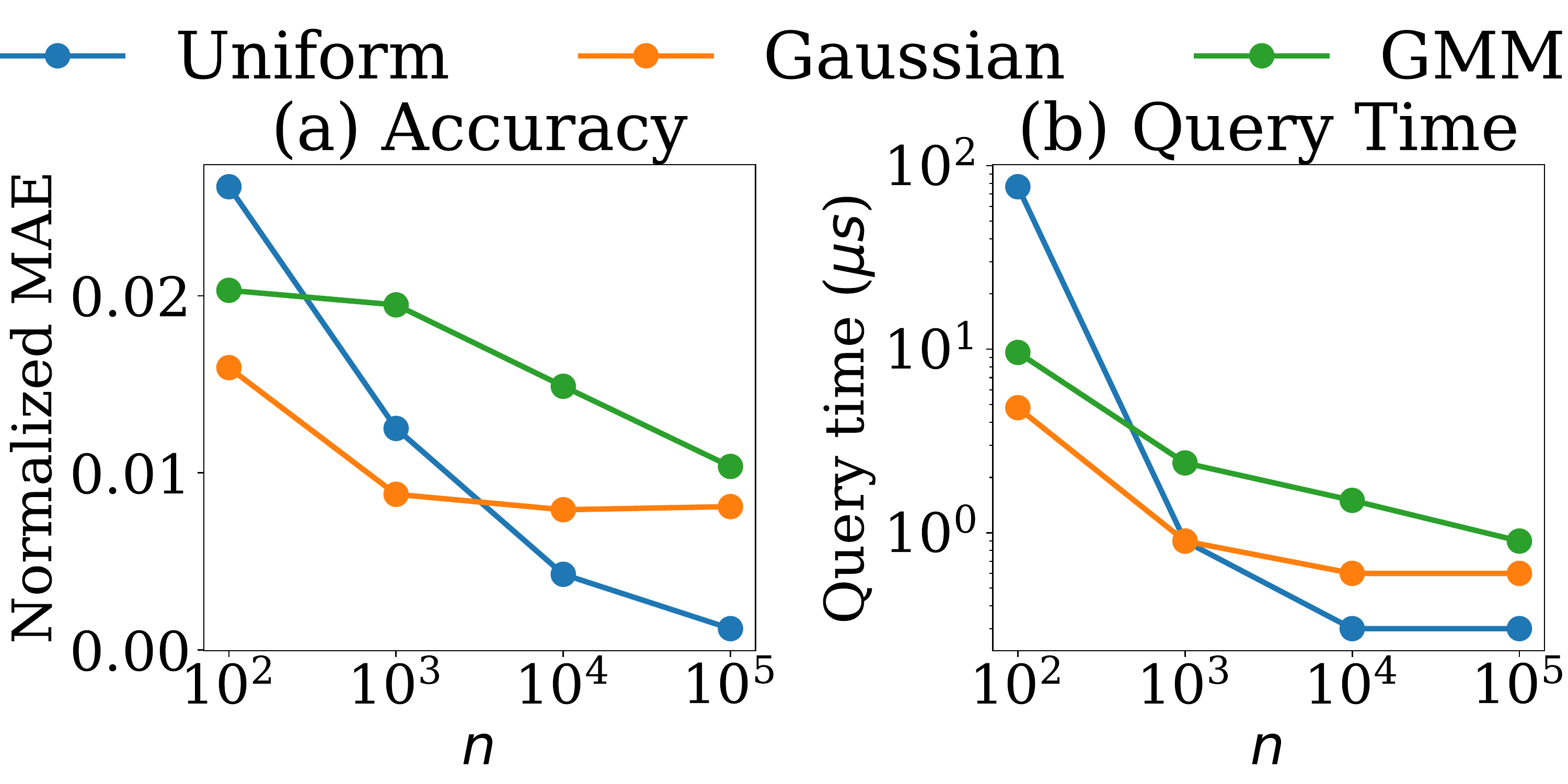}
    \caption{DQD Bound on Synthetic Datasets}
    \label{fig:motivating_exp}
\end{minipage}
\end{figure*}

\revision{Fig.~\ref{fig:exp:generalization} studies  generalization ability of NeuroSketch from train to test queries across across datasets. The results are for a NeuroSketch with tree height 0 (i.e., no partitioning), neural network depth 5 and with neural network widths of 30 and 120.  Fig.~\ref{fig:exp:RAQ_training} (a) shows that training size of about 100,000 sampled query points is sufficient for both architectures to achieve close to their lowest error. Furthermore, when sample size is very small, smaller architecture generalizes better, while the larger neural network improves performance when enough samples are available. }

\revision{In Fig.~\ref{fig:exp:RAQ_training} (b), we plot the average Eucleadian distance from test queries to their nearest training query, refered to as dist. NTQ. To compare across datasets, datasets are scaled to be in [0, 1] for this plot,  and the difference in dist. NTQ values is due to different data dimensionality and number of active attributes in the queries. We ensure none of the test queries appear in the training set, but as the number of training samples increases, dist. NTQ decreases. Nonetheless, when model size is small, eventhough increasing number of samples beyond 100,000 decreaes dist. NTQ, model accuracy does not improve. This suggests that for small neural networks, the error is due to the capacity limit of the model to learn the query function, and not lack of training data.}

\begin{table}[t]
 \begin{tabular}{c|c|c|c}
    \textbf{Dataset}  & \begin{tabular}{@{}c@{}}\textbf{Normalized} \\ \textbf{AQC STD}\end{tabular}& \begin{tabular}{@{}c@{}}\textbf{\% Improved} \\ \textbf{(Merging)}\end{tabular} & \begin{tabular}{@{}c@{}}\textbf{\% Improved} \\ \textbf{(No Merging)}\end{tabular}\\\hline
        VS &  1.02&	47.6&	44.1\\
        PM &0.30&	22.8&	18.6\\
        TPC1& 0.17	&23.5	&6.7\\
        G5&   0.41	&12.0	&13.2\\
        G10&  0.10	&6.8	&6.8\\
        G20&  0.07	&14.6	&14.6\\
        \bottomrule
        \multicolumn{2}{l}{\textbf{Correlation with STD}} & 0.87 & 0.94
        \end{tabular}
	\caption{Improvement of partitioning over no partitioning}
	\label{tab:partitioning}
\end{table}

\vspace{-0.3cm}
\subsection{Ablation Study of Partitioning}\label{exp:ablation}
We study the impact of merging in the prepossessing step of NeuroSketch. Recall that we set the tree height to $4$, so that the partitioning step creates 16 partitions that are merged using AQC, after which 8 partitions remain. We compare this approach with two alternatives. (1) We perform no partitioning and train a single neural network to answer any query. (2) We set the tree height to $3$ so that we obtain 8 partitions without performing any merging. Table~\ref{tab:partitioning} shows the result of this comparison. It shows that performing partitioning, either with merging or without merging is better than no partitioning across all datasets. Second, for almost all datasets, merging provides better or equal performance compared with no merging. Thus, in practice, using AQC as an estimate for function complexity to merge nodes is beneficial. 

In fact, we observed a correlation coefficient of 0.61 between AQC and the error of trained models, which quantifies the benefits of using AQC as an estimate for function complexity. It also implies that AQC can be used to decide whether a query function is too difficult to approximate. For instance, in a database system, the query optimizer may build NeuroSketches for query functions with smaller AQC, and use a default query processing engine to answer query functions with larger AQC.

Furthermore, Table~\ref{tab:partitioning} shows that the benefit of partitioning is dataset dependent. We observed a strong correlation between the standard deviation of AQC estimates across leaf nodes of the kd-tree and the improvement gain from partitioning. Specifically, Let $R=\{AQC_N, \forall \text{ leaf } N\}$, as calculated in line~~\ref{alg:merge:measure_complexity} of Alg.~\ref{alg:maerge}. We calculate $\frac{\text{STD}(R)}{\text{AVG}(R)}$ as the normalized AQC STD for each dataset. This measurement is reported in the second column of Table~\ref{tab:partitioning}. The last row of the table shows the correlation of the improvement for the partitioning methods with this measure. The large correlation suggests that when the difference in the complexity of approximation for different parts of the space is large, partitioning is more beneficial. This matches our intuition for using partitioning, where our intention is to allow specialized models to focus on the complex parts of the query space. It shows that partitioning is beneficial if there are parts of the space that are more complex than others.

\begin{figure*}
\begin{minipage}{0.47\textwidth}
    \centering
    \includegraphics[width=\columnwidth]{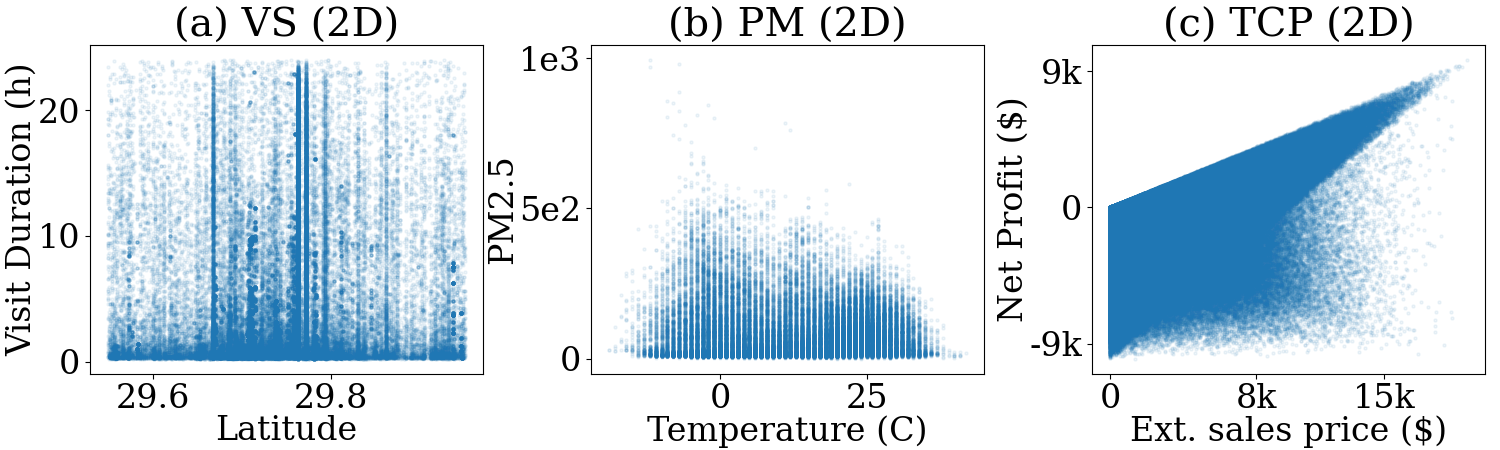}
    \caption{2D data subsets}
    \label{fig:exp:2d_data}    
\end{minipage}
\hfill
\begin{minipage}{0.47\textwidth}
    \centering
    \includegraphics[width=\columnwidth]{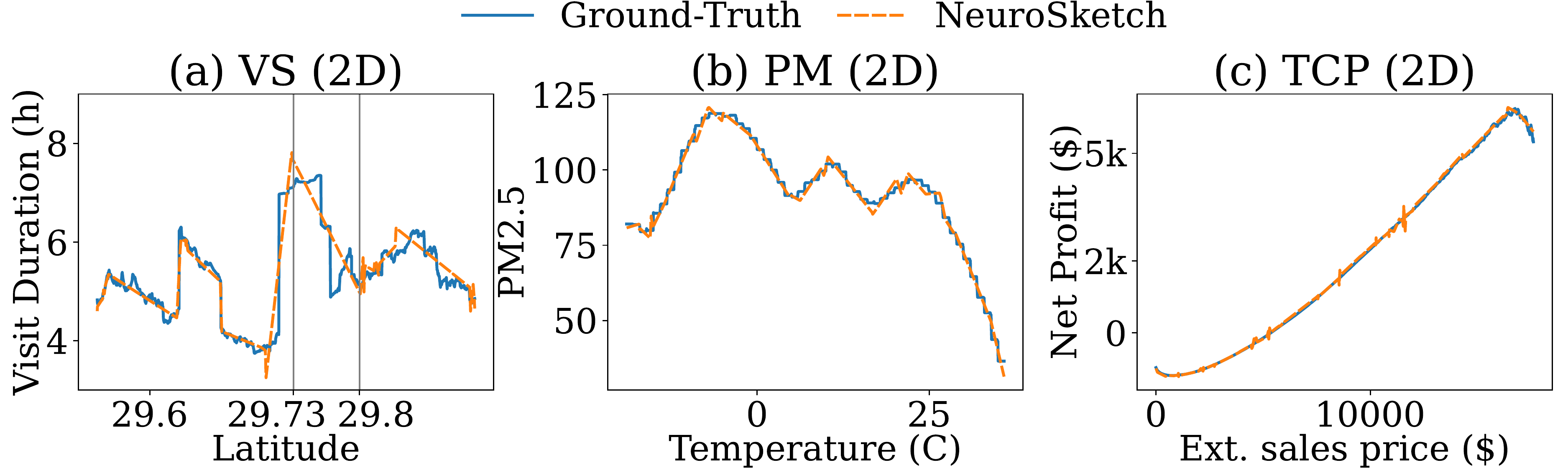}
    \caption{Learned and True Query Functions on 2D Datasets}
    \label{fig:exp:2d_funcs}
\end{minipage}
\end{figure*}

\vspace{-0.3cm}
\subsection{NeuroSketch Preprocessing Time Analysis}\label{sec:exp:preprocessing}
 \revision{\textbf{Training Set Generation}. Fig.~\ref{fig:exp:RAQ_training} (a) shows the time it takes to generate the training set of 100,000 queries is at most 60 seconds, with most datasets taking only a few seconds. The reported results are obtained by answering the queries in parallel on GPU. The queries are answered by scanning all the database records per query and with no indexing. We expect faster training set generation by building indexes.}
 
 \revision{\textbf{Achitecture Search}. Fig.~\ref{fig:exp:RAQ_training} (b) shows the time to perform architecture search for each dataset. We use Optuna \cite{optuna}, a tool that uses baysian optimization to perform hyperparameter search. We use the query time and space requirement (to solve Problem~\ref{prob:RAQ}), to limit maximum number of neural network parameters. Then, we use Optuna to find the width and depth of the neural network that minimizes error. We run Optuna for a total of one hour and set model size limit to be equal to the nueral network size in our default setting. For a point in time, $t$  we report the ratio of error of the best model found by Optuna upto time $t$ divided by error of our default model architecture. This ratio over time is plotted in Fig.~\ref{fig:exp:RAQ_training} (b). The figure shows that Optuna find a model that provides accruacy within 10\% of our default architecture in around 20 minutes. It also finds a better architecture for VS dataset than our default, showing that NeruoSketch accuracy can be improved by performing dataset specific parameter optimization.  Optuna trains models in parallel (multiple models fit in a single GPU), and also stops training early if a setting is not promissing, so that more than 300 parameter settings are evaluated in the presented one hour for each dataset.}
 
 \revision{\textbf{Training Time}. Fig.~\ref{fig:exp:RAQ_training} (c) shows the accuracy of neural networks during training. Models converge within 5 minutes of training across datasets, and error fluctuates when training for longer. Models with larger width converge faster.}

\vspace{-0.3cm}
\subsection{Confirming DQD Bound with NeuroSketch}\label{sec:theory_example}

\textbf{Model Size and DQD}. We revisit Fig~\ref{fig:exp:RAQ_tradeoff} in the context of our DQD bound. First, unsurprisingly, we observe that the overall trend of improved accuracy for larger models matches DQD. More interestingly, we further observe that Fig~\ref{fig:exp:RAQ_tradeoff} shows increase in data size increases accuracy, but only up to a certain point, after which increasing model size has little impact. This also matches DQD, where, in Theorem~\ref{thm:sum_count_all}, increasing size which reduces $\varepsilon_1$ only reduces total error (i.e., $\varepsilon_1+\varepsilon_2$) up to when $\varepsilon_1=0$. After $\varepsilon_1=0$, error cannot be reduced further by increasing number of parameters. As discussed in Sec.~\ref{sec:dqd:statement}, this is because $f_D$, unlike $f_{\chi}$, may be a discontinuous function, so error of a neural network is not guaranteed to ever go to zero (i.e. Theorem.~\ref{thm:nn_appx_error} doesn't apply to $f_D$).

\textbf{Data Size, LDQ and DQD}. We corroborate the observations made in the DQD bound with NeuroSketch using synthetic datasets, so that we can calculate the corresponding LDQs. We sample $n$ points from uniform, Gaussian and two-component GMM distributions (see Sec.~\ref{sec:dqd:dist_dep} on how to calculate their LDQs) and answer RAQs with \texttt{COUNT} aggregation function on the sampled datasets, varying the value of $n$. 
We train NeuroSketch with partitioning disabled to isolate the neural network ability to answer queries. 

Fig.~\ref{fig:motivating_exp} shows the result of this experiment. In Fig.~\ref{fig:motivating_exp} (a), we fix the neural network architecture so that query time and space complexity is fixed (we use one hidden layer with 80 units) and train NeuroSketch for different data sizes and distributions. We observe that, as DQD bound suggests, the error decreases for larger data sizes. Furthermore, uniform distribution, which has a smaller LDQ, achieves the lowest error, then Gaussian whose LDQ is larger and finally GMM which has the largest LDQ. Fig.~\ref{fig:motivating_exp} (b) shows similar observations, but with accuracy fixed to 0.01 and space and time complexity allowed to change. Specifically, we perform a grid search on model width, where we train NeuroSketch for different model widths and find the smallest model width where the error is at most 0.01. We report query time of the model found with our grid search in Figs.~\ref{fig:motivating_exp} (b). As DQD bound suggests, the query time and space consumption decrease when data size increases. Moreover, the same observations hold for storage cost, where we haven't plotted the results as they look identical to that of Figs.~\ref{fig:motivating_exp} (b) (both storage cost and query time are a  constant multiple of the number of parameters of the neural network, so both storage cost and query time are constant multiples of each other). 

Interestingly, for small data sizes, the difficulty of answering queries across distributions does not follow their LDQ order, where uniform distribution is harder when $n=100$ compared with a Gaussian distribution. 
When data size is small, a neural network has to memorize the location of all the data points, which can be more difficult with uniform distribution as the observed points may not follow any recognizable pattern. Nonetheless, as data size increases, as suggested by DQD bound, the error, query time and space complexity improve, and the difficulty of answering queries from different distributions depends on the LDQ.

\revision{\textbf{DQD and Real/Benchmark Distributions}. To further investigate impact of data distribution on accuracy, we visualize 2D subsets of PM, VS and TPC1. We perform RAQs that ask for \texttt{AVG} of the measure attribute where predicate column falls between $c$ and $c+r$, where $r$ is fixed to 10\% of column range and $c$ is the query variable (and input to the query function). Fig.~\ref{fig:exp:2d_data} plots the datasets.  Fig.~\ref{fig:exp:2d_funcs} shows the corresponding true query functions and the function learned by NeuroSkech (without partitioning). Sharp changes in the VS dataset caues difficulties for NeuroSketch, leading to inaccuracies around such sharp changes. This is reflected in both AQC and MAE values shown in Table~\ref{tab:real_dist_analysis} (Norm. AQC is AQC of the functions after they are scaled to [0, 1] to allow for comparions across datasets), where PM and TPC which have less such changes have smaller AQC and MAE. }

\revision{We use Fig.~\ref{fig:exp:2d_funcs} (a) to illustrate why abrupt changes (i.e., large LDQ) make function approximation difficult. Observe in Fig.~\ref{fig:exp:2d_funcs}~(a) such an abrupt change in query function where lat. is between 29.73 and 29.8 (the begning and end of the linear piece are marked in the figure with vertical lines). We see that a single linear piece is assigned to approximate the function in that range (recall that ReLU neural networks are piece-wise linear functions). Such a linear piece has high error, as it cannot capture the (non-linear) change in the function. The error resuling from this approximation grows as the magnitude of the abrupt change in the true function increases. Alternatively, more linear pieces are needed to model the change in the function, which results in a larger neural network.}

\begin{table}[t]
    \centering
      \begin{tabular}{c|c|c|c}
        \textbf{Dataset}& VS (2D) & PM (2D) & TPC (2D) \\\hline
        \textbf{Norm. MAE} &  0.035& 0.014 &  0.0029  \\\hline
        \textbf{Norm. AQC} &  1.28& 0.95 &  0.77
    \end{tabular}    
    \caption{DQD Bound on 2D Real/Benchmark Datasets}
    \label{tab:real_dist_analysis}
\end{table}

\vspace{-0.3cm}
\section{Related Work}\label{sec:rel_work}
\noindent\textbf{Answering RAQs}. The methods for answering RAQs can be divided into sampling-based methods \cite{hellerstein1997online, agarwal2013blinkdb, chaudhuri2007optimized, park2018verdictdb} and model-based methods \cite{graham2012synopses, schmidt2002propolyne, ma2019dbest, thirumuruganathan2019approximate, hilprecht2019deepdb, zeighami2022neural, ahuja2022neural}. Sampling-based methods use different sampling strategies  (e.g., uniform sampling, \cite{hellerstein1997online}, stratified sampling \cite{chaudhuri2007optimized, park2018verdictdb}) and answer the queries based on the samples.  Model-based methods develop a model of the data that is used to answer queries. The models can be of the form of histograms, wavelets, data sketches (see \cite{graham2012synopses} for a survey) or learning based regression and density based models \cite{ma2019dbest, thirumuruganathan2019approximate, hilprecht2019deepdb}. These works create a model of the data and use the data models to answer queries. 

In the case of learned models, a model is created that learns the data, in contrast with NeuroSketch that predicts the query answer. That is, regression and density based models of \cite{ma2019dbest}, generative model of \cite{thirumuruganathan2019approximate} and the sum-product network of \cite{hilprecht2019deepdb} are models of the data created independent of potential queries. We experimentally showed that our modeling choice allows for orders of magnitude performance improvement. Secondly, data models can answer specific queries, (e.g. \cite{ma2019dbest} answers only COUNT, SUM, AVG, VARIANCE,
STDDEV and PERCENTILE aggregations) while, our framework can be applied to any aggregation function. 
Finally, our theoretical analysis for using a learned model is novel, in that it studies why and when a neural network can perform well. Such a study is missing across all existing learning based methods.

Furthermore, learned cardinality estimation \cite{kipf2018learned, wu2021unified,hu2022selectivity,yang2019deep,yang2020neurocard} is related to our work, in that it answers \texttt{COUNT} queries. However, we consider general aggregation functions and such methods do not apply (we also observed that modifying a representative of such approaches, \cite{yang2020neurocard}, to answer RAQs performed worse than DeepDB in practice). \cite{kipf2018learned} uses neural networks for cardinality estimation and thus our theoretical results are applicable to justify their success. Furthermore, \cite{hu2022selectivity} theoretically studies training size needed to learn selectivity function, which is orthogonal to our work. 


\noindent\textbf{Neural Network Approximation}. To approximate a function $f$ with a neural network, similar to Theorem~\ref{thm:nn_appx_error} but under different settings, existing work \cite{petersen2018optimal,bolcskei2019optimal,lu2021deep,shen2020deep,yarotsky2017error,yarotsky2020phase,yarotsky2018optimal,changcun2020relu, shen2019nonlinear} characterize neural network size, $s$, in terms of its error, $\varepsilon$, in the form $s=C_1\varepsilon^{-dC_2}$, where $C_1$ and $C_2$ depend on properties of $f$. The works differ in their notions of \textit{size} and assumptions on $f$, leading to different $C_1$ and $C_2$ values. Closest to our setting, \cite{changcun2020relu,shen2020deep,shen2019nonlinear, petersen2018optimal} bound approximation error for Lipschitz functions for a given \textit{number of neural network parameters}, but don't consider the storage cost. 
Storage cost cannot be related to the number of parameters if the magnitude of the parameters are unbounded, as is the case in \cite{changcun2020relu,shen2020deep,shen2019nonlinear}. \cite{petersen2018optimal} also does not explicitly bound the storage cost, but analyzing their construction yields a bound that, compared to our result, is exponentially worse in $\rho$ and polynomially worse in $d$. 


\vspace{-0.2cm}
\section{Conclusion}\label{sec:conc}
\vspace{-0.1cm}
We presented the first DQD bound for an ML method when answering RAQs. Our DQD bound shows how the error of a neural network relates to the data distribution, data size and the query function. 
Based on our DQD bound, we introduced NeuroSketch, a neural network framework for efficiently answering RAQs, with orders of magnitude improvement in query time over the state-of-the-art algorithms. A NeuroSketch trained for a query function is typically much smaller than the data and answers RAQs without accessing the data. This is beneficial for efficient release and storage of data. For instance, location data aggregators (e.g., SafeGraph \cite{safegraph}) can train a NeuroSketch to answer the average visit duration query, and release it to interested parties instead of the dataset. This improves storage, transmission and query processing costs for all parties. 
Future work can focus on DQD bounds for high dimensions and studying approximation error for separate function classes. Our Lipschitz assumption is very generic (only assumes a bound on the function derivative magnitude), and can yield a loose bound in high dimensions or for some functions classes (e.g., linear functions that can have large derivative magnitude but are easy to approximate). Additionally, modeling impact of query workload on neural network accuracy, as well as studying
parallelism and model pruning methods \cite{blalock2020state} to remove \textit{unimportant} model weights for faster evaluation time. \revision{Support for dynamic data is another interesting future direction. One approach is to frequently test NeuroSketch, and re-train the neural networks whose accuracy fall below a certain threshold. We conjecture that DQD can be used to decide how often retraining is required.}


\section*{Acknowledgement} 
This research has been funded in part by NSF grants IIS-1910950, CNS-2125530, and IIS-2128661, NIH grant 5R01LM014026, and an unrestricted cash gift from Microsoft Research. Vatsal Sharan was supported by NSF CAREER Award CCF-2239265 and an Amazon Research Award. Any opinions, findings, conclusions or recommendations expressed in this material are those of the author(s) and do not necessarily reflect the views of any of the sponsors such as the NSF.

\bibliographystyle{ACM-Reference-Format}
\balance
\bibliography{references} 

\appendix

\section{Proofs}\label{appx:proof}
\subsection{Proof of Theorem~\ref{thm:nn_appx_error}}

To bound the approximation error, we first establish that the memorization is correct at the vertices of all the cells. Then, we ensure that the change in the neural network is bounded within the cell. Since the neural network is exactly accurate at vertices of the cells and it doesn't change too much within each cell, its error within each cell is bounded. We present a sequence of lemmas to formally establish this argument, the proofs of which are deferred Sec.~\ref{sec:lemma:proofs}. 

First we establish the correct memorization property.

\begin{lemma}[Memorization]\label{lemma:correct_mem}
For all, $\vect{p}\in P$, $|f(\vect{p})-\hat{f}(\vect{p})|=0$.
\end{lemma}

Next, we bound neural network change in the following Lemma. For the purpose of the lemma, define $C^*_i=\{\vect{x}\in \mathbb{R}^d: x_r\in[\frac{\pi^i_r}{t}, \frac{\pi^i_r}{t}+\frac{1}{t}]\forall 1\leq r\leq d\}$, which is the subset of the input space that falls in the $i$-th cell. Also define $C_i=\{\vect{x}\in \mathbb{R}^d: x_r\in[\frac{\pi^i_r}{t}, \frac{\pi^i_r}{t}+(\frac{1}{t}-\frac{1}{Mt})]\forall 1\leq r\leq d\}$, which is a subset of $C^*_i$ and let $C'_i=C^*_i\setminus C_i$. The lemma divides each cell into two regions, $C_i$ and $C'_i$ and bounds the neural network change in each region. When $d\leq 3$, we are able to prove a tighter bound on the neural network change, which helps prove the tighter  bound in low dimensions of Theorem~\ref{thm:nn_appx_error}.
\begin{lemma}[Bounded Change]\label{prop:change}
For any, $i\in\{0, ..., (t+1)^d-1\}$, 
\begin{enumerate}[label=(\alph*)]
    \item For all $\vect{x}\in C_i$, we have $\hat{f}(\vect{x})=f(\frac{\vect{\pi}^i}{t})$.
    \item For all $\vect{x}\in C_i, \vect{x}'\in C'_i$, and for all $\vect{x}\in C'_i, \vect{x}'\in C'_i$ , we have $|\hat{f}(\vect{x})-\hat{f}(\vect{x}')|\leq \frac{kd^3\rho 2^{d-1}}{t}$
    \item If $d\leq 3$, for all $\vect{x}, \vect{x}'\in C^*_i$, we have $|\hat{f}(\vect{x})-\hat{f}(\vect{x}')|\leq 36\frac{\rho d}{t}$
\end{enumerate}
\end{lemma}

Using the above lemma, together with the $\rho$-Lipschitz property of $f$ and triangle inequality, integrating over $\mathbf{x}$ to obtain 1-norm, or taking the $\infty$-norm for $d\leq 3$ gives the following bound on neural network error.

\begin{lemma}[Bounded Error]\label{prop:error} The neural network error is bounded as follows.

\begin{enumerate}[label=(\alph*)]
    \item $\normx{\hat{f}-f}_1\leq \frac{3\rho d}{t}$
    \item If $d\leq 3$, $\normx{\hat{f}-f}_\infty\leq \frac{37\rho d}{t}$
\end{enumerate}
\end{lemma}

Furthermore, we bound the space and time complexity of the neural network as follows.

\begin{lemma}[Space and Time Complexity]\label{lemma:space}
Number of bits needed to store the neural network parameters is $O(kd\log \rho+d\log d+\log k)=\Tilde{O}(kd)$ and a neural network forward pass requires $O(kd)$ operations. 
\end{lemma}


Theorem~\ref{thm:nn_appx_error} follows by setting $\varepsilon_1=\frac{\varkappa\rho d}{t}$, for $\varkappa=37$, and recalling that $k=(t+1)^d$ so that $k=(\varkappa\rho d \varepsilon_1^{-1}+1)^d$. Thus, Error is bounded by $\varepsilon_1$, and space and time complexity are $\Tilde{O}(d(\varkappa\rho d \varepsilon_1^{-1}+1)^d)$.\qed

\subsection{Proof of Technical Lemmas for Theorem~\ref{thm:nn_appx_error}}\label{sec:lemma:proofs}

\subsubsection{\textbf{Proof of Memorization Lemma \ref{lemma:correct_mem}}}
\begin{figure}[t]
    \centering
    \includegraphics[width=0.5\columnwidth]{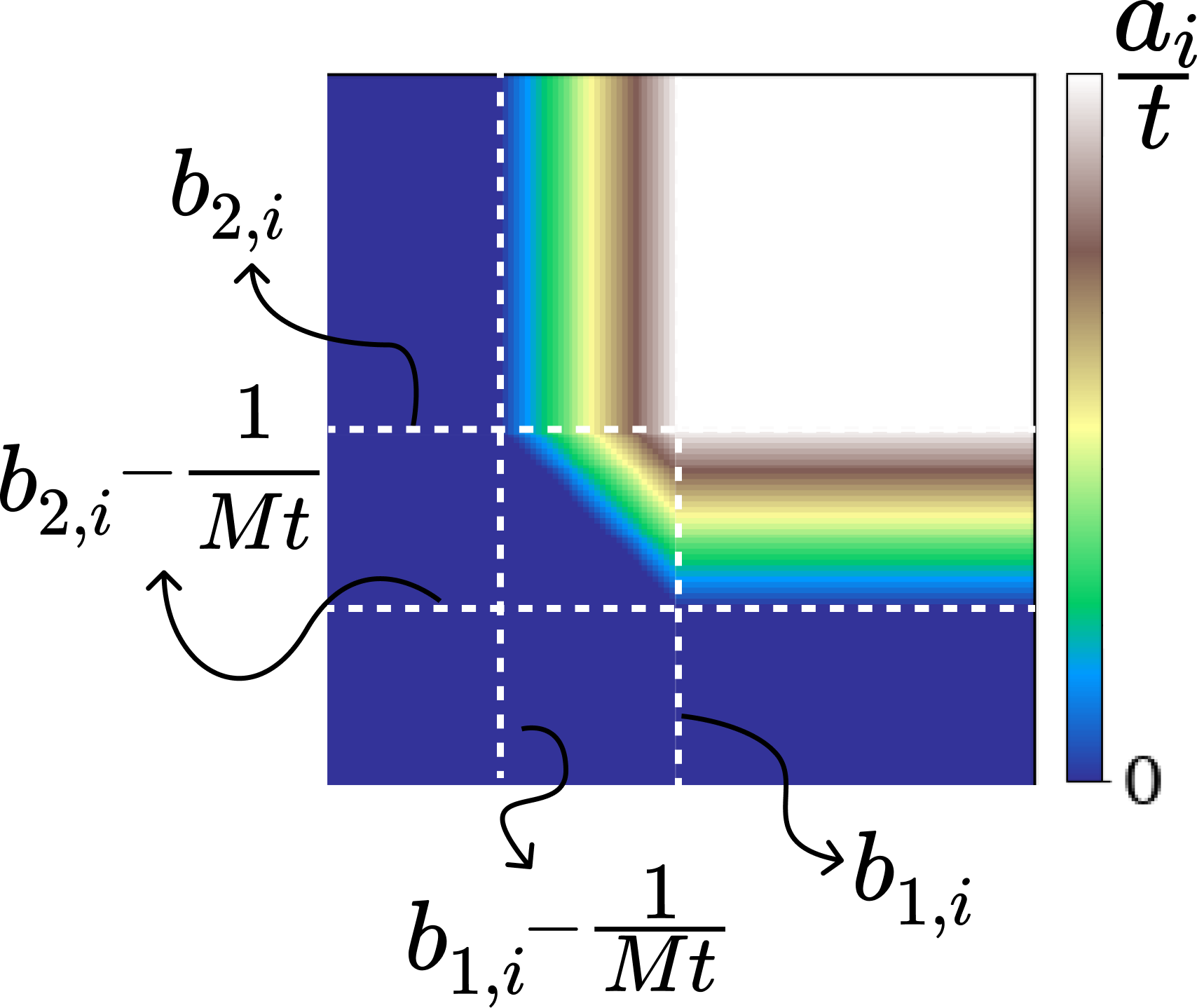}
    \caption{Function surface of $\hat{g}_i(\vect{x})$ for a 2-dimensional $\vect{x}$}
    \label{fig:gunit}
\end{figure}

Intuitively, the memorization property follows based on the construction of g-units, as shown in Fig.~\ref{fig:gunit}. As the figure shows, g-units are non-zero only for a quadrant of the space, the location of which can be controlled with g-unit parameters. This ensures that when the construction iteratively memorizes new points, the neural network will not forget the value of the previously memorized points. The proof formalizes this idea.

The following proposition first establishes some properties of the construction. 

\begin{proposition}\label{prop:gunit_properties}
Based on the construction in Alg.~\ref{alg:construction}, the following properties hold.
\begin{enumerate}[label=(\alph*)]
\item For any $i, j\in\{0, ..., (t+1)^d-1\}$, we have 
\begin{align*}
    \hat{g}_j(\frac{\vect{\pi}^i}{t})=
    \begin{cases}
    \frac{a_j}{t} & \text{if}\;\;\forall r,  \pi^j_r\leq\pi^i_r\\
    0              & \text{otherwise}
\end{cases}
\end{align*}
\item At the $i$-th iteration of Alg.~\ref{alg:construction}, we have $b+\sum_{j=1}^{i}\hat{g}_{i}(\vect{\pi}^i/t)=f(\vect{\pi}^i/t)$.
\end{enumerate}
\end{proposition} 
\textit{Proof of Prop.~\ref{prop:gunit_properties}.} To prove part (a) 
First, note that based on the construction, a g-unit can be written as

\begin{align}\label{eq:gunit_filled}
\hat{g}_j(\vect{x})= a_j\sigma(\sum_{r=1}^{d}-M\sigma( -x_r+\frac{\pi^j_r}{t})+\frac{1}{t})    
\end{align}

Assume for some $r$, we have $\frac{\pi^j_r}{t}-\frac{\pi^i_r}{t}>0$, so $\sigma(-\frac{\pi^i_r}{t}+\frac{\pi^j_r}{t})=-\frac{\pi^i_r}{t}+\frac{\pi^j_r}{t}$. Together with Eq.~\ref{eq:gunit_filled} we get
\begin{align*}
\hat{g}_j(\vect{x})= a_j\sigma(M(\frac{\pi^i_r}{t}-\frac{\pi^j_r}{t})+\frac{1}{t}+\sum_{r'=1, r'\neq r}^{d}-M\sigma( -\frac{\pi^i_{r'}}{t}+\frac{\pi^j_{r'}}{t}))    
\end{align*}

$\pi^i_r$ and $\pi^j_r$ are integers so we have $\pi^i_r\leq \pi^j_r-1$ and recall that $M\geq 1$. Thus, $M(\frac{\pi^i_r}{t}-\frac{\pi^j_r}{t})+\frac{1}{t}\leq 1(\frac{\pi^j_r-1}{t}-\frac{\pi^j_r}{t})+\frac{1}{t}= 0$. Given that $\sum_{r'=1, r'\neq r}^{d}-M\sigma( -\frac{\pi^i_{r'}}{t}+\frac{\pi^j_{r'}}{t})\leq 0$, we have $\sum_{r'=1}^{d}-M\sigma( -\frac{\pi^i_{r'}}{t}+\frac{\pi^j_{r'}}{t})+\frac{1}{t}\leq 0$ and thus $\hat{g}_j(\frac{\vect{\pi}^i}{t})=0$. If $\forall r$,  $\pi^j_r\leq \pi^i_r$, then $\sigma(-\frac{\pi^i_r}{t}+\frac{\pi^j_r}{t})=0$ for all $r$, So $\hat{g}_j(\frac{\vect{\pi}^i}{t})=\frac{a_j}{t}$

To prove part (b), by line \ref{alg:construction:set_a} of the algorithm, $$\frac{a_i}{t}=f(\frac{\vect{\pi}^i}{t})-(b+\sum_{j=1}^{i-1}\hat{g}_{j}(\frac{\vect{\pi}^i}{t})).$$ The result follows using part (a). \qed

Next, to prove Lemma~\ref{lemma:correct_mem}, by Prop.\ref{prop:gunit_properties}~(b), for any $\vect{p}\in P$, where $\vect{p}=\frac{\vect{\pi}^i}{t}$ for some $i$, at the $i$-th iteration of Alg.~\ref{alg:construction}, we ensure that $\sum_{j=1}^{i}\hat{g}_{i}(\vect{p})+b=f(\vect{p})$. For the $j$-iteration, $j>i$, we have that $\pi^j_r>\pi^i_r$ for some $r$. Thus, by Prop.~\ref{prop:gunit_properties}~(a), $\hat{g}_j(\vect{p})=0$. So, $\hat{f}(\vect{p})=\sum_{j=1}^{k}\hat{g}_{j}(\vect{p})+b=f(\vect{p})$. \qed

\subsubsection{\textbf{Proof of Bounded Change Lemma \ref{prop:change}}}
First, we prove the following Lemma that bounds the magnitude of the weights of the neural network.

\begin{lemma}\label{lemma:a_mag}
For any $i$, $|a_i|\leq 2^{d-1} d\rho$.
\end{lemma}

\noindent\textit{Proof.} For convinience, define $\hat{g}_0(\vect{x})=b$ Then, by line \ref{alg:construction:set_a} of Alg.~\ref{alg:construction}, we have that 

\begin{align}\label{eq:a_mag}
    |a_i|=t|f(\frac{\vect{\pi}^i}{t})-\sum_{j=0}^{i-1}\hat{g}_{j}(\frac{\vect{\pi}^i}{t})|
\end{align}

Consider $\sum_{j=0}^{i-1}\hat{g}_{j}(\frac{pi^i}{t})$. By Prop.~\ref{prop:gunit_properties}~(a), we have that $\sum_{j=0}^{i-1}\hat{g}_{j}(\frac{\vect{\pi}^i}{t})=\sum_{j\in \mathcal{I}_{i}}\frac{a_{j}}{t}$, where
$\mathcal{I}_{i}=\{j\in\mathbb{Z}: 0\leq \pi^j_r\leq \pi^i_r\}\setminus\{i\}$. Define $\mathcal{I}_{i}^{r}=\{j\in\mathbb{Z}: 0\leq \pi^j_{r'}\leq \pi^i_{r'} \;\forall r'\neq r, 0\leq \pi^j_r\leq \pi^i_r-1\}$. Clearly, $\cup_{r=1}^d\mathcal{I}_{i}^{r}=\mathcal{I}_{i}$. Thus, we use the inclusion-exclusion principle to rewrite $\sum_{j\in \mathcal{I}_{i}}\frac{a_{i}}{t}$, making sure each term in the sum is present exactly once. We have 

$$\sum_{j\in \mathcal{I}_{i}}\frac{a_{j}}{t}=\sum_{\emptyset\neq S\subseteq\{1,..., d\}}(-1)^{|S|+1}\sum_{j\in\cap_{r\in S}\mathcal{I}_i^r}\frac{a_j}{t}$$

For any $S$ consider the index $j_S$ such that $\pi^{j_S}_r = \pi^i_r$ if $r\not\in S$ and $\pi^{j_S}_r = \pi^i_r-1$ otherwise. Observe that $\cap_{r\in S}\mathcal{I}_i^r=\{j\in\mathbb{Z}: 0\leq \pi^j_{r'}\leq \pi^i_{r'}, 0\leq \pi^j_r\leq \pi^i_{r}-1, \forall r\in S, r'\not\in S\}=\mathcal{I}_{j_S}$ so that $\sum_{j\in\cap_{r\in S}\mathcal{I}_i^r}\frac{a_j}{t}=\sum_{j\in\mathcal{I}_{j_S}}\frac{a_j}{t}$. Then, by Prop.~\ref{prop:gunit_properties}~(a) and Lemma~\ref{lemma:correct_mem}, $\sum_{j\in\mathcal{I}_{j_S}}\frac{a_j}{t}=\hat{f}(\vect{\pi}^{j_S}/t)=f(\vect{\pi}^{j_S}/t)$. 

Putting this in Eq.~\ref{eq:a_mag}, we get

\begin{align*}
    |a_i|=&t|f(\frac{\vect{\pi}^i}{t})-\sum_{\emptyset\neq S\subseteq\{1,..., d\}}(-1)^{|S|+1}f(\vect{\pi}^{j_S}/t)|\\
    =&t|\sum_{S\subseteq\{1,..., d\}}(-1)^{|S|}f(\vect{\pi}^{j_S}/t)|\\
    &\leq 2^{d-1}d\rho
\end{align*}

Where the last inequality follows from the $\rho$-Lipschitz property of $f$ and that every two points $\vect{\pi}^{j_S}$ and $\vect{\pi}^{j_{S'}}$ for $S, S'\subseteq\{1, ..., d\}$ are at most $\frac{d}{t}$ apart and that there are $2^{d-1}$ positive and negative terms in the summation.\qed

Next, we provide the following lemma to bound the change in a piece-wise linear function.
\begin{lemma}\label{lemma:piecewise_bound}
For a piece-wise linear function $\hat{f}$ where the magnitude of the gradient of each piece is bounded by $B$, and for two points $\vect{x}'$ and $\vect{x}^*$ in the domain of the function, we have $|\hat{f}(x')-\hat{f}(x^*)|\leq B\normx{\vect{x}^* - \vect{x}'}$.
\end{lemma}
Let $h(\alpha)=\hat{f}(\alpha \vect{x}'+(1-\alpha)\vect{x}^*)$, for $\alpha\in[0, 1]$, so we are interested in $|h(0)-h(1)|$. Let $\alpha_1$, ..., $\alpha_l$ be the points of non-linearity of $\hat{f}$ on the line $\{\vect{x}\in\mathbb{R}^d:\vect{x}=\alpha \vect{x}'-(1-\alpha)\vect{x}^*$, $0\leq \alpha\leq 1$\} (i.e., where $\nabla\hat{f}$ does not exist). So $h(\alpha_i)-h(\alpha_{i+1})=\hat{f}(\alpha_i\vect{x}^*+(1-\alpha_i)\vect{x}')-\hat{f}(\alpha_{i+1}\vect{x}^*+(1-\alpha_{i+1})\vect{x}')=\vect{m}^i(\alpha_i\vect{x}^*+(1-\alpha_i)\vect{x}'-(\alpha_{i+1}\vect{x}^*+(1-\alpha_{i+1})\vect{x}'))$ for a vector $\vect{m}^i$ which is the gradient of the $i$-th linear piece of $\hat{f}$. Letting $\alpha_0=0$ and $\alpha_{l+1}=1$, we have $|h(0)-h(1)|=|\sum_{i=0}^{l}h(\alpha_i)-h(\alpha_{i+1})|\leq \sum_{i=0}^{l}|h(\alpha_i)-h(\alpha_{i+1})|$. Since $\normx{\vect{m}^i}\leq B$ for all $i$ we have

\begin{align*}
&\sum_{i=0}^l|h(\alpha_i)-h(\alpha_{i+1})|\\
\leq&\sum_{i=0}^l\normx{\vect{m}^i}\normx{\alpha_i\vect{x}^*+(1-\alpha_i)\vect{x}'-(\alpha_{i+1}\vect{x}^*+(1-\alpha_{i+1})\vect{x}')}\\
\end{align*}
\begin{align*}
\leq &B\sum_{i=0}^l\normx{\alpha_i\vect{x}^*+(1-\alpha_i)\vect{x}'-(\alpha_{i+1}\vect{x}^*+(1-\alpha_{i+1})\vect{x}')}\\
= &B\sum_{i=0}^l(\alpha_{i+1}-\alpha_{i})\normx{\vect{x}^* - \vect{x}'}\\
=&B\normx{\vect{x}'-\vect{x}^*}.
\end{align*}
\qed

Finally, we are ready to prove Lemma~\ref{prop:change}.

\textit{Proof of Part (a)}. For any $i$, we study the behaviour of $\hat{g}_j(\vect{x})$ for $\vect{x}\in C_i$ and all $j$. Note that $x_r$, the $r$-th dimension of $\vect{x}$ can be written as $x_r=\frac{\pi^i_r}{t}+z_r$, where $0\leq  z_r\leq \frac{1}{t}-\frac{1}{Mt}$ for all $r$. If $\exists r$ where $x_r < \frac{\pi^j_r}{t}$, then $\frac{\pi^i_r}{t}<\frac{\pi^j_r}{t}$ so that $x_r-\frac{\pi^j_r}{t}\leq\frac{\pi^j_r-1}{t}+\frac{1}{t}-\frac{1}{Mt}-\frac{\pi^j_r}{t}=-\frac{1}{Mt}$. So $M(x_r-\frac{\pi^j_r}{t})+\frac{1}{t}\leq0$. Given that $\sum_{r'=1, r'\neq r}^{d}-M\sigma( -x_{r'}+\frac{\pi^j_{r'}}{t})\leq 0$, we have $\sum_{r'=1}^{d}-M\sigma( -x_{r'}+\frac{\pi^j_{r'}}{t})+\frac{1}{t}\leq 0$ and thus $\hat{g}_j(\vect{x})=0$. If $\forall r$,  $x_r\geq \frac{\pi^j_r}{t}$, then $\sigma(-x_r+\frac{\pi^j_r}{t})=0$ for all $r$, So $\hat{g}_i(\vect{x})=\frac{a_i}{t}$. Thus, for all $j$, $\hat{g}_j(\vect{x})$ is constant for $\vect{x}\in C_i$, which implies the neural network is constant. Given that $\hat{f}(\frac{\vect{\pi}^i}{t})=f(\frac{\vect{\pi}^i}{t})$, and that $\frac{\vect{\pi}^i}{t}\in C_i$, we have $\hat{f}(\vect{x})=f(\frac{\vect{\pi}^i}{t})$ for all $\vect{x}\in C_i$.

\textit{Proof of Part (b).}  Note that $\hat{f}$ is a piece-wise linear function. 

If $\vect{x}\in C'_i$, let $\vect{x}^*=\vect{x}$. Otherwise, let $\vect{x}^*$ be the closest point in $C_i$ to $\vect{x}'$. Since $\hat{f}$ is constant in $C_i$, $|\hat{f}(\vect{x})-\hat{f}(\vect{x}')|=|\hat{f}(\vect{x}^*)-\hat{f}(\vect{x}')|$, so we only need to prove the result for $\vect{x}^*$. Note that $\normx{\vect{x}^* - \vect{x}'}\leq \frac{d}{Mt}$. Using 
Lemma~\ref{lemma:piecewise_bound},  we have that 
\begin{align}\label{eq:change_first_bound}
    |\hat{f}(\vect{x})-\hat{f}(\vect{x}')|\leq B\frac{d}{Mt}.
\end{align}

It remains to find $B$, the bound on the magnitude of the gradient of $\hat{f}$ for all linear pieces. Note that the derivative in every direction is bounded by $\sum_{j=1}^kM|a_i|$, so that the gradient norm is at most 
\begin{align}\label{eq:gradbound_first}
B\leq d\sum_{j=1}^kM|a_i|.
\end{align}

Combining Eq.~\ref{eq:change_first_bound} with Eq.~\ref{eq:gradbound_first} and Lemma~\ref{lemma:a_mag} we obtain  

\begin{align*}
    |\hat{f}(\vect{x})-\hat{f}(\vect{x}')|&\leq d\sum_{j=1}^kM|a_i|\frac{d}{Mt}\\
    &\leq  2^{d-1} d^3\rho \frac{k}{t}
\end{align*}

which completes the proof.\qed

\textit{Proof of part (c)}. For ease of discussion, we first provide this elementary lemma used to bound the derivative of a linear function.
\begin{lemma}\label{lemma:derivative_linear}
Consider a linear function $f:\mathbb{R}^d\rightarrow \mathbb{R}$, and two points $\vect{x},\vect{x'}\in \mathbb{R}^d$ where $\vect{x}'=\vect{x}+h\vect{x}^i$ for $h>0$ and $\vect{x}^i\in\mathbb{R}^d$, where $x^i_j=1$ for $i=j$ and $x^i_j=0$. The derivative of $f$ in the direction of $\vect{x}^i$, is $\frac{|f(\vect{x})-f(\vect{x'})|}{h}$.
\end{lemma}
\textit{Proof}. Follows trivially from definition of derivative and that $f$ is linear.\qed

To prove part (c), we set $M=1$. We say $f(\vect{x})$ is linear at $\vect{x}$ \textit{in the direction of} $\vect{u}$ if there exists an $\varepsilon>0$ and constants $\vect{a}$ and $b$ such that for all $0<\varepsilon'<\varepsilon$, $f(\vect{x}+\varepsilon' \vect{u})= \vect{a}\vect{x}+b$. As before, we say $f(\vect{x})$ is linear at $\vect{x}$ if it is linear \textit{for all directions} at $\vect{x}$ and that $\vect{x}$ is a point of non-linearity if $f(\vect{x})$ is not linear at $\vect{x}$. 

Observe that non-linearities happen only when the input to a ReLU unit is zero. We first study the non-linearities created by the  $i$-th g-unit, for any $i\in\{0, ..., t^d-1\}$. The first layer ReLU units create non-linearities when $x_r=\frac{\pi^i_r}{t}$. 
The second layer ReLU units create  non-linearities where $\sum_{r=1}^d-\sigma(-x_r+\frac{\pi^i_r}{t})+\frac{1}{t}=0$. Thus, the non-linearities are where $\vect{x}\in\{\vect{x}:\sum_{r\in S}x_r-\frac{\pi^i_r}{t}+\frac{1}{t}=0, \emptyset\neq S\subseteq\{1, ...d\}\}$. Hence, the set of all non-linearirites of the neural network is $\{\vect{x}:\sum_{r\in S}x_r=\frac{1}{t}(\sum_{r\in S}\pi^i_r-1), \emptyset\neq S\subseteq\{1, ...d\}, 0\leq i\leq t^d-1\}$

Consider a cell created by first-layer non-linearities with it's maximum corner at $\frac{\vect{\pi}^i}{t}$. 

For any $S$ and $i$-th and $j$-th g-unit for $j\neq i$, observe that hyperplanes $\sum_{r\in S}x_r=\frac{1}{t}(\sum_{r\in S}\pi^i_r-1)$ and $\sum_{r\in S}x_r=\frac{1}{t}(\sum_{r\in S}\pi^j_r-1)$, $j\neq i$, are parallel. Specifically, they either define the same hyperplane if $\pi^j_r=\pi^i_r$, $\forall r\in S$ and, otherwise, they are at least $\frac{1}{t}$ apart. Thus, consider the uniform partitioning of the space into cells with width $\frac{1}{t}$, done in the construction of the neural network. We see that hyperplanes passing through the $i$-th cell are $\{\vect{x}:\sum_{r\in S}x_r-\frac{\pi^i_r}{t}+\frac{1}{t}=0, \emptyset\neq S\subseteq\{1, ...d\}\}$. These points of non-linearity partition each cell into linear pieces. We consider the $i$-th cell, and bound the error for each linear piece. Furthermore, if $|S|=1$, points of non-linearity overlap borders of the cell which are the same points of non-linearity of the first layer ReLU units. So we consider $\{\vect{x}:\sum_{r\in S}x_r-\frac{\pi^i_r}{t}+\frac{1}{t}=0, \emptyset  S\subseteq\{1, ...d\}, |S|\geq 2\}$

\textit{The case of $d=2$}.  There is only one non-lineararity hyperplane, $\sum_{r=1}^2x_r-\frac{\pi^i_r}{t}+\frac{1}{t}=0$, in each cell. Thus, each cell consists of two linear pieces,  $\mathcal{L}_1=\{\vect{x}: \sum_{r=1}^2x_r-\frac{\pi^i_r}{t}+\frac{1}{t}\geq 0\}$ and $\mathcal{L}_2=\{\vect{x}: \sum_{r=1}^2x_r-\frac{\pi^i_r}{t}+\frac{1}{t}\leq 0\}$. For a set $I$ of integers, we define $\vect{\pi}^{i, I}$ as the vector such that for $r\in I$, $\pi^{i, I}_r=\pi^i_r-1$ and $\pi^{i, I}_r=\pi^i_r$ otherwise. Thus, the set $C=\{\frac{\vect{\pi}^{i, I}}{t}, I\subseteq\{1, 2\}\}$, is the set of all the four corners of the $i$-th cell.
By memorization property of the construction, $\forall \vect{p} \in C$, $f(\vect{p})=\hat{f}(\vect{p})$.  We define $I_1=\{1\}$, $I_2=\{2\}$ and $I_3=\{1, 2\}$.

Since $\hat{f}$ is linear in $\mathcal{L}_i$, we can write $\hat{f}(\vect{x})=\vect{m}^i\vect{x}+b_i$ for $\vect{x}\in \mathcal{L}_i$. To bound $\normx{\vect{m}^1}$, we apply Lemma~\ref{lemma:derivative_linear} to points $\frac{\vect{\pi}^i}{t}$ and $\frac{\vect{\pi}^{i,I_1}}{t}$ and again to $\frac{\vect{\pi}^i}{t}$ and $\frac{\vect{\pi}^{i, I_2}}{t}$. Note that since memorization is exact at these points, the change in $\hat{f}$ is at most $\frac{\rho}{t}$ for every pair of points, and each pair are $\frac{1}{t}$ apart. So $\normx{\vect{m}^1}\leq 2\rho$.
Replacing $\vect{\pi}^i$ with $\vect{\pi}^{i, I_3}$ and repeating the same argument, we also bound $\normx{\vect{m}^2}$ by $2\rho$. This bounds the gradient for each linear piece. Thus, using Lemma~\ref{lemma:piecewise_bound} with $B=2\rho$, and observing that every pair of points in a cell are at most $\frac{2}{t}$ apart, we obtain
$$
|(\hat{f}(\vect{x})-\hat{f}(\vect{x}'))|\leq \frac{4\rho}{t}
$$
which proves the lemma in this case.

\begin{figure}
    \centering
    \includegraphics[width=0.7\columnwidth]{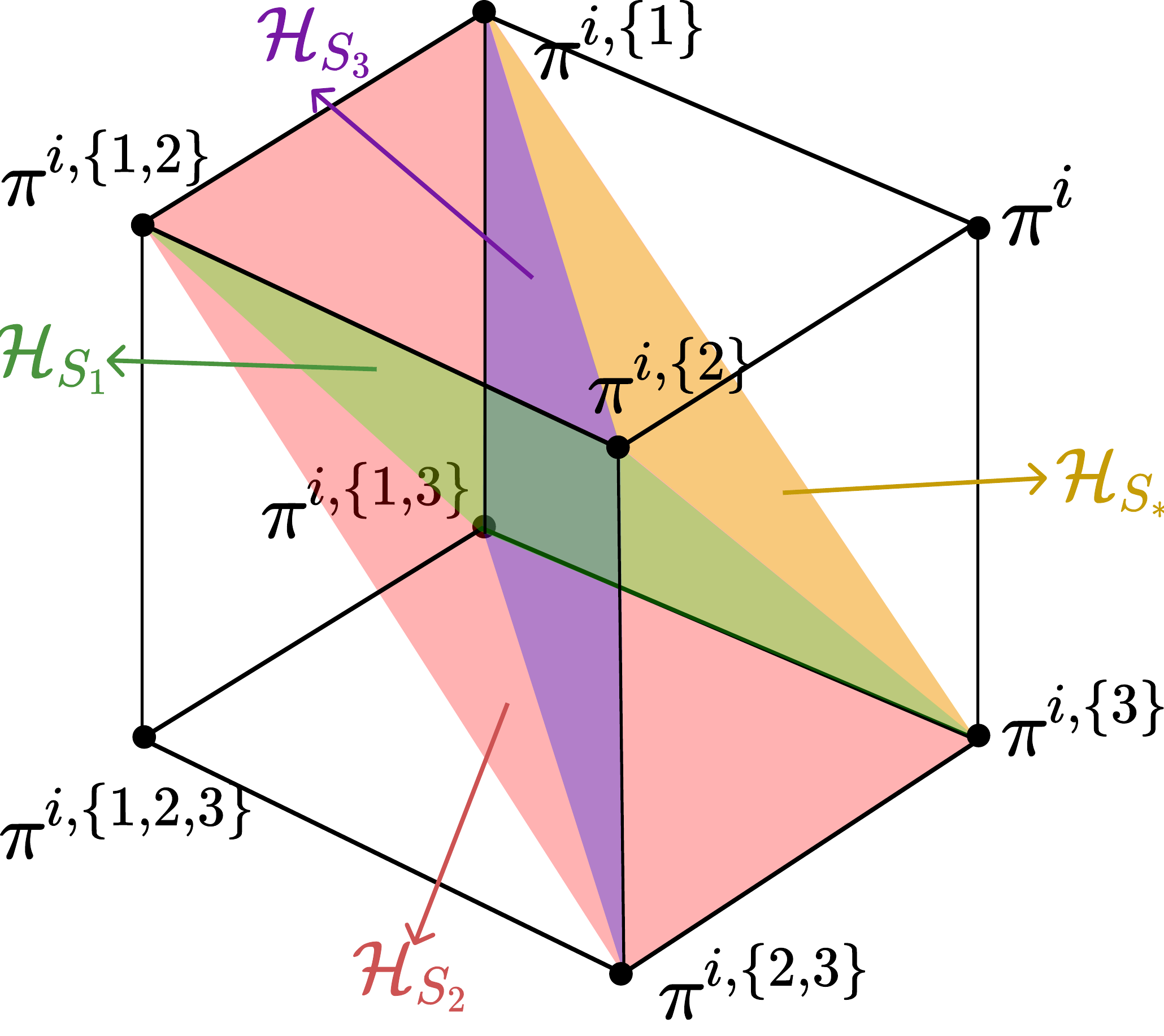}
    \caption{Non-linarities in 3 dimensions. Figure shows the input space portioned by hyperplanes corresponding to points of non-linearity.}
    \label{fig:nonlinearities}
\end{figure}

\textit{The case of $d=3$}. The argument is similar to $d=2$, but now there are four hyperplanes partitioning a cell, three corresponding to $|S|=2$ and one for $|S|=3$. Let $S_1=\{2, 3\}$, $S_2=\{1, 3\}$, $S_3=\{1,2\}$ and  $S_*=\{1, 2, 3\}$. For any such set $S$, we define $\mathcal{H}_S=\{\vect{x}: \sum_{r\in S}x_r-\frac{\pi^i_r}{t}+\frac{1}{t}= 0\}$, $\mathcal{H}_S^+$ as the set of points above $\mathcal{H}_S$ and $\mathcal{H}_S^-$ as the set of points below $\mathcal{H}_S$. Together, with the cell boundaries, the cell is partitioned into polytopes, where within each polytope $\hat{f}$ is a linear function. We bound the error of each of the linear pieces one by one. Note that the polytopes can be defined by which side of each hyperplane they fall on. Fig.~\ref{fig:nonlinearities} shows how the $i$-th cell is partitioned based on the hyperplanes discussed above. 

(1) The linear piece below all the hyperplanes, $\cap_S\mathcal{H}_S^-$, contains the points $\frac{\vect{\pi}^{i, I}}{t}$ when $I\subseteq\{1, ..., 3\}$ and $|I|\geq 2$. 
Thus, applying Lemma~\ref{lemma:piecewise_bound} three times we bound the derivative in each direction by $\rho$.

(2) For any $j$, $1\leq j\leq 3$, let $\mathcal{S}=\{S_*\}\cup\{S_z: 1\leq z\leq 3, z\neq j\}$ and consider the polytope where $C=(\cap_{S\in\mathcal{S}}\mathcal{H}_S^-)\cap \mathcal{H}_{S_j}^+$.
Note that the derivative of function in this linear piece w.r.t., $x_j$ is the same as case (1), because $\mathcal{H}_{S_j}$ does not depend on $x_j$. To bound the derivative w.r.t $x_z$ for $z\neq j$, without loss of generality assume $j=1$ and observe that
$\frac{\pi^{i, I}}{t}\in C$ for $I=\{1, 3\}$, $\{1, 2\}$ and $\{1\}$. So applying Lemma~\ref{lemma:piecewise_bound} twice bounds the derivative w.r.t $x_2$ and $x_3$ by $\rho$. 

(3) For any $j$, $1\leq j\leq 3$, let $\mathcal{S}=\{S_z: 1\leq z\leq 3, z\neq j\}$ and consider the polytope where $C=(\cap_{S\in\{S_*, S_j\}}\mathcal{H}_S^-)\cap (\cap_{S\in\mathcal{S}}\mathcal{H}_S^+)$.
Without loss of generality, assume $j=1$. 
Now, derivative w.r.t $x_2$ is the same as when $(\cap_{S\in\{S_*, S_1, S_2\}}\mathcal{H}_S^-)\cap (\mathcal{H}_{S_2}^+)$ and 
and derivative w.r.t $x_3$ is the same as when $C_3=(\cap_{S\in\{S_*, S_1, S_3\}}\mathcal{H}_S^-)\cap (\mathcal{H}_{S_2}^+)$.

Hence, we only need to bound derivative w.r.t $x_1$. Consider some points $\vect{p}$ on the hyperplane $\mathcal{H}_{S_3}$ 
and take the derivative in the direction of $\vect{u}=(1/\sqrt{2}, -1/\sqrt{2}, 0)$, written as $D_{\vect{u}}^{\vect{p}}$. Note that $D_{\vect{u}}^{\vect{p}}$ is defined because $\hat{f}$ is a linear at $\vect{p}$ in the direction of $\vect{u}$, since $\vect{p}+\varepsilon \vect{u}\in \mathcal{H}_S$ for small enough positive $\varepsilon$. Furthermore, since $\vect{p}\in C$ and $\vect{p}\in C_3$ and that both $C$ and $C_3$ are linear pieces, for any point $\vect{p}'\in C\cup C_3$, $D_{\vect{u}}^{\vect{p}'}=D_{\vect{u}}^{\vect{p}}$. This shows that the directional derivative in the direction of $\vect{u}$ is the same for all points in both $C$ and $C_3$. Thus, bounding $D_{\vect{u}}^{\vect{p}}$ with gradient of $\hat{f}$ in $C_3$, we get $D_{\vect{u}}^{\vect{p}}\leq 3\rho$. At the same time, for points $\vect{p}'$ in $C$ we can write $|D_{\vect{u}}^{\vect{p}'}|=|\nabla_{\vect{p}'}\cdot \vect{u} |\leq \frac{1}{\sqrt{2}}|\partial_{x_1}-\partial_{x_2}|$. Therefore, $\frac{1}{\sqrt{2}}||\partial_{x_1}|-|\partial_{x_2}||\leq |\partial_{x_1}-\partial_{x_2}|\leq 3\rho$. So that $|\partial_{x_1}|\leq 3\rho+|\partial_{x_2}|$. Given that $|\partial_{x_2}|\leq \rho$, we have that derivative w.r.t $x_1$ is at most $4\rho$. 

(4) Let $\mathcal{S}=\{1, 2 , 3\}$ and consider the polytope $(\cap_{S\in\mathcal{S}}\mathcal{H}_S^+)\cap \mathcal{H}_{S_*}^-$.  The derivative w.r.t $x_1$ is the same as when $(\cap_{S\in\{S_*, S_1\}}\mathcal{H}_S^-)\cap (\cap_{S\in\{S_2, S_3\}}\mathcal{H}_S^+)$, and the derivative w.r.t $x_2$ and $x_3$ can similarly be calculated based on previously bounded derivatives.

(5) Finally, note that when $\vect{x}\in\mathcal{H}_{S_*}^+$, we have that $\vect{x}\in\mathcal{H}_{S_i}^+$ for $1\leq i\leq 3$ and thus all cases are considered. In this final case, the polytope contains the points $\frac{\vect{\pi}^{i, I}}{t}$ when $I\subseteq\{1, ..., 3\}$ and $|I|\leq 2$, so the gradient is bounded by $3\rho$ by applying Lemma~\ref{lemma:derivative_linear} three times. 

Putting all cases together, the magnitude of the gradient is at most $12\rho$. Thus, using Lemma~\ref{lemma:piecewise_bound} with $B=12\rho$ and observing that every pair of points in a cell are at most $\frac{3}{t}$ apart, we obtain
$$
|(\hat{f}(\vect{x})-\hat{f}(\vect{x}'))|\leq \frac{36\rho}{t}
$$
which proves the lemma in this case.

\qed

\subsubsection{\textbf{Proof of Bounded Error Lemma~\ref{prop:error}}}
Lemma~\ref{prop:error} (b) directly follows from   Lemma~\ref{prop:change}~(c). For $\mathbf{x}\in C^*_i$ for any $i$,
\begin{align*}
 |\hat{f}(\vect{x})-f(\vect{x})|&=|\hat{f}(\vect{x})-\hat{f}(\frac{\vect{\pi}^i}{t})-(f(\vect{x})-\hat{f}(\frac{\vect{\pi}^i}{t}))|\\
                  &\leq |\hat{f}(\vect{x})-\hat{f}(\frac{\vect{\pi}^i}{t})| + |f(\vect{x})-\hat{f}(\frac{\vect{\pi}^i}{t}))|\\
                  &= |\hat{f}(\vect{x})-\hat{f}(\frac{\vect{\pi}^i}{t})| + |f(\vect{x})-f(\frac{\vect{\pi}^i}{t}))|\\
                  &\leq 36\frac{\rho d}{t} + \frac{\rho d}{t}\\
\end{align*}

Next, we prove Lemma~\ref{prop:error} (a). 

By Lemma.~\ref{prop:change}~(a) and~(b), $\hat{f}$ is either constant or non-constant. We bound the error separately for each part of the space.

In the constant region, that is $\vect{x}\in C_i$ for any $i$, by Lemma.~\ref{prop:change}~(a), $\hat{f}(\vect{x})=f(\frac{\vect{\pi}^i}{t})$, so
\begin{align*}
 |\hat{f}(\vect{x})-f(\vect{x})|&=|f(\frac{\vect{\pi}^i}{t})-f(\vect{x})|\\
                  &\leq \rho \normx{\vect{x}-\frac{\vect{\pi}^i}{t}}\\
                  &\leq \frac{\rho d}{t}
\end{align*}

Next, consider an $\vect{x}\in C'_i$ for any $i$ and let $\vect{x}^*$ be the closest point in $C_i$ to $\vect{x}$. We have

\begin{align*}
 |\hat{f}(\vect{x})-f(\vect{x})|&=|\hat{f}(\vect{x})-\hat{f}(\vect{x}^*)-(f(\vect{x})-\hat{f}(\vect{x}^*))|\\
                  &\leq |\hat{f}(\vect{x})-\hat{f}(\vect{x}^*)| + |f(\vect{x})-\hat{f}(\vect{x}^*)|\\
\end{align*}

First, consider $|f(\vect{x})-\hat{f}(\vect{x}^*)|$. We have 
\begin{align*}
 |f(\vect{x})-\hat{f}(\vect{x}^*)|&=|f(\vect{x})-f(\vect{x}^*)-(\hat{f}(\vect{x}^*)-f(\vect{x}^*))|\\
                  &\leq |f(\vect{x})-f(\vect{x}^*)| + |\hat{f}(\vect{x}^*)-f(\vect{x}^*)|\\
                  &\leq \frac{2\rho d}{t}
\end{align*}

Moreover, $|\hat{f}(\vect{x})-\hat{f}(\vect{x}^*)|\leq \frac{kd^3\rho2^{d-1}}{t}$ by Lemma~\ref{prop:change}~(b), so that 
\begin{align*}
 |f(\vect{x})-\hat{f}(\vect{x})|&\leq\frac{kd^3\rho 2^{d-1}}{t}+\frac{2\rho d}{t}\\
                  &=\frac{\rho d}{t}(kd^2 2^{d-1}+2)
\end{align*}

Thus, the 1-norm error is

\begin{align*}
    \int_{\vect{q}}|f(\vect{q})-\hat{f}(\vect{q})|=&\int_{\vect{q}\in \cup_iC_i}|f(\vect{q})-\hat{f}(\vect{q})|+\int_{\vect{q}\in \cup_iC'_i}|f(\vect{q})-\hat{f}(\vect{q})|\\
    \leq&\int_{\vect{q}\in \cup_iC_i}\frac{\rho d}{t}+\int_{\vect{q}\in \cup_iC'_i}\frac{\rho d}{t}(kd^2 2^{d-1}+2)\\
    =&\frac{\rho d}{t}\int_{\vect{q}\in \cup_iC_i}+\frac{\rho d}{t}(kd^2 2^{d-1}+2)\int_{\vect{q}\in \cup_iC'_i}\\
    =&\frac{\rho d}{t}(1-\frac{1}{M})^d+\frac{\rho d}{t}(kd^2 2^{d-1}+2)(1-(1-\frac{1}{M})^d)\\
    =&\frac{\rho d}{t}((1-\frac{1}{M})^d+(kd^2 2^{d-1}+2)(1-(1-\frac{1}{M})^d))
\end{align*}
Finally, we set $M$ so that 
\begin{align*}
    kd^2 2^{d-1}(1-(1-\frac{1}{M})^d)=1\\
\end{align*}
and thus
$$
M=\frac{1}{1-(1-\frac{1}{kd^2 2^{d-1}})^\frac{1}{d}}
$$

Which yields $\int_{\vect{q}}|f(\vect{q})-\hat{f}(\vect{q})|\leq \frac{\rho d}{t}((1-\frac{1}{M})^d+1+2(1-(1-\frac{1}{M})^d))\leq \frac{3\rho d}{t}$.

\qed

\subsubsection{\textbf{Proof of Space/Time Complexity Lemma~\ref{lemma:space}}}
Number of operations for a forward pass is proportional to the number of neural network parameters, which is $O(kd)$. Next we study space complexity.

Note that we only need to store $a_i$, for $1\leq i\leq k$, $b$ and $M$. Assuming a number $C$ can be stored in $O(\log C)$ number of bits and using Lemma\ref{lemma:a_mag} to bound the magnitude of $a_i$, the total space consumption is $k\log(2^{d-1}d\rho)+\log(M)+kd+\log(f(0))=O(kd(\log\rho)+\log M)$.

To study $\log M$, Note that $kd^2 2^{d-1}\leq(d^2t)^d$ for $d\geq 2$. So we study
\begin{align*}
\frac{1}{1-(1-(\frac{1}{td^2})^d)^{1/d}}&=\frac{1}{1-(\frac{(td^2)^d-1}{(td^2)^d})^{1/d}}\\
                                        &=\frac{1}{1-\frac{((td^2)^d-1)^{1/d}}{td^2}}\\
                                        &=\frac{td^2}{td^2-((td^2)^d-1)^{1/d}}    
\end{align*}
 So 
$$\log M\leq\log(td^2)+\log(\frac{1}{td^2-((td^2)^d-1)^{1/d}})$$

Next, for ease of notation we consider, $\frac{1}{x-(x^d-1)^{1/d}}$ for $x=td^2$. Assume $d = 2^s$ for an integer $s$ (or otherwise increase $d$ by a constant factor so that it can be written as a power of 2). By repeated multiplication of numerator and denominator we have 

\begin{align*}
    \frac{1}{x-(x^d-1)^{1/d}}=&\frac{x+(x^d-1)^{1/d}}{(x+(x^d-1)^{1/d})(x-(x^d-1)^{1/d})}\\
    =&\frac{x+(x^d-1)^{1/d}}{x^2-(x^d-1)^{2/d}}\\
    =&\frac{(x^2+(x^d-1)^{2/d})(x+(x^d-1)^{1/d})}{(x^2+(x^d-1)^{2/d})(x^2-(x^d-1)^{2/d})}\\
    =&\frac{(x^2+(x^d-1)^{2/d})(x+(x^d-1)^{1/d})}{x^4-(x^d-1)^{4/d}}\\
    &\vdotswithin{=} \\
    =&\frac{\Pi_{i = 0}^{s-1}(x^{2^i}+(x^d-1)^{2{^i}/d})}{1}
\end{align*}

Taking the log, we obtain 

\begin{align*}
    \log(\frac{1}{x-(x^d-1)^{1/d}})=&\sum_{i = 0}^{s-1}\log(x^{2^i}+(x^{d}-1)^{2{^i}/d})\\
    \leq&\sum_{i = 0}^{s-1}\log(2x^{2^i})\\
    =&\sum_{i = 0}^{s-1}(2^i\log x+\log 2)\\
    =&\log x\sum_{i = 0}^{s-1}(2^i)+s\log 2\\
    \leq&2^s\log x+s\log2\\
    =&\log(d)\log(2)+\log(d^2t)d\\
\end{align*}

So $\log M \leq O(d\log dt)=O(d\log d+\log k)$. Thus, the total size is $O(kd\log \rho+d\log d+\log k)=\Tilde{O}(kd)$. \qed

\subsection{Proof of Theorem~\ref{cor:sum_count_err}}\label{sec:proof:sum_count_err}
Consider a query with \texttt{COUNT} aggregation function. Define the indicator function $h$ as 
\begin{align*}
    h_{\vect{c}, \vect{r}}^C(\vect{p})=
    \begin{cases}
    1 & \text{if}\;\;\forall i,  c_i\leq p_i < c_i+r_i\\
    0              & \text{otherwise}.
\end{cases}
\end{align*}
We can write $f_D(\vect{c}, \vect{r})=\sum_{\vect{p}\in D}h_{\vect{c}, \vect{r}}^C(\vect{p})$, and that 
$$f_\chi(\vect{c}, \vect{r})=n\mathds{E}_{\vect{p}\sim \chi}[h_{\vect{c}, \vect{r}}(\vect{p})]$$
Thus, to study the error $\frac{1}{n}|f_D(\vect{c}, \vect{r})-f_\chi(\vect{c}, \vect{r})|$, we consider
\begin{align*}
\sup_{\vect{c}, \vect{r}}|\frac{1}{n}\sum_{\vect{p}\in D}h_{\vect{c}, \vect{r}}^C(\vect{p})-\mathds{E}_{\vect{p}\sim \chi}[h_{\vect{c}, \vect{r}}^C(\vect{p})]|.
\end{align*}
We define the class of function $\mathcal{H}^C=\{h_{\vect{c}, \vect{r}}^C, \forall \vect{c}, \vect{r}\}$ and rewrite the above expression as 
\begin{align}\label{eq:bound2}
\sup_{h\in\mathcal{H^C}}|\frac{1}{n}\sum_{\vect{p}\in D}h(\vect{p})-\mathds{E}_{\vect{p}\sim \chi}[h(\vect{p})]|
\end{align}

Now, we can bound the above error in terms of properties of $\mathcal{H^C}$. Observe that we can repeat the procedure for \texttt{SUM} aggregation function. Assume we would like to take the sum of the attribute at location $*$, and define

\begin{align*}
    h_{\vect{c}, \vect{r}}^S(\vect{p})=
    \begin{cases}
    p_* & \text{if}\;\;\forall i,  c_i\leq p_i < c_i+r_i\\
    0              & \text{otherwise}
\end{cases}
\end{align*}

Observe that $f_D(\vect{c}, \vect{r})=\sum_{\vect{p}\in D}h_{\vect{c}, \vect{r}}^S(\vect{p})$, and define $\mathcal{H}^S=\{h_{\vect{c}, \vect{r}}^S, \forall \vect{c}, \vect{r}\}$. Thus, we can similarly write the error for the \texttt{SUM} aggregation function as in Eq.~\ref{eq:bound2} by replacing $\mathcal{H}^C$ with $\mathcal{H}^S$. Note that $\mathcal{H}^C$ and $\mathcal{H}^S$ depend both on the aggregation function and the range predicates.

Next, we present some definition and results from VC theory that allows us to provide the required bounds.

\begin{definition}[Pseudo-shattering \cite{anthony1999neural}]\label{def:shatter} Let $I$ be a countable subset of $[0, 1]^d$. $I$ is said to be pseudo-shattered by $\mathcal{H}$ if for some function $g:I\rightarrow \mathbb{R}$, for every $J \subseteq I$ there exists $h_J \in \mathcal{H}$ such that
$$
h_J(\vect{x}) \leq g(\vect{x}) \quad \text { for } \vect{x} \in J, \quad h_J(\vect{x}) > g(\vect{x}) \quad \text { for } \vect{x} \in I \setminus J .
$$
\end{definition}

\begin{definition}[Pseudo-dimension \cite{anthony1999neural}] The pseudo-dimension of $\mathcal{H}$ is defined as $\operatorname{vc}(\mathcal{H})=\sup \{|I|: I$ is pseudo-shattered by $\mathcal{H}\}$.
\end{definition}
\begin{theorem}[VC-Theorem \cite{anthony1999neural}]\label{thm:vc}
For a class of functions, $\mathcal{H}$, where $h:\mathbb{R}^d\rightarrow [0, 1]$ for all $h\in\mathcal{H}$, and a set $D$ consisting of $n$ i.i.d samples from a distribution $\chi$, then 
$$
\mathds{P}\left[{\sup}_{h\in \mathcal{H}}\abs{\frac{1}{n}{\textstyle\sum}_{\vect{p}\in D} h(\vect{p})-\mathop{\mathds{E}}_{\vect{p}\sim\chi}h(\vect{p})}\geq \varepsilon\right]\leq 8ed\left(\nicefrac{32e}{\varepsilon}\right)^de^{-\frac{\varepsilon^2n}{32}}
$$
Where $d=\operatorname{vc}(\mathcal{H})$.
\end{theorem}
We are interested in bounding Eq.~\ref{eq:bound2}, which can readily be done using the above VC-Theorem, after finding $\operatorname{vc}({\mathcal{H}})$. 
This is done in the following lemma.
\begin{lemma}\label{prop:q_func_vc}
For $\mathcal{H}^S$ and $\mathcal{H}^C$ defined as above, $\operatorname{vc}(\mathcal{H}^S)\leq 2d$ and $\operatorname{vc}(\mathcal{H}^C)\leq 2d$.
\end{lemma}
\noindent\textit{Proof}. We note that $\mathcal{H}^C$ is the class of axis-parallel rectangle classifiers, whose VC-dimension is well-known to be $2d$ \cite{shalev2014understanding}. Our proof below uses a similar but slightly more general argument to account for both $\mathcal{H}^C$ and $\mathcal{H}^S$.

We show that no set of size $2d+1$ can pseudo-shatter $\mathcal{H}^S$. Let $I=\{\vect{p}^1, ..., \vect{p}^{2d+1}\}$. First, note that if $p^i_{*}=0$ for some $i$, the set cannot be pseudo-shattered. To see this, consider $J_2=\{\vect{p}^i\}$ and $J_1=I\setminus J_2$. For any $h\in \mathcal{H}^S$, $h(\vect{p}^i)=0$. Now, for some $g$, we need to have that $h^{J_1}(\vect{p}^i)> g(\vect{p}^i)$ and that $h^{J_2}(\vect{p}^i)\leq g(\vect{p}^i)$. Implying $h^{J_1}(\vect{p}^i)>h^{J_2}(\vect{p}^i)$, which is a contradiction because $h^{J_1}(\vect{p}^i)=h^{J_2}(\vect{p}^i)=0$.


Define $S=\{\vect{p}:\exists i, p_i=\min_{p'\in I} p'_i\text{ or }p_i=\max_{p'\in I} p'_i \}$. 
Note that $1\leq |S|\leq 2d$. For the purpose of contradiction, assume that there exists some $g$ that satisfies the conditions of Def.~\ref{def:shatter}. Specifically, that there exists some $g$ such that conditions are satisfied for $J_1 = S$ and $J_2=I\setminus S$ simultaneously. 

Note that by definition, $h(\vect{p})$ is either zero or $p_{*}$ for $h\in \mathcal{H}^S$. Since, $|S|\leq 2d$, $|J_2\cap I|\geq 1$ so let $\vect{p}'\in J_2\cap I$. We have that $h^{J_1}(\vect{p}')> g(\vect{p}')$, and that  $h^{J_2}(\vect{p}')\leq g(\vect{p}')$, so that $h^{J_1}(\vect{p}')>h^{J_2}(\vect{p}')$. Since $0<p'_{*}\leq 1$ (and specifically $p'_{*}$ is positive), the only solution to the inequality is $h^{J_1}(\vect{p}')=p'_{*}$ and $h^{J_2}(\vect{p}')=0$. A similar argument for all $\vect{p}\in J_1$ shows that $h^{J_1}(\vect{p})=0$ and $h^{J_2}(\vect{p})=p_{*}$. Now since $h^{J_2}(\vect{p})=p_{*}$ is true $\forall \vect{p}\in J_1$, it must be true that $h^{J_2}(\vect{p}')=p'_{*}$ (this is because if a range predicate contains all the points in $S$ it must contain all the points in $I$). However. this contradicts $h^{J_2}(\vect{p}')=0$, which completes the proof for $\operatorname{vc}(\mathcal{H}^S)$. To bound $\operatorname{vc}(\mathcal{H}^C)$, repeat the same argument with $p_*=1$.
\qed


Theorem~\ref{cor:sum_count_err} follows directly from the above lemma and the VC-theorem.\qed

\subsection{Proof of Lemma~\ref{lemma:avg_bound_error}}\label{sec:proof:avg_bound_error}
Let $f^{C}_D(\vect{q})=f^{C}_\chi(\vect{q})+\varepsilon_c^{\vect{q}}$ and $f^{S}_D(\vect{q})=f^{S}_\chi(\vect{q})+\varepsilon_s^{\vect{q}}$. Then, for any $\vect{q}$,

\begin{align*}
    |\bar{f}^{A}_\chi(\vect{q})-f^{A}_D(\vect{q})|&=|\frac{f^{S}_\chi(\vect{q})}{f^{C}_\chi(\vect{q})}-\frac{f^{S}_\chi(\vect{q})+\varepsilon_s^{\vect{q}}}{f^{C}_\chi(\vect{q})+\varepsilon_c^{\vect{q}}}|\\
    &=|\frac{\varepsilon_c^{\vect{q}} f^{S}_{\chi}(\vect{q})-\varepsilon_s^{\vect{q}} f^{C}_{\chi}(\vect{q})}{f^{C}_{\chi}(\vect{q})(f^{C}_{\chi}(\vect{q})+\varepsilon_c^{\vect{q}})}|\\
    &\leq|\frac{f^{S}_{\chi}(\vect{q})}{f^{C}_{\chi}(\vect{q})}||\frac{\varepsilon_s^{\vect{q}}}{f^{C}_{\chi}(\vect{q})+\varepsilon_c^{\vect{q}}}|+|\frac{\varepsilon_s^{\vect{q}}}{(f^{C}_{\chi}(\vect{q})+\varepsilon_c^{\vect{q}})}|\\
    &=|\frac{f^{S}_{\chi}(\vect{q})}{f^{C}_{\chi}(\vect{q})}||\frac{\varepsilon_s^{\vect{q}}}{f^{C}_{D}(\vect{q})}|+|\frac{\varepsilon_s^{\vect{q}}}{f^{C}_{D}(\vect{q})}|\\
\end{align*}

For any $\varepsilon$, by Theorem~\ref{cor:sum_count_err} and union bound, $\mathds{P}[\sup_{\vect{q}}|\varepsilon_c^{\vect{q}}| \geq \varepsilon\text{ or }\sup_{\vect{q}}|\varepsilon_c^{\vect{q}}|\geq \varepsilon]\leq 16ed\left(\nicefrac{32e}{\varepsilon}\right)^de^{-\frac{\varepsilon^2n}{32}}$. Define $\mathcal{Q}_\xi=\{\vect{q}, f^{C}_{\chi}(\vect{q})\geq \xi\}$. Note that the event $A=\{\forall \vect{q}\in \mathcal{Q}_\xi, |\varepsilon_c^{\vect{q}}|< \varepsilon$ and $|\varepsilon_c^{\vect{q}}|< \varepsilon\}$ implies  $\forall \vect{q}\in\mathcal{Q}_\xi, f^{C}_{D}(\vect{q})> \xi-\varepsilon$ and thus the event $\{\forall \vect{q}\in\mathcal{Q}_\xi, |\frac{\varepsilon_c^{\vect{q}}}{f^{C}_{D}(\vect{q})}|< \frac{\epsilon}{\xi-\epsilon}$ and $|\frac{\varepsilon_s^{\vect{q}}}{f^{C}_{D}(\vect{q})}|< \frac{\epsilon}{\xi-\epsilon}\}$. Therefore, event $A$ implies the event $B=\{\forall \vect{q}\in\mathcal{Q}_\xi, |\frac{f^{S}_{\chi}(\vect{q})}{f^{C}_{\chi}(\vect{q})}||\frac{\varepsilon_c^{\vect{q}}}{f^{C}_{D}(\vect{q})}|+|\frac{\varepsilon_s^{\vect{q}}}{f^{C}_{D}(\vect{q})}|< |\frac{f^{S}_{\chi}(\vect{q})}{f^{C}_{\chi}(\vect{q})}|\frac{\varepsilon}{\xi-\varepsilon}+\frac{\varepsilon}{\xi-\varepsilon}\}$. So $\mathds{P}[A]\leq \mathds{P}[B]$. Considering the complement of events $A$ and $B$, we obtain
\begin{align*}
        \mathds{P}[\exists{\vect{q}\in \mathcal{Q}_\xi}:|\frac{f^{S}_{\chi}(\vect{q})}{f^{C}_{\chi}(\vect{q})}||\frac{\varepsilon_c^{\vect{q}}}{f^{C}_{D}(\vect{q})}|+|\frac{\varepsilon_s^{\vect{q}}}{f^{C}_{D}(\vect{q})}|\geq |&\frac{f^{S}_{\chi}(\vect{q})}{f^{C}_{\chi}(\vect{q})}|\frac{\varepsilon}{\xi-\varepsilon}+\frac{\varepsilon}{\xi-\varepsilon}]\\\leq &16ed\left(\nicefrac{32e}{\varepsilon}\right)^de^{-\frac{\varepsilon^2n}{32}}.
\end{align*}
Therefore, 
\begin{align*}
    \mathds{P}[\sup_{\vect{q}\in \mathcal{Q}_\xi}\frac{|\bar{f}^{A}_\chi(\vect{q})-f^{A}_D(\vect{q})|}{|\bar{f}^{A}_\chi(\vect{q})|+1}&\geq \frac{\varepsilon}{\xi-\varepsilon}]\leq 16ed\left(\nicefrac{32e}{\varepsilon}\right)^de^{-\frac{\varepsilon^2n}{32}}\\
    \mathds{P}[\sup_{\vect{q}\in \mathcal{Q}_\xi}\frac{|\bar{f}^{A}_\chi(\vect{q})-f^{A}_D(\vect{q})|}{|\bar{f}^{A}_\chi(\vect{q})|+1}&\geq \varepsilon]\leq 16ed\left(32e\frac{1+\varepsilon}{\xi\varepsilon}\right)^de^{-\frac{(\xi\varepsilon)^2n}{(1+\varepsilon)^232}}.
\end{align*}
\qed
\subsection{Utilizing Construction in Practice}\label{exp:construct}
We study the benefits of using the theoretical construct of Sec.~\ref{sec:theory:appx} in practice. We consider two variations. First, referred to as CS, we use the construct exactly as in Sec.~\ref{sec:theory:appx}. Second, referred to as CS+SGD, we consider the construct as an initialization for the SGD algorithm. That is, we first construct the neural network and further optimize its parameters using the SGD algorithm. This replaces line \ref{alg:sgd:init} of Alg.~\ref{alg:sgd} with calling Alg.~\ref{alg:construction} to initialize the parameters. 

\begin{figure}[t]
    \centering
    \includegraphics[width=\columnwidth]{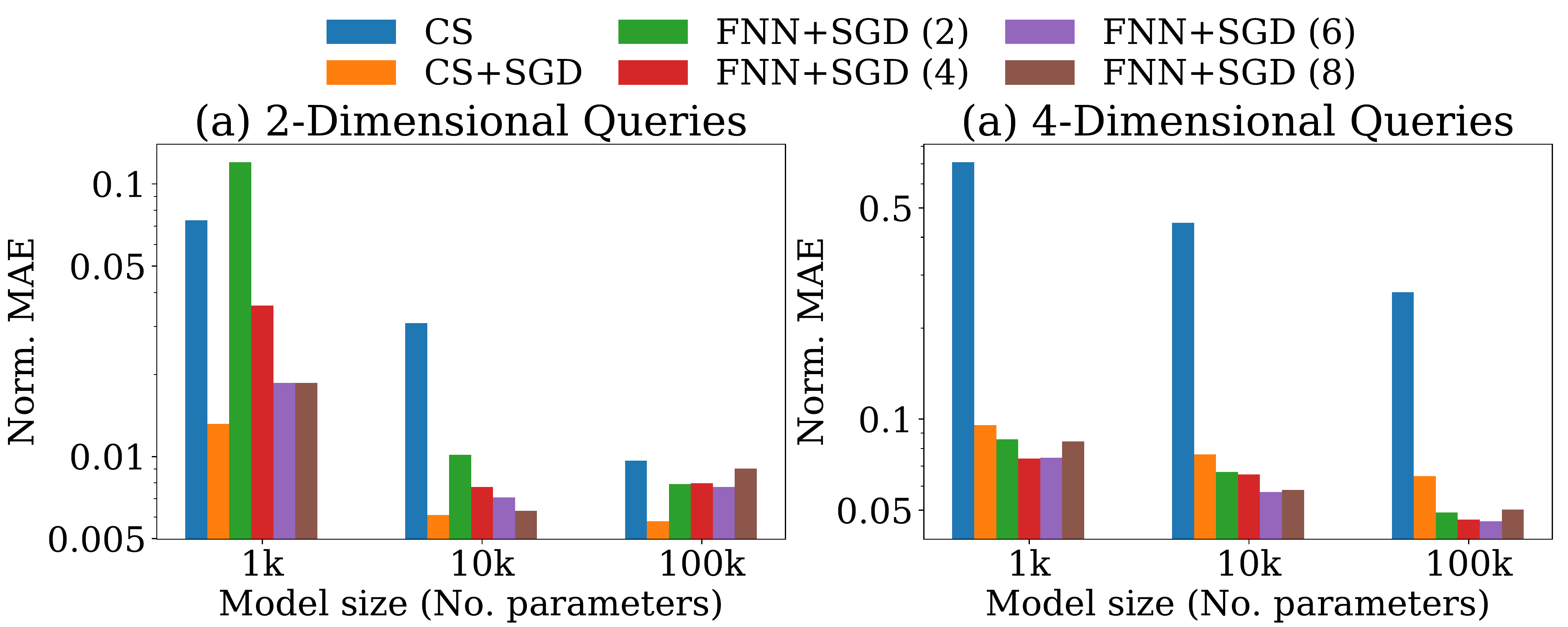}
    \caption{Construction vs. SGD.}
    \label{fig:const_vs_sgd}
\end{figure}

Fig.~\ref{fig:const_vs_sgd} shows how the above two algorithms compare with training fully connected neural networks with different depths. Lines labeled FNN+SGD ($x$) refer to a randomly initialized fully connected neural network (FNN) of depth $x$ trained with SGD. Number of parameters per model is fixed for each setting, so that as depth increases the width of the FNNs decreases. We consider 2 and 4 dimensional queries in this experiment. The 2-dimensional query asks average visit duration for the fixed range of 0.2. Thus, the query function only takes latitude and longitude as inputs, and outputs average visit duration. The 4-dimensional query is the usual query of average visit duration, where the query function takes minimum and maximum latitude and longitude as its 4 inputs, and outputs average visit duration. None of the algorithms use partitioning.

Fig.~\ref{fig:const_vs_sgd} shows that for the 2-dimensional query, CS+SGD performs better than all other architectures, while CS's accuracy is close to FNNs. However, for the 4-dimensional queries, CS is much worse than FNNs and although CS+SGD performs similar to FNNs, it is always outperformed by them. This shows that for low dimensional queries, CS can be useful in practice as an initialization for SGD.

\end{document}